\documentclass[11pt,a4paper]{article} 
\pdfoutput=1
\usepackage{style}
\usepackage{amsmath}
\usepackage{amssymb}
\usepackage{amsthm}
\usepackage{psfrag}
\usepackage{graphicx}
\usepackage{hyperref}
\usepackage{feynmp}
\usepackage{color}
\usepackage{bm}
\usepackage{multirow}

\newcommand{\be}{\begin{equation}}
\newcommand{\ee}{\end{equation}}
\newcommand{\beqa}{\begin{eqnarray}}
\newcommand{\eeqa}{\end{eqnarray}}
\newcommand{\bsm}{\begin{smallmatrix}}
\newcommand{\esm}{\end{smallmatrix}}

\newcommand\m{\mu}
\newcommand\p{{\bf p}}
\renewcommand\O{\Omega}

\newcommand\D{\Delta}
\newcommand\G{\Gamma}

\newcommand\s{\sigma}
\renewcommand\t{\theta}
\renewcommand\a{\alpha}
\renewcommand\b{\beta}

\newcommand{\T}{\Theta}

\newcommand{\ve}{\varepsilon}

\newcommand\z{{\bf z}}
\renewcommand\k{{\bf k}}
\newcommand\q{{\bf q}}
\newcommand{\e}{\eta}
\newcommand{\bea}{\begin{eqnarray}}
\newcommand{\eea}{\end{eqnarray}}
\newcommand{\nn}{\nonumber}

\def\d{\partial}
\newcommand{\bseq}{\begin{subequations}}
\newcommand{\eseq}{\end{subequations}}

\renewcommand{\tanh}{\mathop{\rm th}\nolimits}

\renewcommand{\ln}{\mathop{\rm ln}\nolimits}

\renewcommand{\S}{\mathcal{S}}
\renewcommand{\D}{\mathcal{D}}
\newcommand{\E}{\mathcal{E}}
\newcommand{\F}{\mathcal{F}}

\renewcommand{\P}{\bar P}
\newcommand{\hiddensubsection}[1]{
    \stepcounter{subsection}
    \subsection*{\Alph{section}.\arabic{subsection}\hspace{1em}{#1}}
}

\relax

\title{Time-Sliced Perturbation Theory II: 
Baryon Acoustic Oscillations and Infrared Resummation}

\author[a]{Diego Blas\footnote{diego.blas@cern.ch}} 
\author[a]{Mathias Garny\footnote{mathias.garny@cern.ch}}
\author[b,c,d]{Mikhail M. Ivanov\footnote{mikhail.ivanov@cern.ch}}
\author[a,b,d]{ Sergey Sibiryakov\footnote{sergey.sibiryakov@cern.ch}}

\affiliation[a]{Theoretical Physics Department, 
CERN, CH-1211 Gen\`eve 23, Switzerland}
\affiliation[b]{FSB/IPHYS/LPPC, \'Ecole Polytechnique F\'ed\'erale de Lausanne, \normalsize\it CH-1015, Lausanne, Switzerland}
\affiliation[c]{Faculty of Physics, Moscow State University,\normalsize \it Vorobjevy Gory, 119991 Moscow, Russia}
\affiliation[d]{Institute for Nuclear Research of the
Russian Academy of Sciences, \\ 
\normalsize \it  60th October Anniversary Prospect, 7a, 117312
Moscow, Russia}

\abstract{We use
\emph{time-sliced perturbation theory} (TSPT) to give an accurate
description of the infrared non-linear effects affecting the baryonic
acoustic oscillations (BAO) present in the distribution of matter at
very large scales. 
In TSPT this can be done via a systematic resummation that has a
simple diagrammatic representation and does not involve uncontrollable
approximations. 
We discuss the power counting rules and derive explicit expressions
for the resummed
matter power spectrum up to next-to leading order and the bispectrum
at the leading order.
The two-point correlation function agrees well with $N$-body data at
BAO scales. The systematic approach also allows to reliably assess
the shift of the baryon acoustic peak due to non-linear effects.
}

\begin{document}

\begin{flushright}
CERN-TH-2016-059\\
INR-TH-2016-009
\end{flushright}

\maketitle

\section{Introduction}

The imprint of the baryonic acoustic oscillations (BAO) on the distribution of matter at large scales and at different redshifts
 gives precise information about the expansion history and constituents of the Universe \cite{Eisenstein:1998tu, Eisenstein:2005su, Percival:2009xn, Anderson:2013zyy, Delubac:2014aqe}.
In the near future galaxy surveys will measure the two-point correlation function at BAO scales with (sub-)percent precision.
Moreover, the baryon acoustic feature has also been detected in the three-point function in BOSS data \cite{Slepian:2015hca}.
It is well-known that the shape and the position of the baryon
acoustic peak are affected by non-linear evolution. 
These non-linearities should be understood as accurate as possible
in order to exploit the full potential of 
future precision data.

Apart from numerical efforts, techniques based on cosmological 
perturbation theory have contributed to the qualitative and also quantitative
understanding of non-linear effects relevant for the BAO peak. Since the characteristic scale of the BAO (around $1/k_{osc} \equiv 110\ \mathrm{Mpc}/h$ in comoving coordinates) is much larger than the non-linear scale, one may a priori expect that perturbative methods can be applicable. Nevertheless, it has been observed
long ago that the leading non-linear correction computed in Standard
Perturbation Theory (SPT) \cite{Bernardeau:2001qr} fails to
reproduce the behavior seen in $N$-body simulations or data. 
The source for this disagreement is the effect of bulk flows on the
BAO
\cite{Eisenstein:2006nj,Eisenstein:2006nk,Crocce:2007dt,Baldauf:2015xfa}. In
fact, the displacement of long modes with wavenumber $q$ produces a sweeping effect (or bulk motion) on  shorter modes $k$ due to the non-linear coupling. In Eulerian perturbation theory, this sweeping
is responsible for an enhancement of non-linear interactions $\propto
k/q$ involving a long-wavelength mode $q$. For equal-time correlation
functions, the equivalence principle 
implies that this infrared (IR) enhancement largely 
cancels out when summing all perturbative contributions at a fixed
order in perturbation theory 
\cite{Jain:1995kx, Scoccimarro:1995if, Peloso:2013zw, Kehagias:2013yd, Blas:2013bpa, Sugiyama:2013gza}.
However, the cancellation is incomplete
if the matter power spectrum has a component that varies with a
characteristic scale 
$k_{osc} \lesssim q \ll k$ \cite{Senatore:2014via,Baldauf:2015xfa}.

Different frameworks have been proposed to deal with these bulk motions: on the one hand, a data-driven method is to first measure the large scale bulk motions and use them to \emph{reconstruct} 
the BAO feature
\cite{Eisenstein:2006nj,Noh:2009bb,Tassev:2012hu,Burden:2014cwa}. On
the other hand, precise determination of the cosmological parameters
may require more theoretical insight.
The
bulk motion can in principle be efficiently treated by moving from the
Eulerian to the Lagrangian picture
\cite{Matsubara:2007wj,Padmanabhan:2008dd}.  In particular, the linear
Lagrangian perturbation theory, or Zel'dovich approximation, gives a
rather accurate description of the BAO peak. However, it is unclear
how to improve systematically 
over the
Zel'dovich approximation within the Lagrangian picture. Finally, one
can stick to the Eulerian picture and try to identify the physical
contributions of the bulk flows with the idea to
resum the latter at all orders in perturbation theory
\cite{Crocce:2007dt,Senatore:2014via,Baldauf:2015xfa}.  

In this work, we develop a systematic approach to describe non-linear
effects on the BAO feature in equal-time correlation functions based 
on  
\emph{time-sliced perturbation theory} (TSPT) \cite{Blas:2015qsi}. The
latter is a proposal to describe the statistical properties of  the
large-scale structure based on the evolution of the distribution
function, as opposed to SPT where the individual field variables are
evolved. A major advantage of this description is that it eliminates 
spurious IR contributions from the beginning, and therefore
allows for a transparent description of the physical effects of bulk
motion on the BAO feature. 
On the other hand, TSPT is free from the difficulties of 
higher-order Lagrangian perturbation theory. Our main
result is a systematic technique 
to identify and resum enhanced infrared contributions affecting the
BAO feature. It admits a simple diagrammatic representation within TSPT
and allows to compute and assess higher-order corrections in a
systematic way. 

The main idea of TSPT is to disentangle time-evolution from
statistical ensemble averaging. In a first step, the probability
distribution ${\cal P}$ for the perturbations is evolved from the
initial time
to a finite redshift and expressed in terms of an expansion in
powers of the density- and velocity divergence field at this
redshift. In a 
second step, the statistical averages are computed perturbatively. The
latter step can be conveniently represented by a diagrammatic series,
where the quadratic cumulant represents a propagator, and the higher cumulants
--- $n$-point vertices $\Gamma_n$. In \cite{Blas:2015qsi} it has been
shown that these vertices are IR safe, 
\emph{i.e.} free from spurious enhancements $\propto k/q$ when any of
the wavenumbers become small. 

In order to identify enhanced contributions related to the BAO, we
split the initial power spectrum into a smooth component $P_s$ and an
oscillatory (`wiggly') contribution $P_w$.
Then the TSPT three-point vertex expanded for $q\ll k$ and to first order in $P_w$ is given by
\be
 \Gamma_3({\bm k}, {\bm q}, {\bm q'}) \to \delta^{(3)}({\bm k}+{\bm q}+{\bm q'})\frac{{\bm k}\cdot{\bm q}}{{q}^2}\left(\frac{P_w(|\k+\q|)-P_w(q)}{P_s(k)^2}\right)\;.
\ee
In the limit $q\to 0$ the difference of the 
two power spectra in the numerator goes to zero and cancels 
the $1/q$ enhancement from the vertex, as required by the equivalence principle.
However, as emphasized in \cite{Baldauf:2015xfa}, the Taylor expansion
of $P_w(|\k+\q|)$ becomes unreliable for $k_{osc} \lesssim q \ll k$. This means
that 
non-linear corrections to the correlation functions at scale $k$ receive
large corrections from IR modes $q$ within this range. In this work we
identify these 
contributions for all $\Gamma_n$ vertices, and establish a power 
counting scheme to compute corrections to the most enhanced terms. The
leading 
contributions to the oscillatory part of the power spectrum are given
by 
a set of `daisy' diagrams, and their resummation is represented
diagrammatically 
in the following form (see Sec.~\ref{sec:leading} for details), 
\begin{align}
\label{daisygraphsintro}
	&P^{IR\;res,LO}_{w}(\e;k)	~=~   
	\begin{fmffile}{w-ps_a}
	    \begin{gathered}
        \begin{fmfgraph*}(80,80)
        \fmfpen{thick}
        \fmfleft{l1}
        \fmfright{r1}
		\fmf{wiggly,label=$ $}{l1,r1}
	    \end{fmfgraph*}
	     \end{gathered}
	    \end{fmffile} ~+~
	\begin{fmffile}{w-vertex4_3_a}
	    \begin{gathered}
          \begin{fmfgraph*}(80,60)
        \fmfpen{thick}
        \fmfkeep{1loop}
        \fmfleft{l1,l2,l3}
        \fmfright{r1,r2,r3}
        \fmfv{d.sh=circle,d.filled=shaded,d.si=.15w,label=$ $,l.a=-90,l.d=.0w}{b1}
		\fmf{plain,label=$\bar \G^w_4$}{b1,l2}
		\fmf{plain,label=$ $}{b1,r2}
\fmf{phantom}{l1,u1}
		\fmf{phantom}{u2,r1}
		\fmf{phantom}{l3,v1}
		\fmf{phantom}{v2,r3}
	    \fmf{phantom,right=0.5,tension=0.01,l.side=left}{b1,u1}
	     \fmf{phantom,right=0.5,tension=2,label=$ $}{u1,u2}
	   \fmf{phantom,left=0.5,tension=0.01}{b1,u2}
	   	    \fmf{plain,left=0.5,tension=0.01,l.side=right}{b1,v1}
	     \fmf{plain,left=0.5,tension=2,label=$ $}{v1,v2}
	   \fmf{plain,right=0.5,tension=0.01}{b1,v2}
	    \end{fmfgraph*}
 	     \end{gathered}   
  \end{fmffile}\\
  \nonumber
  &~+~
	    \begin{fmffile}{w-vertex6}
     \begin{gathered}
        \begin{fmfgraph*}(80,60)
        \fmfpen{thick}
        \fmfkeep{2loop}
        \fmfleft{l1,l2,l3}
        \fmfright{r1,r2,r3}
        \fmfv{d.sh=circle,d.filled=shaded,d.si=.15w,label=$ $,l.a=155,l.d=.0w}{b1}
		\fmf{plain,label=$\bar \G^w_6$}{b1,l2}
		\fmf{plain,label=$ $}{b1,r2}
		\fmf{phantom}{l1,u1}
		\fmf{phantom}{u2,r1}
		\fmf{phantom}{l3,v1}
		\fmf{phantom}{v2,r3}
	    \fmf{plain,right=0.5,tension=0.01,l.side=left}{b1,u1}
	     \fmf{plain,right=0.5,tension=2,label=$ $}{u1,u2}
	   \fmf{plain,left=0.5,tension=0.01}{b1,u2}
	   	    \fmf{plain,left=0.5,tension=0.01,l.side=right}{b1,v1}
	     \fmf{plain,left=0.5,tension=2,label=$ $}{v1,v2}
	   \fmf{plain,right=0.5,tension=0.01}{b1,v2}  
	    \end{fmfgraph*}
	    	    	    \end{gathered}
	    	  \quad  	    + 
	    \end{fmffile}
	    \begin{fmffile}{w-vertex8}
	    \begin{gathered}
        \begin{fmfgraph*}(100,80)
        \fmfpen{thick}
        \fmfkeep{3loop}
        \fmfleft{l1,l2,l3}
        \fmfright{r1,r2,r3}
        \fmfv{d.sh=circle,d.filled=shaded,d.si=.12w,label=$ $,l.a=90,l.d=.00w}{b1}
		\fmf{plain,l.side=left,label=$\bar \G^w_8$}{l2,b1}
		\fmf{plain,l.side=left,label=$ $}{b1,r2}
		\fmf{phantom}{l1,u1}
		\fmf{phantom}{u2,r1}
		\fmf{phantom,tension=2}{l3,v1}
		\fmf{phantom,tension=2}{v1,r3}
	    \fmf{plain,right=0.55,tension=0.01,l.side=left}{b1,u1}
	   \fmf{plain,left=0.55,tension=0.01,l.side=right,label=$ $}{b1,u1}
	   	    \fmf{plain,right=0.55,tension=0.01,l.side=left,label=$ $}{b1,u2}
	   \fmf{plain,left=0.55,tension=0.01}{b1,u2}
	     \fmf{plain,right=0.7,tension=0.00001,label=$ $}{b1,v1}
	   \fmf{plain,left=0.7,tension=0.00001}{b1,v1}
	    \end{fmfgraph*}
	    	    	    \end{gathered}
	    \end{fmffile}
	  ~+~ \begin{fmffile}{w-vertex10}
	    \begin{gathered}
        \begin{fmfgraph*}(100,80)
        \fmfpen{thick}
        \fmfkeep{3loop}
        \fmfleft{l1,l2,l3}
        \fmfright{r1,r2,r3}
        \fmfv{d.sh=circle,d.filled=shaded,d.si=.14w,label=$ $,l.a=90,l.d=.0w}{b1}
		\fmf{plain,l.side=right,label=$ $,l.d=.15w}{b1,l2}
		\fmf{plain,l.side=left,label=$ $}{b1,r2}
		\fmf{phantom}{l1,u1}
		\fmf{phantom}{u2,r1}
		\fmf{phantom}{l3,v1}
		\fmf{phantom}{v2,r3}
	    \fmf{plain,right=0.55,tension=0.01,l.side=left}{b1,u1}
	   \fmf{plain,left=0.55,tension=0.01,l.side=right,label=$ $}{b1,u1}
	   	    \fmf{plain,right=0.55,tension=0.01,l.side=left,label=$ $}{b1,u2}
	   \fmf{plain,left=0.55,tension=0.01}{b1,u2}   
	      \fmf{plain,right=0.55,tension=0.01,l.side=left}{b1,v1}
	     \fmf{plain,left=0.55,tension=0.01,l.side=right,label=$ $}{b1,v1}
	   	 \fmf{plain,right=0.55,tension=0.01,l.side=left,label=$ $}{b1,v2}
	   	   \fmf{plain,left=0.55,tension=0.01}{b1,v2}
	    \end{fmfgraph*}
	    	    	    \end{gathered}
	    \end{fmffile}
	    +...
	    \end{align}
This diagrammatic representation is straightforwardly extended to
IR-enhanced contributions into higher correlation functions.

\begin{figure}
\begin{center}
  \includegraphics[width=0.6\textwidth]{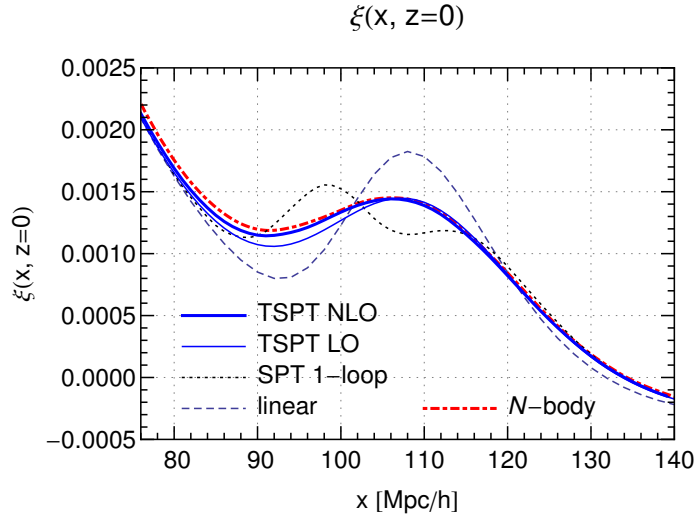}
\end{center}
\caption{\label{fig:summary}
Matter two-point correlation function $\xi(x)$ at redshift $z=0$. The
thin (thick) blue solid line shows the infrared-resummed result obtained
in TSPT at leading order (next-to-leading order).
At BAO scales the perturbative expansion within the TSPT framework
converges well and agrees with $N$-body results from large-scale
numerical simulations \cite{Kim:2011ab} (red dot-dashed line).
For comparison, we also show the linear (black dashed) and SPT 1-loop
(black dotted) results.
}
\end{figure}

At leading order (LO) for the power spectrum
the resummation reproduces the result found in \cite{Baldauf:2015xfa},
which consists in a broadening of the BAO peak.
Using TSPT we systematically compute next-to leading order (NLO)
corrections. Apart from being quantitatively important in order to
achieve good agreement with the  
results of large-scale $N$-body simulations at BAO scales, these NLO
contributions are crucial for a reliable determination of the shift of the 
BAO peak.
Furthermore, they are sensitive to the non-dipole corrections and
consequently capture deviations from the Zel'dovich approximation. 
This may also be helpful to assess potential biases introduced in the
data-driven technique of BAO reconstruction, where effectively the
Zel'dovich approximation 
is used for the backwards evolution. 

Our numerical results for the matter correlation function $\xi(x)$ at
redshift $z=0$ are summarized in Fig.\,\ref{fig:summary}, where we
compare the TSPT 
results at LO and NLO with $N$-body data, and also show the naive SPT
1-loop result for comparison. 
It is worth noting that the results are obtained from first principles
without adjusting any free parameters. We further find that the
difference between the NLO correlation function computed in TSPT and
that obtained within the Zel'dovich approximation is about 5\% in the
region of the BAO peak (see Sec.~\ref{sec:Nbody}). While small, this
difference is above the estimated uncertainty in the TSPT calculation
and the expected ultimate precision required to analyze the data of
future surveys. Importantly, the TSPT framework can be systematically
extended to 
 take into account the next-order contributions, as well
as the corrections due to the short-distance dynamics. 

The paper is organized as follows. 
In Sec.~\ref{sec:prel} we outline the basic formalism.
In Sec.~\ref{sec:PC} we describe how to identify the enhanced
IR-effects and establish the power counting rules.
The resummation of LO contributions is performed in
Sec.~\ref{sec:leading} for the power spectrum and bispectrum.
In Sec.~\ref{sec:hard} we extend the resummation to diagrams with
loops of short modes. Next-to-leading IR contributions are resummed 
in Sec.~\ref{sec:NLO} and 
a concise formula for the resummed correlation functions is derived.
In Sec.~\ref{sec:pract} we discuss the practical implementation of our
procedure, compare our result to $N$-body data and
discuss the BAO shift. Section~\ref{sec:conclusions} is devoted to
conclusions and discussion of future directions.
Appendices~\ref{app:rec}---\ref{app:shift} contain details of the
calculations, whereas in Appendices~\ref{app:SPT_TSPT}, \ref{app:ZA} 
we present an alternative derivation of IR resummation of
the power spectrum in SPT and
compare our results with the exact formulas in the Zel'dovich
approximation.

\section{TSPT and wiggly-smooth decomposition}\label{sec:prel}

In this section we first briefly remind the basic elements of the TSPT approach to large-scale structure formation
 (see \cite{Blas:2015qsi} for a detailed presentation)
and then discuss our strategy to identify IR enhanced effects on the BAO peak by decomposing the matter power spectrum
as well as the TSPT vertices into smooth and oscillatory components.
 
\subsection{Brief review of TSPT}\label{sec:tspt}

We are interested in the time evolution of correlation functions of
the overdensity field $\delta=(\rho-\bar \rho)/\bar\rho$
and the velocity divergence field $\Theta \propto \nabla \cdot {\bm
  u}$, whose time-evolution is
governed by the continuity and Euler equations for
the peculiar flow velocity ${\bm u}$,
\bseq
\begin{align}
 & \frac{\d\delta}{\d t} + \nabla \cdot [(1+\delta){\bm u}] = 0 \,, \\
 & \frac{\d{\bm u}}{\d t} + {\cal H}{\bm u} + ({\bm u}\cdot \nabla) {\bm u} =  -\nabla\Phi \,,
\end{align}
\eseq
where $\nabla^2\Phi=\frac{3}{2}{\cal H}^2\Omega_m\delta$ and ${\cal
  H}=aH$. 
Here $t$ is conformal time and $\Omega_m$ is the matter
density fraction. It is well-known \cite{Bernardeau:2001qr} that in
the case of an Einstein--de Sitter universe these equations can be cast
in a form free from any explicit time dependence by introducing  
the time parameter $\eta=\ln D$, where $D$ is the linear growth
factor, and appropriately rescaling the velocity divergence
\be
\Theta = \frac{\nabla \cdot {\bm u}}{{\cal H}f}
\ee
with $f=d\ln D/d\ln a$. For the realistic $\Lambda$CDM cosmology,
the above substitution leaves a mild residual time dependence which,
however, has little effect on the dynamics. Following conventional
practice we will neglect this explicit time dependence in the
equations, although none of our findings crucially depend on this
restriction. With a slight abuse of language we will refer to this
setup as `exact dynamics' (ED). 
For comparison, we also consider Zel'dovich approximation (ZA)
obtained by replacing the Poisson equation by
$\nabla^2\Phi_{ZA}=\frac{3}{2}{\cal H}^2\Omega_m\Theta$. The linear
growth factor $D$ plays the role of the expansion parameter in TSPT.
In order to emphasize this, and in analogy to notation used
in quantum field theory, we denote it by
\be\label{coupling}
  g(\eta) \equiv e^\eta = D(z)\,.
\ee
We also use the short-hand notation $\delta_\e \equiv \delta(\e;\k)$,
and analogously for $\T$. 

The main idea of the TSPT approach
is to substitute the time evolution of the overdensity and velocity
divergence fields, $\delta$ and $\Theta$, by that of the their time
dependent probability  
distribution functional.
For adiabatic initial conditions
only one of the two fields is statistically 
independent.
We choose it to be the velocity
divergence field $\T$, and its distribution functional is denoted by
$\mathcal{P}[\T;\e]$.
At any moment in time, the field $\delta_\e$ can be expressed in terms of  $\T_\e$
as 
\be
\label{eq:psi1}
\delta_\e(\k)\equiv\delta[\T_\e;\e,\k]= \sum_{n=1}^\infty\frac{1}{n!}
\int [dq]^n K_n(\q_1,...,\q_n)\,\delta^{(3)}\Big(\k-\sum_{i=1}^n
\q_i\Big)
\prod_{j=1}^n\T_\e(\q_j)  \,,
\ee
where the kernels $K_n$ can be found  in Appendix~\ref{app:rec} and we
introduced the notation $[dq]^n = d^3q_1 \cdots d^3q_n$.

Equal-time correlation functions for $\T$ or $\delta$ can be obtained by
taking functional derivatives with respect to the external sources $J$ or
$J_\delta$, respectively, of the following generating functional,
\be
\label{eq:ztfp}
Z[J,J_\delta;\e]=\int [\mathcal{D}\T]\;{\mathcal P}[\T;\e]\;
\exp\bigg\{\int [dk] \T(\k) J(-\k)+\int[dk] \delta[\T;\e,\k]J_{\delta}(-\k)\bigg\}\,.
\ee
For example, the matter power spectrum is given by
\be
  P(\e;\k)\,\delta(\k+\k') = 
\frac{\delta^2 Z}{\delta J_\delta(-\k)\delta J_\delta(-\k')}\bigg|_{J=J_\delta=0}\,.
\ee
It is useful to expand  $\mathcal{P}[\T;\e]$ as a series 
in powers of $\T$,
\be
\label{eq:statweight}
\mathcal{P}[\T;\e]=\mathcal{N}^{-1}\exp\Bigg\{-\sum_{n=1}^{\infty}\frac{1}{n!}\int [dk]^n\;
(\G_n+C_n)(\e;\k_1,...,\k_n)\; \prod^n_{j=1} \T(\k_j)\Bigg\} \,,
\ee
where ${\cal N}$ is a normalization factor.
The $\G_n$ vertices have the physical meaning of 1-particle irreducible contributions to the tree-level correlators with amputated external propagators, 
and $C_n$ are {\it counterterms}, whose role is to cancel divergences
in the loop corrections \cite{Blas:2015qsi}. Both satisfy a hierarchy
of equations which replace the dynamical equations of SPT.  
For the purposes of this paper the expressions for $C_n$ will not be needed.
In the case of 
Gaussian initial conditions, the $\G_n$ vertices have universal
dependence on time both in ZA and ED, 
\be
\label{treeact}
 \G_n(\e;\k_1,...,\k_n)=\frac{1}{g^2(\e)} \bar\G_n(\k_1,...,\k_n)\;,
\ee
where the TSPT {\it coupling constant} is defined by \eqref{coupling} and 
$\bar \G_n(\k_1,...,\k_n)$ are time - independent functions
generated by recursion relations given in Appendix~\ref{app:rec}. Due
to momentum conservation, the latter are proportional to a 
$\delta$-function of the sum of their arguments;  
in what follows we use primes to denote the
quantities stripped off such $\delta$-functions, 
\be
\label{eq:tildeGC}
\bar\G_n={\bar\G}'_n(\k_1,\ldots,\k_n)\,\delta^{(3)}\Big(\sum^n_{i=1}
\k_i\Big)\,.
\ee 
The Gaussian part  of the integral \eqref{eq:ztfp}
\be
\label{eq:initial}
\bar \Gamma_2(\k_1,\k_2)= \frac{\delta^{(3)}(\k_1+\k_2)}{\bar P(k_1)}
\ee
is the seed in the recursion relations \eqref{recurGZA}, (\ref{Gned}) 
so the $\bar \G_n$
vertices can be seen as functionals of the initial power spectrum $\bar P(k)$.

\begin{figure}
\begin{align*}
\begin{fmffile}{example-ps}
\parbox{90pt}{
\begin{fmfgraph*}(80,80)
\fmfpen{thick}
\fmfleft{l1}
\fmfright{r1}
\fmf{plain,label=${\bf k}$}{l1,r1}
\end{fmfgraph*}}
\end{fmffile}&=~g^2(\e)\bar P(k), ~~
\begin{fmffile}{example-bisp_f}
\parbox{100pt}{
\begin{fmfgraph*}(90,70)
\fmfpen{thick}
\fmfleft{l1,l2}
\fmfright{r1}
\fmf{plain,label=${\bf k}_1$,label.side=right}{l1,b1}
\fmf{plain,label=${\bf k}_2$,label.side=left}{l2,b1}
\fmf{plain,label=${\bf k}_3$}{b1,r1}     
\end{fmfgraph*}}
\end{fmffile}=-g^{-2}(\e)\frac{\bar\G_3(\k_1,\k_2,\k_3)}{3!}\\
\begin{fmffile}{example-C1}
\parbox{80pt}{
\begin{fmfgraph*}(75,75)
\fmfpen{thick}
\fmfleft{l1}
\fmfright{r1}
\fmf{plain,label=${\bf k}$}{l1,r1}
\fmfv{d.sh=cross,d.si=.15w}{l1}
\end{fmfgraph*}}
\end{fmffile}&=-C_1(\k),~~~~~
\begin{fmffile}{composite}
\parbox{80pt}{
\begin{fmfgraph*}(75,75)
\fmfpen{thick}
\fmfleft{l1}
\fmfright{r5,r6}
\fmfv{d.sh=circle,d.filled=full,d.si=3thick}{b1}
\fmf{phantom_arrow,tension=4,label=$\k$,l.s=left}{l1,b1}
\fmf{plain}{b1,v6}
\fmf{plain,label=$\q_1$,l.s=left}{v6,r6}
\fmf{plain}{b1,v5}
\fmf{plain,label=$\q_2$,l.s=left,l.d=4}{v5,r5}
\end{fmfgraph*}}
\end{fmffile}=\frac{K_2(\q_1,\q_2)}{2!}\,\delta^{(3)}\Big(\k-\sum_{i=1}^2\q_i\Big)
\end{align*}
\caption{Example of TSPT Feynman rules.\label{fig:feynmanrules}}
\end{figure}
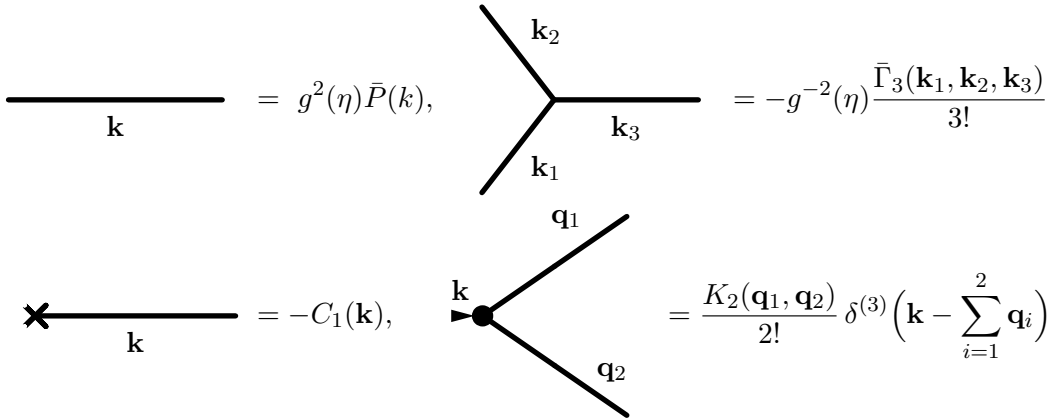

The TSPT perturbative expansion is organized by expanding the generating functional \eqref{eq:ztfp} over the Gaussian part of $\mathcal{P}$,
which is equivalent to an expansion
in the coupling constant $g(\e)$. This calculation can be represented
as a sum of Feynman diagrams, whose first elements are summarized in 
Fig.~\ref{fig:feynmanrules}: $\G_2$ is represented by a line
(propagator), the different elements  $\G_n$ (with $n>2$) and $C_n$
correspond to vertices, and $K_n$ are depicted as vertices with an
extra arrow. To compute an $n$-point correlation
function of the velocity divergence $\T$
one needs to draw all diagrams with $n$ external legs. For the correlators of the density field $\delta$ one has to add diagrams with external arrows.   

\subsection{TSPT in terms of wiggly and smooth elements}\label{sec:BAO}

\begin{figure}
\begin{center}
  \includegraphics[width=0.6\textwidth]{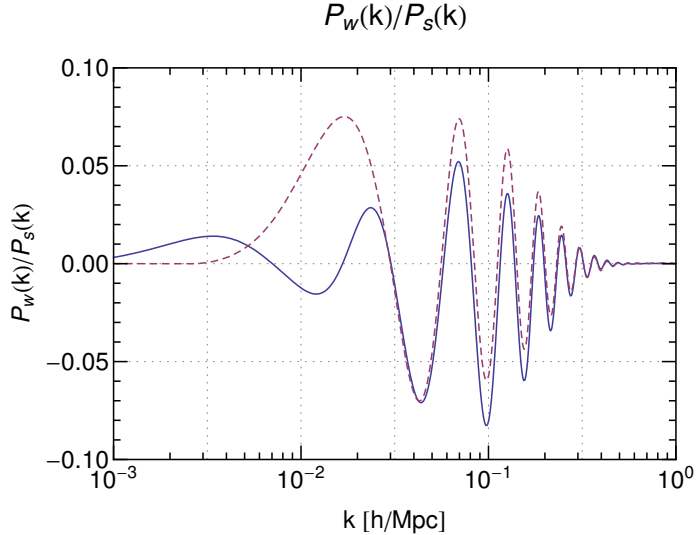}
\end{center}
\caption{\label{fig:Pwiggly} 
Ratio of oscillatory (wiggly) part $P_w$ of the linear power spectrum
to the smooth part $P_s$ obtained using two separation prescriptions.
The $\Lambda$CDM
cosmological parameters have been chosen as in \cite{Kim:2011ab}.
The solid curve corresponds to the decomposition used in numerical
computations in Sec.~\ref{sec:pract}.
The alternative
decomposition (dashed curve) is used for cross-checks.
}
\end{figure}

The initial power spectrum that sources the different elements of TSPT can be decomposed into 
a {\it smooth} part (with the maximum at $k=k_{eq}\sim
0.02\,h/\mathrm{Mpc}$ corresponding to 
the matter-radiation equality) and an oscillatory
part (or  {\it wiggly power spectrum}) that describes the impact of the BAO,
\be
\label{wigsm}
P_{lin}(\e, k) = g(\eta)^2\bar P(k)=g(\eta)^2\left(\bar P_s(k)+\lambda \bar P_w(k) \right)\,.
\ee
Here we have factored out the time-dependence given by the growth factor $g=D(z)$
and 
introduced a book-keeping parameter $\lambda$ to count the powers of
$\bar P_w$ in various expressions.
As for the vertices, the bar denotes the time-independent power spectra.
The amplitude of the 
wiggly power spectrum is 
suppressed compared to that of the smooth power spectrum in a realistic cosmological model.
Its value can be estimated 
as \cite{Gorbunov:2011zzc} (see also \cite{Seo:2007ns}),
\be
\label{wigsup}
\frac{\bar P_w}{\bar P_s}= O\left(e^{-(\e_{rec}-\e_{eq})}\frac{\O_b}{\O_m}\right) \sim  0.05\,.
\ee
The wiggly power spectrum $\bar P_w$ 
can be parametrized as \cite{Seo:2007ns,Gorbunov:2011zzc},
\be
\label{pwfit}
\bar P_w(k)\propto \sin \left({k}/{k_{osc}}\right)
\exp\left[-(k/k_{Silk})^2\right] T(k)\,,
\ee
where $1/k_{osc} \simeq 110\ \mathrm{Mpc}/h$, the Silk damping scale
is 
$k_{Silk}\sim 0.2\ h/\mathrm{Mpc}$ and
$T(k)$ is the dark matter transfer function which is slowly
varying\footnote{It tends to 1 at $k< k_{eq}$
  and behaves as $\ln k/k^2$ at higher $k$.}
with $k$. 
For our numerical analysis we do not use \eqref{pwfit}, but extract
the wiggly part by fitting a smooth multi-parameter 
template to the linear power spectrum for a given cosmological
model. The details of this procedure are outlined in
Appendix~\ref{app:wiggly}. The corresponding (time-independent) ratio
$P_w/P_s$ is shown by the solid curve in Fig.~\ref{fig:Pwiggly} 
for the case of the reference cosmological model used later on. 
Notice that this ratio vanishes both at low and high wavenumbers.

To
check that the results do not depend on the precise prescription for
separating the total power spectrum into smooth and wiggly components,
we have also used an alternative decomposition (see
Appendix~\ref{app:wiggly})
with 
$P_w/P_s$ depicted by
the dashed curve in Fig.~\ref{fig:Pwiggly}. We find
that the difference in the final results for the total power spectrum
and the correlation function obtained using the two forms of $P_w$
are at the sub-percent level, below the uncertainties
introduced by other approximations. This is to be expected: the
ambiguity in the wiggly-smooth decomposition is relevant at large
scales, $k\lesssim 0.03\,h/{\rm Mpc}$, which are essentially
unaffected by the non-linear IR dynamics. As will become clear later,
in the physical observables at these scales the smooth and wiggly
components are simply summed back, and the ambiguity
disappears. Similarly, an overall vertical off-set between the two curves in
Fig.~\ref{fig:Pwiggly} at $k\gtrsim 0.1\,h/{\rm Mpc}$ does not
contribute into the BAO feature.

The decomposition \eqref{wigsm} can be extended to all the $\bar \Gamma_n$ vertices, since they are functionals of the initial power spectrum.
Let us start with $\bar\G_2$ from \eqref{eq:initial},
\be
\label{decG2}
\begin{split}
\bar \G'_2(\k,-\k)=\frac{1}{\bar P(k)}&= \frac{1}{\bar P_s(k)}-\lambda \frac{\bar P_w(k)}{\bar P^2_s(k)}
+O(\lambda^2)\\
&\equiv \bar \G'^s_2(\k,-\k)+\lambda \, \bar\G'^w_2(\k,-\k)+O(\lambda^2)
\,.
\end{split}
\ee 
Given that the vertex $\bar \G'_2$ 
generates all the higher vertices by the recursion relation
\eqref{Gned}, 
one can introduce a similar decomposition for all the $\bar\G'_n$ vertices
\be
\label{decG}
\bar \G'_n=\bar \G'^{s}_n+\lambda \, \bar \G'^{w}_n+O(\lambda^2)\,.
\ee
The $\bar \G'^{w}_n \,(\bar \G'^{s}_n)$ vertices are computed using $\bar \G'^w_2 \, (\bar \G'^{s}_2)$ as an input in \eqref{Gned}.  

	    \begin{figure}[h]
\begin{align*}
	 \begin{fmffile}{w-ps}
	    \begin{gathered}
        \begin{fmfgraph*}(80,80)
        \fmfpen{thick}
        \fmfleft{l1}
        \fmfright{r1}
		\fmf{wiggly,label=$\textbf{k}$}{l1,r1}
	    \end{fmfgraph*}
	     \end{gathered}
	    \end{fmffile}~~=~g^2(\eta) \bar P_w(k) \,,
	    &
	     ~~~ \begin{fmffile}{w-bs}
	    \begin{gathered}
        \begin{fmfgraph*}(100,80)
        \fmfpen{thick}
        \fmfleft{l1,l2}
        \fmfv{d.sh=circle,d.filled=shaded,d.si=.13w,label=$ $,l.a=-90,l.d=.00w}{b1}
        \fmfright{r1}
		\fmf{plain,label=${\bf k}_1$}{l1,b1}
		\fmf{plain,label=${\bf k}_2$}{l2,b1}
		\fmf{plain,label=${\bf k}_3$}{b1,r1}   
	    \end{fmfgraph*}
	    \end{gathered}
	    \end{fmffile}~~=~-\frac{g^{-2}(\eta)}{3!}\bar\Gamma^w_3(\k_1,\k_2,\k_3)\,,\\
	     \\
	    	  \begin{fmffile}{s-ps}
	    \begin{gathered}
        \begin{fmfgraph*}(80,80)
        \fmfpen{thick}
        \fmfleft{l1}
        \fmfright{r1}
		\fmf{plain,label=$\textbf{k}$}{l1,r1}
	    \end{fmfgraph*}
	     \end{gathered}
	    \end{fmffile}~~=~g^2(\eta) \bar P_s(k) \,,
	    &
	     ~~~ \begin{fmffile}{s-bs}
	    \begin{gathered}
        \begin{fmfgraph*}(100,80)
        \fmfpen{thick}
        \fmfleft{l1,l2}
        \fmfv{d.sh=circle,d.filled=full,d.si=.04w,label=$ $,l.a=-90,l.d=.00w}{b1}
        \fmfright{r1}
		\fmf{plain,label=${\bf k}_1$}{l1,b1}
		\fmf{plain,label=${\bf k}_2$}{l2,b1}
		\fmf{plain,label=${\bf k}_3$}{b1,r1}   
	    \end{fmfgraph*}
	    \end{gathered}
	    \end{fmffile}~~=~-\frac{g^{-2}(\eta)}{3!}\bar\Gamma^s_3(\k_1,\k_2,\k_3)\,.
\end{align*}
	    \caption{Example of Feynman rules for wiggly and smooth elements.}    \label{feynmanruleswiggly}
\end{figure}
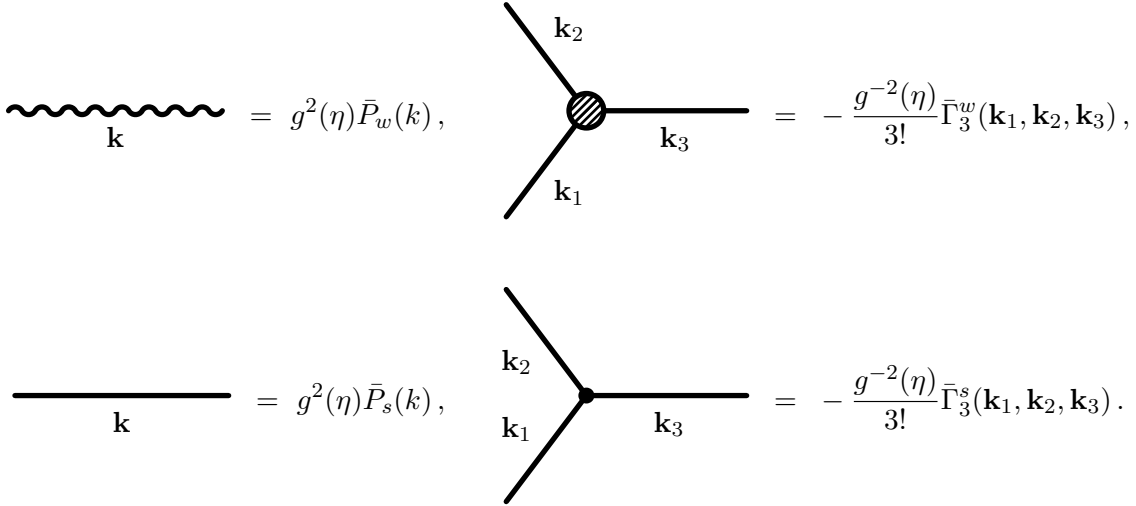

The decomposition \eqref{decG} can be introduced back in the partition function \eqref{eq:statweight}.
Since the $C_n$ and $K_n$ kernels 
are not functionals of the linear power spectrum \cite{Blas:2015qsi}
(see also Appendix~\ref{app:rec}), 
they are not subject to the wiggly-smooth decomposition.
The leading order $O(\lambda^0)$ corresponds to 
the {\it smooth} correlation functions. The 
 $O(\lambda)$ results include the {\it wiggly} contribution. In terms of diagrams, they can be summarized in the form of a new 
  wiggly propagator $g^2 P_w$ (represented by a wiggly line) 
and by wiggly vertices $\G'^w_n$ (represented by a dashed circle), 
see Fig.~\ref{feynmanruleswiggly}. 
We use small dots to depict the smooth vertices and straight lines to
represent the smooth power spectrum. The terms $O(\lambda^2)$ are
quadratic in $P_w$ and will be neglected.

The graphs with the wiggly elements are  
loosely referred to as wiggly graphs. For instance, the tree-level wiggly bispectrum is given by the following four wiggly graphs\footnote{
In terms of SPT kernels the result of \eqref{eq:amputated} can be rewritten as
\be 
\begin{split}
&\langle \T_\e(\k_1)\T_\e(\k_2)\T_\e(\k_3)\rangle^\mathrm{tree}_w\\
&=2\lambda g^4(\e)G_2(\k_1,\k_2)(\bar P_w(k_1)\bar P_s(k_2)+\bar P_w(k_2)\bar P_s(k_1))+\text{permutations}\,.
\end{split}
\ee
},
\begin{align}
\label{eq:amputated}
&\langle \T_\e(\k_1)\T_\e(\k_2)\T_\e(\k_3)\rangle^\mathrm{tree}_w
=\begin{fmffile}{example-bisp-small}
\parbox{50pt}{
\begin{fmfgraph*}(50,40)
\fmfpen{thick}
\fmfleft{l1,l2}
\fmfright{r1}
\fmf{plain}{l1,b1}
\fmf{plain}{l2,b1}
\fmf{plain}{b1,r1}   
\fmfv{d.sh=circle,d.filled=shaded,d.si=.14w,label=$ $,l.a=90,l.d=.35w}{b1}  
\end{fmfgraph*}}
\end{fmffile}
+
\begin{fmffile}{example-bisp-small1}
\parbox{50pt}{
\begin{fmfgraph*}(50,40)
\fmfpen{thick}
\fmfleft{l1,l2}
\fmfright{r1}
\fmf{wiggly}{l1,b1}
\fmf{plain}{l2,b1}
\fmf{plain}{b1,r1}     
\end{fmfgraph*}}
\end{fmffile}
+
\begin{fmffile}{example-bisp-small2}
\parbox{50pt}{
\begin{fmfgraph*}(50,40)
\fmfpen{thick}
\fmfleft{l1,l2}
\fmfright{r1}
\fmf{plain}{l1,b1}
\fmf{wiggly}{l2,b1}
\fmf{plain}{b1,r1}     
\end{fmfgraph*}}
\end{fmffile}
+
\begin{fmffile}{example-bisp-small3}
\parbox{50pt}{
\begin{fmfgraph*}(50,40)
\fmfpen{thick}
\fmfleft{l1,l2}
\fmfright{r1}
\fmf{plain}{l1,b1}
\fmf{plain}{l2,b1}
\fmf{wiggly}{b1,r1}     
\end{fmfgraph*}}
\end{fmffile} \\
\nonumber
 &=-\lambda g^4(\e)\left(\prod_{i=1}^3 \bar P_s(k_i) \;\bar\G^w_3(\k_1,\k_2,\k_3)
+\sum_{j=1}^{3}\bar P_w(k_j)\prod_{\begin{smallmatrix}i=1,\\i\neq j
\end{smallmatrix}}^3 \bar P_s(k_i) \;\bar\G^s_3(\k_1,\k_2,\k_3)\right)\,.
\end{align}
To obtain the matter density bispectrum one has 
to add to the previous expression six more graphs with the vertex $K_2$,
\begin{align}
\label{eq:bispcomposite}
\Delta \langle \delta_\e(\k_1)& \delta_\e(\k_2) \delta_\e(\k_3)\rangle^\mathrm{tree}_w
=\begin{fmffile}{example-bisp-comp-1}
\parbox{50pt}{
\begin{fmfgraph*}(50,40)
\fmfpen{thick}
\fmfleft{l1,l2}
\fmfright{r1}
\fmf{phantom_arrow}{l1,b1}
\fmf{wiggly}{l2,b1}
\fmf{plain}{b1,r1}   
\fmfv{d.sh=circle,d.filled=full,d.si=.1w,label=$ $,l.a=90,l.d=.35w}{b1}  
\end{fmfgraph*}}
\end{fmffile}
~~+~~
\text{permutations} \\
\nonumber
 =&
 -\lambda g^4(\e)\sum_{i<j=1}^{3}
 (\bar P_w(k_j) \bar P_s(k_i)+\bar P_s(k_j) \bar P_w(k_i)) \;K_2(\k_i,\k_j)\,\delta^{(3)}\big(\sum_l \k_l\big) \,.
\end{align}
In what follows we set $\lambda=1$ whenever there is no possible confusion.

\section{IR enhanced diagrams and power counting}\label{sec:PC}

One of the advantages of the TSPT approach is that all of its building blocks
are free of spurious infrared enhancements. In particular, the $\bar \Gamma_n$
vertices are finite in the limit $\{q\}\to0$, where $\{q\}$ 
is any subset of the arguments of $\bar \G_n$ \cite{Blas:2015qsi}. In contrast,
within SPT individual vertices have poles in $q$ which cancel only after summing all contributions
at a given order. Nevertheless, as mentioned previously, for a linear
power spectrum with an oscillatory behavior the cancellation of
enhanced terms is incomplete for $q\gtrsim k_{osc}$. 
In this section we show how to extract these enhanced contributions
within TSPT and then discuss power 
counting rules. These are helpful to organize the resummation of enhanced contributions, and later on to develop
a perturbative expansion for taking sub-leading corrections to the
resummed result into account.

\subsection{IR enhanced vertices}

We consider a TSPT $n$-point vertex $\bar\G_n(\k_1,\dots\k_n)$ evaluated with arguments $\k_i$ that
have magnitudes given schematically by two different scales: a hard scale denoted by $k$ and a
soft scale denoted by $q$, with
\be
\label{eq:soft}
  q/k \ll 1\,.
\ee 
Let us first analyze the three-point vertex $\bar \G'^{w}_3$.
Using \eqref{recurGZA} or (\ref{Gned}) we find,
\be 
\label{preG3}
\bar \G'^{w}_3(\textbf{k},-\textbf{k}-\textbf{q},\textbf{q})=I_2(\k,\q)\frac{\bar P_w(|\k+\q|)}{\bar P^2_s(|\k+\q|)}
+I_2(-\k-\q,\q)\frac{\bar P_w(k)}{\bar P^2_s(k)}+I_2(-\k-\q,\k)\frac{\bar P_w(q)}{\bar P^2_s(q)}\,,
\ee
where $I_2$ is given in (\ref{I2}).
In the limit \eqref{eq:soft} the rightmost term in \eqref{preG3} is
smaller than  
$O(q/k)$ and will be neglected  in what follows.
The other terms provide an expression of the form
\begin{align}
\bar \G'^w_3(\k,-\k-\q,\q)=   & \frac{\textbf{k}\cdot \textbf{q}}{q^2}\left(\frac{\bar P_w(|\k+\q|)-\bar P_w(k)}{\bar P_s(k)^2}\right)+O\left((q/k)^0\right)\,,
\label{eq:asg3}
\end{align}
where we have used that the derivatives of the smooth component scale
as $1/k$ and expanded $\bar
P_s(|\k+\q|)=\bar P_s(k)+O(q/k)$. 
In contrast, we kept the finite difference for the wiggly component
that varies substantially for $q\gtrsim k_{osc}$.

It is convenient to use compact notations by introducing the linear
operator 
\begin{align}
\label{finitediff}
\D_{\q}[\bar P_w(k)]=
\frac{\k \cdot \q}{q^2}\Big(\bar P_w(|\k+\q|)-\bar P_w(k)\Big) & =
\frac{\k \cdot \q}{q^2}\left(e^{\textbf{q}\cdot
    \nabla_{\textbf{k}'}}-1\right)\bar P_w(k')\Big|_{k'=k}\;.
\end{align}
This operator will play a central role in the following, and therefore
we elaborate on some of its properties. 
Consider first its action on a purely oscillatory
function $e^{ik/k_{osc}}$, where we are interested in the case $k\gg
k_{osc}$. 
Expanding the exponent in the small parameter $q/k$ we obtain,
\be
\label{powcount}
  \D_{\q}[e^{ik/k_{osc}}] \approx 
\frac{\k \cdot \q}{q^2}
\left(e^{i q \cos(\k,\q)/k_{osc}}-1\right)e^{ik/k_{osc}}\,,
\ee
where we introduced $\cos(\k,\q)=\frac{\k\cdot\q}{kq}$. For $q\gtrsim
k_{osc}$ the expression in the brackets is of order one, whereas the
prefactor is enhanced by $k/q$. On the other hand, if $q\ll k_{osc}$,
Eq.~(\ref{powcount}) reduces to 
\be
\label{powcount1}
\D_\q[e^{ik/k_{osc}}] \approx \frac{ik}{k_{osc}}\cos^2(\k,\q)\; e^{ik/k_{osc}}\;,
\ee
so that the enhancement is given by $k/k_{osc}$.
In a more realistic case the wiggly power spectrum can be viewed as an
oscillating function that is modulated by a 
smooth envelope, $P_w(k) \sim f_{env}(k)e^{i(k/k_{osc}+\varphi)} +
$c.c. with\footnote{Strictly speaking, due to existence of the Silk damping,  
 $\nabla f_{env}(k) \sim O\big(\max(1/k,1/k_{Silk})\big)
f_{env}(k)$. However, we use
the simpler estimate from above since in practice we do not consider 
values of $k$ that are parametrically larger than $k_{Silk}$.} 
$\nabla f_{env}(k) \sim O(1/k) f_{env}(k)$. 
For example, the parametrization \eqref{pwfit}
is of this form. 
Inserting this parameterization into (\ref{finitediff}) one observes
that the derivatives acting on the envelope are suppressed compared to
those acting on the oscillating part. They must be taken into account
only when looking at the sub-leading corrections. We conclude that
\be
\label{Dorder}
 \D_\q[\bar P_w(k)] \sim {O}(\varepsilon^{-1})\bar P_w(k)\,,
\ee
where we introduced the small parameter
\be\label{defeps}
 \varepsilon \equiv {\rm max}\left(q/k, k_{osc}/k\right)\,.
\ee
As we are going to see shortly, the enhancement by $1/\varepsilon$ is the
reason why the naive SPT loop expansion breaks down for the
BAO. The sub-leading corrections coming from the derivatives of the
envelope and higher terms in the expansion of the oscillating part
correspond to contributions of 
order $\varepsilon^0$ (or higher powers of $\varepsilon$)
that are suppressed relative to the leading $1/\varepsilon$ enhancement.
This will be important to establish systematic power counting rules.

It turns out to be useful to extend the action of $\D_\q$ to any wiggly
element. This is done by recalling that the latter are linear
expressions in $P_w$ with smooth $k$-dependent coefficients. Then, by
definition, $\D_\q$ acts on any occurrence of $P_w$ according to
(\ref{finitediff}), leaving the smooth coefficients intact. For
example,
\be
\D_\q [\bar \G'^{w}_3(\k_1,\k_2,\k_3)]\Big|_{\k_3=-\k_1-\k_2} \equiv
\left(\frac{\D_\q[\bar P_w(k_1)]}{\bar
    P^2_s(k_1)}I_2(\k_2,\k_3)+\text{perm.} \right)
\bigg|_{\k_3=-\k_1-\k_2} \;,
\ee
and similarly for other $\bar\G'^w_n$. Note that an immediate
consequence of this definition is that $\D_\q$ commutes with itself,
\be
\D_{\q_1}\D_{\q_2}\bar\G'^w_n=\D_{\q_2}\D_{\q_1}\bar\G'^w_n
\ee

The result (\ref{eq:asg3}) for the 3-point vertex can be generalized
by induction to arbitrary $n$-point vertices with $m$ hard wavenumbers
$\k_i$ and $n-m$ wavenumbers $\q_j$ going uniformly to zero. In
Appendix~\ref{app:TSPT} we prove the following formula,
\be
\label{IRnm}
\bar
\G'^w_n\Big(\k_1,...,\k_m\!-\!\sum_{j=1}^{n-m}\!\q_j,
\q_1,...,\q_{n-m}\Big)\!= 
(-1)^{n-m}\left( \prod_{j=1}^{n-m} \D_{\q_j}\right)
[\bar \G'^w_{m}(\k_1,...,\k_m)] \times\big(1+O(\varepsilon)\big)\,,
\ee
where $\k_m=-\sum_{i=1}^{m-1}\k_i$ due to momentum conservation.
Note that the leading IR enhancement $\propto (1/\varepsilon)^{n-m}$ 
is equal to the number of soft arguments. The maximal enhancement
happens for the case of $n-2$ soft wavenumbers where we have
\be 
\label{asympPs}
\begin{split}
\bar \G'^{w}_n\Big(\k,-\k&\!-\!\sum_{i=1}^{n-2}\!\q_i,\q_1,...,\q_{n-2}\Big)
=(-1)^{n-2}\left(\prod_{i=1}^{n-2}\D_{\q_i}\right)[\G'^w_2(\k,-\k)] 
\times \big( 1 + O(\varepsilon)\big)\,\\
&=(-1)^{n-1}\left[\prod_{i=1}^{n-2}\frac{(\k\cdot
    \q_i)}{q_i^2}(e^{\q_i\cdot \nabla_{k'}}-1)\right]\frac{\bar
  P_w(k')}{\bar P_s^2(k)}\Big|_{\k'=\k} 
\times \big( 1 + O(\varepsilon) \big)\,,
\end{split}
\ee
which scales as $O(\varepsilon^{-n+2})$. Clearly, the sensitivity of
the vertices $\bar\G'^{w}_n$ to the large parameter $1/\varepsilon$
grows with $n$. 
In the subsequent sections we 
show how these large enhancement factors can be resummed within a
systematic approach.

\subsection{Leading diagrams and power counting rules}\label{sec:PCsub}

Consider a loop diagram containing a wiggly TSPT vertex $\Gamma_n^w$ with $m$ external legs  
and $(n-m)$ legs attached to the loops. As we saw above, this vertex is enhanced by powers $1/\ve$ in the limit where its arguments $\q_j$ flowing in the loops become soft compared to the external wavenumbers $\k_i$.  
In order to extract the corresponding enhancement of the loop diagrams,
we split all loop integrations into
a soft part with $q<k_S$ and a hard part with $q>k_S$. The scale $k_S$ is in principle arbitrary, and
observables do not depend on it when computed exactly. Nevertheless, this splitting allows us
to separately treat the IR and UV parts of loop integrations, and resum the large IR loops.
Any residual dependence on $k_S$ should be taken as an estimate of the
theoretical uncertainty, that should become smaller and smaller when computing at higher orders.
In practice, the BAO feature is mostly affected by the modes with $q$ between $k_{osc}$ and $k_{Silk}$, so the 
range $k_{osc} < k_S < k_{Silk}$ can be expected to lead to good
convergence properties. We will return to the choice of $k_S$ in Sec.~\ref{sec:pract}.

To account for the IR enhancement, we
identify the expansion parameter with (cf.~(\ref{defeps}))
\be
 \varepsilon = \frac{\langle q\rangle}{k}\,,
\ee
where $k_{osc}<\langle q\rangle <k_S$ is the characteristic scale giving the dominant contributions into the IR loop integrals. As will become clear below by inspection of the eventual expressions (\ref{Slead}), (\ref{Sigmadefk}) for the IR enhanced loops, the integrand in them peaks roughly at the maximum of the smooth power spectrum implying $\langle q\rangle\sim k_{eq}$.

In addition to $\ve$, the relevant parameter that controls the loop expansion is given by the variance of the input linear power spectrum. The latter is dominated by the smooth component $P_s$. Due to the splitting into  
an IR (`soft') and UV (`hard') parts we can discriminate
two variances
\bea
  \sigma_S^2 &\equiv& g^2\int_{q<k_S} d^3q\, \bar P_s(q) \,, \nn\\
  \sigma_h^2 &\equiv& g^2\int_{q>k_S} d^3q\, \bar P_s(q) \,. \label{eq:sigmah}
\eea
For example, for a realistic $\Lambda$CDM model one has $\sigma_S^2\sim 0.16g^2$ (recall that $g\equiv D(z)$)
for the choice $k_S=0.1h/$Mpc, whereas $\sigma_h^2$ is formally
UV divergent. 
In practice, the hard part of the loop corrections remains finite due to additional suppression of the actual integrands in the UV. Still, these corrections are UV dominated and their reliable calculation requires proper renormalization of the contribution due to very short modes. On the other hand, while the importance of the UV counterterms increases for high wavenumbers and at higher orders of the perturbation theory, they are not essential for the calculation of hard one-loop corrections to the power spectrum at the BAO performed in this paper. The latter corrections are well-behaved and are of order few$\times 10\%$ at $z=0$.
We postpone the study of the UV counterterms in TSPT for future work and focus in this paper on the IR loops. $\s_h^2$ will be used in what follows as a formal counting parameter for the number of hard loops. 

Although $\sigma_S^2$ seems to be
rather small, we will now argue that soft loops are enhanced by a factor $1/\varepsilon^2\sim O(10)$ for the
wiggly observables. Therefore they are proportional to the product $\sigma_S^2 \times 1/\varepsilon^2$ which 
is $O(1)$ at
low redshift, implying that the corresponding soft loops need to be resummed.
Consider, for example, 1-loop corrections to the wiggly $\T \T$ power spectrum, given by the following 
TSPT diagrams,
\vspace{0.5cm}
\begin{align}
\nonumber
         P_{w,\T\T}^{1-loop}(\e;k)= ~~&
      \begin{fmffile}{w-vertex3_3_0}
	    \begin{gathered}
        \begin{fmfgraph*}(60,20)
        \fmfpen{thick}
        \fmfkeep{1loop_2}
         \fmfleft{l1}
        \fmfright{r1}
    \fmfv{d.sh=circle,d.filled=full,d.si=.01w,label=$\bar \G^s_3$,l.a=135,l.d=.06w}{v1}
        \fmfv{d.sh=circle,d.filled=full,d.si=.01w,label=$\bar \G^s_3$,l.a=45,l.d=.06w}{v2}
		\fmf{plain,label=$ $}{l1,v1}
		\fmf{plain,label=$ $}{v2,r1}
	    \fmf{wiggly,left=1.0,tension=0.5,label=$ $}{v1,v2}
	    \fmf{plain,left=1.0,tension=0.5,label=$ $,l.side=left}{v2,v1}
	    \fmfposition
	    \fmfdotn{v}{2}      
	    \end{fmfgraph*}
	    \end{gathered}    
 \end{fmffile}
  ~+~
      \begin{gathered}
   \begin{fmffile}{w-vertex4_0}
        \begin{fmfgraph*}(60,20)
        \fmfpen{thick}
        \fmfkeep{1loop}
        \fmfleft{l1}
        \fmfright{r1}
        \fmfv{d.sh=square,d.filled=empty,d.si=.07w,label=$\bar \G^s_4$,l.a=-90,l.d=.1w}{b1}
		\fmf{plain,label=$ $}{l1,b1}
		\fmf{plain,label=$ $}{b1,r1}
	     \fmf{wiggly,right=0.5,tension=0.7,label=$ $}{b1,b1}
	    \fmfdotn{b}{1}      
	    \end{fmfgraph*}
	         \end{fmffile}
	         	     \end{gathered}  
	     ~+~ 
	     \begin{fmffile}{w-vertex3_1_0}
	    \begin{gathered}
        \begin{fmfgraph*}(60,20)
        \fmfpen{thick}
        \fmfkeep{1loop_2}
         \fmfleft{l1}
        \fmfright{r1}
        \fmfv{d.sh=circle,d.filled=full,d.si=.01w,label=$\bar \G^s_3$,l.a=135,l.d=.06w}{v1}
        \fmfv{d.sh=circle,d.filled=full,d.si=.01w,label=$\bar \G^s_3$,l.a=45,l.d=.06w}{v2}
		\fmf{wiggly,label=$ $}{l1,v1}
		\fmf{plain,label=$ $}{v2,r1}
	    \fmf{plain,left=1.0,tension=0.5,label=$ $}{v1,v2}
	    \fmf{plain,left=1.0,tension=0.5,label=$ $,l.side=left}{v2,v1}
	    \fmfposition
	    \fmfdotn{v}{2}      
	    \end{fmfgraph*}
	    \end{gathered}    
  \end{fmffile}
  ~+~
	     \begin{gathered}
   \begin{fmffile}{w-vertex4_2_0}
        \begin{fmfgraph*}(60,20)
        \fmfpen{thick}
        \fmfkeep{1loop}
        \fmfleft{l1}
        \fmfright{r1}
        \fmfv{d.sh=square,d.filled=empty,d.si=.04w,label=$\bar \G^s_4$,l.a=-90,l.d=.1w}{b1}
		\fmf{wiggly,label=$ $}{l1,b1}
		\fmf{plain,label=$ $}{b1,r1}
	     \fmf{plain,right=0.5,tension=0.7,label=$ $}{b1,b1}
	    \fmfdotn{b}{1}      
	    \end{fmfgraph*}
	         \end{fmffile}
	         	     \end{gathered} 
	     	   \\
  	     \nonumber
  	     \\
  	       \label{1lwig}
  &~+~
     \begin{fmffile}{w-vertex3_5_0}
	    \begin{gathered}
        \begin{fmfgraph*}(60,20)
        \fmfpen{thick}
        \fmfkeep{1loop_2}
         \fmfleft{l1}
        \fmfright{r1}
            \fmfv{d.sh=circle,d.filled=shaded,d.si=.15w,label=$ $,l.a=135,l.d=.00w}{v1}
        \fmfv{d.sh=circle,d.filled=full,d.si=.04w,label=$\bar \G^s_3$,l.a=45,l.d=.08w}{v2}
		\fmf{plain,label=$\bar \G^w_3$,l.d=.08w}{v1,l1}
		\fmf{plain,label=$ $}{v2,r1}
	    \fmf{plain,left=1.0,tension=0.3,label=$ $}{v1,v2}
	    \fmf{plain,left=1.0,tension=0.3,label=$ $,l.side=left}{v2,v1}
	    \fmfposition
	    \end{fmfgraph*}
	    \end{gathered}    
  \end{fmffile}
  ~+~
    \begin{gathered}
	   \begin{fmffile}{w-vertex4_3_0}
          \begin{fmfgraph*}(60,20)
        \fmfpen{thick}
        \fmfkeep{1loop}
        \fmfleft{l1}
        \fmfright{r1}
        \fmfv{d.sh=circle,d.filled=shaded,d.si=.15w,label=$\bar \G^w_4$,l.a=-90,l.d=.15w}{b1}
		\fmf{plain,label=$ $,l.a=-20,l.d=.1w}{l1,b1}
		\fmf{plain,label=$ $}{b1,r1}
	     \fmf{plain,right=0.5,tension=0.7,label=$ $}{b1,b1}
	    \end{fmfgraph*}
  \end{fmffile}
  	     \end{gathered}
 ~~\,.
\end{align}
The loop integration in each diagram can be either hard, $q>k_S$, in which case the diagrams are
counted at order $\sigma_h^2$, or soft, $q<k_S$, and are of order $\sigma_S^2$.
Only the soft contributions can be IR enhanced, so we focus on them for the moment.
The diagrams in the first line of \eqref{1lwig}
are never IR-enhanced, \emph{i.e.} they are at most of order $\varepsilon^0\times \sigma_S^2$,
because they do not contain a wiggly vertex. On the other hand, the diagrams with the wiggly vertices 
do receive an IR enhancement. The first diagram in the second line contains $\G_3^w$ and
is according to \eqref{asympPs}, (\ref{Dorder}) of order $1/\varepsilon\times \sigma_S^2$. The last diagram 
contains $\G_4^w$, and using \eqref{asympPs}, (\ref{Dorder}) we find that it is of order  
$1/\varepsilon^2\times \sigma_S^2$
and thus is the most 
IR-enhanced one-loop
diagram. At leading order in $\ve$ it is given by

\vspace{0.5cm}
\be
\label{1lpw}
\begin{gathered}
	   \begin{fmffile}{w-vertex4_3_1}
          \begin{fmfgraph*}(80,20)
        \fmfpen{thick}
        \fmfkeep{1loop}
        \fmfleft{l1}
        \fmfright{r1}
        \fmfv{d.sh=circle,d.filled=shaded,d.si=.15w,label=$\bar \G^w_4$,l.a=-90,l.d=.15w}{b1}
	\fmf{plain}{l1,b1}
			\fmf{plain}{b1,r1}
	     \fmf{plain,right=0.5,tension=0.8,label=$ $}{b1,b1}
	    \end{fmfgraph*}
  \end{fmffile}
  	     \end{gathered}
  	    ~~ =\frac{g^2}{2} \int_{q\leq k_S} [dq]\bar P_s(q){\cal D}_\q {\cal D}_{-\q} P_w(\e;k) \equiv  - g^2 \S P_w(\e;k)\,,
\ee
where in the last step we defined the operator $\S$ which can be written as
\be 
\label{Slead}
\begin{split}
\S \;  P_w(\e; k)&=  \int_{q\leq k_S}[dq]\bar P_s(q)\frac{(\k \cdot \q)^2}{q^4}\big(1-
\cosh\left( \q\cdot\nabla_{k'} \right)\big)P_w(\e; k')\Big|_{k'=k} \\
\end{split} 
\ee
In terms of our power counting
\be
  \S \;  P_w(\e; k) \sim   O(1/\varepsilon^2\times \sigma_S^2)\;  P_w(\e; k)\;. 
\ee
As discussed previously, the product $1/\ve^2\times \s_S^2$ can be of  $O(1)$ at low redshift and
therefore this one-loop contribution can be comparable to the linear wiggly spectrum.

In order to identify and eventually 
resum such terms we now discuss how to determine the order of
an arbitrary $L$-loop diagram in our power counting.
This is valid for any $n$-point correlation function, with external wavenumbers $k_i$ around the BAO scale,
$k_{Silk}\gtrsim k_i \gg k_S, k_{osc}$. 
Given a TSPT diagram with $L$ loops (\emph{i.e.} scaling as $g^{2L}$), one must 
\begin{enumerate}

\item choose for each propagator and each vertex whether it is smooth or wiggly. Since we are interested in diagrams that contain one power of $P_w$,
at most one element (either propagator or vertex) can be wiggly. 
To obtain the full answer, one eventually needs to sum over all possibilities to choose a vertex or propagator 
to be the wiggly one.

\item assign each loop to be either hard ($q> k_S$) or soft ($q<
  k_S$). Formally, this can be done by splitting the linear input
  spectrum 
into two parts as
$P_{lin}(q)=\theta(q-k_S)P_{lin}(q)+\theta(k_S-q)P_{lin}(q)$, and
calling  
a loop hard if all propagators and vertices along the
loop contain only power spectra of the former type\footnote{
Strictly speaking, the splitting into hard and soft contributions is
only necessary for the loops 
that contain a wiggly vertex. 
For other loops, since $\sigma_S^2+\sigma_h^2 \approx \sigma_h^2$
in a realistic case, it is effectively irrelevant whether we make this
split or not.}. 
The number of hard loops is denoted by $L_h$ and the number of soft loops by $L_s$. Trivially $L=L_h+L_s$,
and the diagram contributes at order $(\sigma_S^2)^{L_s}\times (\sigma_h^2)^{L_h}$.
Again, to obtain the full answer, one needs to sum over all assignments eventually.

\item count the \emph{number of soft lines that are attached to the wiggly vertex}. We call this number $l$. According to \eqref{asympPs}, the IR-enhancement is $1/\varepsilon^l$.
\end{enumerate}
The order in our power counting of a contribution  characterized by the numbers $(L_h, L_s, l)$ is therefore given by\footnote{To be more precise, this provides an upper estimate for the magnitude of the contribution. Specific terms can be further suppressed, as will be seen below.}
\be\label{eq:powerc}
O\left( (\sigma_S^2)^{L_s}\times (\sigma_h^2)^{L_h} \times 1/\varepsilon^l \right)\,.
\ee
If a diagram contains no wiggly vertex then $l=0$ and no IR-enhancement occurs. The most IR-enhanced contributions have
the largest value of $l$. As a single loop cannot contain more than two lines attached to the same vertex, we have the inequality 
$l \leq 2L_s$. This means that the most IR enhanced contributions are of order $(\sigma_S^2/\epsilon^2)^{L_s}\times (\sigma_h^2)^{L_h}$.
As argued before $\sigma_S^2/\epsilon^2$ is ${O}(1)$ at low redshift and therefore it is desirable to resum
all diagrams with the maximal enhancement $l=2L_s$. This is the subject of the next section.

\section{Resummation of leading infrared effects}\label{sec:leading}

Here we perform the resummation of dominant IR enhanced diagrams contributing into wiggly observables. 
We start with the power spectrum, then consider 
the bispectrum and outline the generalization to higher $n$-point functions. We work at the order $(\s_h^2)^0$, i.e. neglecting the hard loop corrections. The task of taking them into account is postponed till Sec.~\ref{sec:hard}.

\subsection{Power spectrum}\label{sec:leading_PS}

The most IR-enhanced contributions correspond to diagrams with $l=2L_s$ and
$L_h=0$, \emph{i.e.} all loops are soft and they contain a wiggly vertex to which $l=2L_s$ soft lines are attached.
At one-loop, the most IR enhanced diagram is the tadpole diagram \eqref{1lpw} with $L_s=1$ soft loop
and $l=2$ soft lines attached to $\bar\Gamma_4^w$ (the two lines that belong to the loop). At two-loop, $L_s=2$,
the most IR-enhanced diagram should contain a wiggly vertex with $l=4$ soft lines attached to it. In addition,
the IR enhancement can only occur if also a hard momentum
flows through the wiggly vertex. For $L_h=0$ this can only be the external momentum. Therefore, the most
IR-enhanced diagram has to contain a wiggly vertex $\bar\G_6^w$ with four soft arguments attached to loops and
two hard arguments that correspond to the two external legs. The only possibility that remains is
a single diagram, given by a double-tadpole. Analogously, at higher loop orders, the most IR-enhanced
diagrams are obtained by attaching more and more loops to the wiggly vertex in the center. 
The first few diagrams that contribute to the wiggly part of the power spectrum are shown in Eq.~(\ref{daisygraphsintro}). It is natural to call them \emph{daisy} diagrams.
Note that the leading IR-enhanced contributions are the same for the density and velocity power spectra. Indeed, within TSPT the $\delta$ power spectrum is obtained from that of $\T$ by adding diagrams with the kernels $K_n$ (see Sec.~\ref{sec:tspt}).  These kernels do not depend on the wiggly power spectrum and are therefore not subject to IR enhancement. Consequently, diagrams involving $K_n$
kernels give  subdominant contributions at
each loop order.

The daisy diagram with $L$ loops, all of which are soft (\emph{i.e.} $L_s=L$), is given by
\be
\begin{split}
\label{leadlooplim}
&P^{L-loop,LO}_{w}(\e;k) = -\frac{1}{(2L+2)!}\cdot (2L+2)(2L+1)(2L-1)!!\\
&~~~\times \prod_{i=1}^{L}\left[\int_{q_i\leq k_S} [dq_i]g^2\bar P_s(q_i)\right]g^4\bar P_s(k)^2g^{-2}\bar \G'^{w,LO}_{2L+2}(\textbf{k},-\textbf{k},\textbf{q}_1,-\textbf{q}_1,...,\textbf{q}_{L},-\textbf{q}_{L})\,.
\end{split}
\ee
The symmetry factor in this formula arises as follows:
there are $(2L+2)(2L+1)$ ways to choose the two external legs, and $(2L-1)\cdot(2L-3)\cdot\ldots\cdot 1$ 
ways to connect all the remaining lines into the loops.
Making use of \eqref{asympPs}, one obtains
\be 
\begin{split}
P^{L-loop,LO}_{w}(\e;k) = &\frac{1}{L!}
\prod_{i=1}^{L}\left[\frac{g^2}{2}\int_{q_i\leq k_S} [dq_i] \bar P_s(q_i){\cal D}_{\q_i} {\cal D}_{-\q_i}  \right] 
 P_w(\e;k) = \frac{(-g^2(\e)\S)^L}{L!}P_w(\e;k) 
 \,,
\end{split}
\ee
where the operator $\S$ has been defined in \eqref{1lpw}. 
The sum over all daisy graphs gives the leading-order IR-resummed wiggly power spectrum 
\begin{align} 
\label{Irressim}
P^{IR\;res,LO}_{w}(\e,k)&=\sum_{L=0}^{\infty}P_w^{L-loop,LO}(\e; k)
=e^{-g^2(\e) \S } P_w(\e;k)\,.
\end{align}
We see that the operator $\S$ exponentiates.

The total power spectrum is obtained by adding the smooth part which, to the required order of accuracy, can be taken at tree level. This yields,
\be
\label{eq:finalLO}
P^{IR\;res,LO}
=P_s(\e;k)+ e^{-g^2(\e)\S } P_{w}(\e;k)\,.
\ee
Let us stress again that this expression holds both for the density and velocity divergence power spectra. Moreover, it is the same in ED and ZA as the expansion (\ref{asympPs}) used in the derivation is valid in both cases. The difference between ED and ZA and between $\delta$ and $\T$ appears for higher correlators 
and for the power spectrum beyond the leading order.

\subsection{Bispectrum and other $n$-point correlation functions}\label{sec:leading_n}

In this section we extend the resummation procedure of the IR-enhanced loop contributions to the
wiggly part of higher order correlation functions. 
We first discuss the bispectrum and then general $n$-point correlation
functions.
 
The wiggly part of the 
tree-level bispectrum for the velocity divergence field is given by the four graphs \eqref{eq:amputated},
whereas for the density bispectrum one has to add the graphs (\ref{eq:bispcomposite}).
At one loop order, and assuming all external wavenumbers are hard, the most IR-enhanced contributions
are obtained by   `dressing' the wiggly 
vertices and propagators in \eqref{eq:amputated}, \eqref{eq:bispcomposite}  with a soft loop ($q<k_S$) attached to the wiggly vertex,
   \begin{align}
   \label{eq:bisp1l}
   &B^{1-loop, LO}_{w,\delta \delta \delta}(\e;\k_1,\k_2,\k_3)=\\
   \nonumber\\
&~~  
	        \nonumber
  \begin{fmffile}{w-bisp6}
	    \begin{gathered}
        \begin{fmfgraph*}(80,60)
          \fmfpen{thick}
        \fmfleft{l1,l2}
        \fmfright{r1}
        \fmfv{d.sh=circle,d.filled=shaded,d.si=.1w,label=$\bar\G^w_5$,l.a=-45,l.d=.12w}{b1}
		\fmf{plain,label=$\textbf{k}_1$,l.side=left}{l1,b1}
		\fmf{plain,label=$\textbf{k}_2$,l.side=right}{l2,b1}
		\fmf{plain,label=$~~~\textbf{k}_3$}{b1,r1}
		\fmf{plain,right=0.5,tension=0.9,label=$\textbf{q}$}{b1,b1}
	    \end{fmfgraph*}
	    \end{gathered}  
	    \end{fmffile}	 ~+~
\begin{fmffile}{w-bisp5}
	    \begin{gathered}
        \begin{fmfgraph*}(80,60)
        \fmfpen{thick}
        \fmfleft{l1,l2}
        \fmfright{r1}
        \fmfv{d.sh=circle,d.filled=full,d.si=.04w,label=$\bar \G^s_3$,l.a=80,l.d=.12w}{b1}
          \fmfv{d.sh=circle,d.filled=shaded,d.si=.1w,label=$\bar \G^w_4$,l.a=-90,l.d=.12w}{c1}
				\fmf{plain,label=$\textbf{k}_1$,l.side=left}{l1,ll1}
		\fmf{plain,label=$\textbf{k}_2$,l.side=right}{l2,ll2}
		\fmf{plain}{ll1,b1}
		\fmf{plain}{ll2,b1}
		\fmf{plain,label=$ $}{b1,c1}
		\fmf{plain,label=$~~\textbf{k}_3$}{c1,r1}
	     \fmf{plain,right=0.7,tension=0.6,label=$\textbf{q}$}{c1,c1}
	    \end{fmfgraph*}
	    \end{gathered} 
	    	    \end{fmffile}	~~  ~+~   	    ~~
	    	    \begin{fmffile}{w-bisp4}
	    \begin{gathered}
        \begin{fmfgraph*}(80,60)
        \fmfpen{thick}
        \fmfleft{l1,l2}
        \fmfright{r1}
        \fmfv{d.sh=circle,d.filled=full,d.si=.07w,label=$ K_2$,l.a=80,l.d=.12w}{b1}
          \fmfv{d.sh=circle,d.filled=shaded,d.si=.1w,label=$\bar \G^w_4$,l.a=-90,l.d=.12w}{c1}
		\fmf{phantom_arrow}{ll1,b1}
		\fmf{phantom,label=$\textbf{k}_1$,l.side=left,l.d=0.1w}{l1,ll1}
		\fmf{plain,label=$\textbf{k}_2$,l.side=right}{l2,ll2}
		\fmf{plain}{ll2,b1}
		\fmf{plain,label=$ $}{b1,c1}
		\fmf{plain,label=$~~\textbf{k}_3$}{c1,r1}
	     \fmf{plain,right=0.7,tension=0.6,label=$\textbf{q}$}{c1,c1}
	    \end{fmfgraph*}
	    \end{gathered} 
	    	    \end{fmffile}~+~\text{perm.}
	    \end{align}	      
Within our power counting, these diagrams
contribute at order $\sigma_L^2 \times 1/\varepsilon^2$ compared to the tree-level
bispectrum. 
Using a similar reasoning as for the power spectrum, one finds that at higher loop orders
the most IR-enhanced corrections are given by daisy diagrams obtained
by attaching more soft loops to the wiggly vertices appearing in each diagram in \eqref{eq:bisp1l}.
Parametrically, these $L$-loop diagrams scale as
\be
B^{L-loop,LO}_{w,\delta \delta \delta}\sim ( \sigma_L^2/\varepsilon^2)^L B^{tree}_{w,\delta \delta \delta}\;,
\ee
and thus need to be resummed.

The daisy diagrams centered on the propagator (like the second and
third terms in (\ref{eq:bisp1l})) are essentially the same as those
appearing in the calculation of the power spectrum from the previous
subsection. They are evaluated using Eq.~(\ref{leadlooplim}) and their
resummation leads to the replacement of the external wiggly
propagators in the tree level expression, 
\be
\label{replace}
\bar P_w\mapsto e^{-g^2\S}\bar P_w\;.
\ee 
A new type of contributions comes
from one-particle-irreducible (1PI) diagrams with soft loops dressing
the 3-point vertex (like the first diagram in (\ref{eq:bisp1l})). In
view of future uses, let us consider the general case of a wiggly
vertex with $n$ hard wavenumbers dressed by $L$ soft loops, 
\be
\begin{split}
V_{w,n}^{L-loop}=&~~
\begin{gathered}
\begin{fmffile}{dressed_n}
\begin{fmfgraph*}(110,100)
\fmfpen{thick}
\fmfleft{l0,l1,l2,l3,l4,l5,l6,l7}
\fmfright{r0,r1,r2,r3,r4,r5,r6,r7}
\fmfv{d.sh=circle,label=$\bar \G^w_{n+2L}$,l.a=90,d.filled=shaded,d.si=5thick,l.d=17}{b1}
\fmf{phantom,tension=5,label={\footnotesize $\bullet$},l.s=left}{r7,b2}
\fmf{phantom,tension=5,label={\footnotesize $\bullet$},l.s=right}{l7,b2}
\fmf{phantom}{b1,b2}
\fmfv{d.sh=circle,d.filled=full,label={\footnotesize $\bullet$},d.si=0.05thick,l.d=0}{b2}
\fmf{phantom}{b1,vv3}
\fmf{phantom,label=$ $,l.s=left,l.d=4}{r3,vv3}
\fmf{plain}{b1,vv6}
\fmf{plain,label=$\k_{n-1}$,l.s=left,l.d=2}{r6,vv6}
\fmf{phantom}{b1,vv5}
\fmf{phantom,label=$ $,l.s=right,l.d=0}{r5,vv5}
\fmf{plain}{b1,vv4}
\fmf{plain,label=$\k_n$,l.s=left,l.d=4}{r4,vv4}
\fmf{plain}{b1,v6}
\fmf{plain,label=$\k_2$,l.s=left,l.d=2}{v6,l6}
\fmf{phantom}{b1,v5}
\fmf{phantom,label=$ $,l.s=left,l.d=1.5}{v5,l5}
\fmf{plain}{b1,v4}
\fmf{plain,label=$\k_1$,l.s=left,l.d=2}{v4,l4}
\fmf{phantom}{b1,v3}
\fmf{phantom,label=$ $,l.s=left,l.d=4}{v3,l3}
\fmf{phantom}{b1,v2}
\fmf{phantom,label=$ $,l.s=right,l.d=0}{v2,l2}
\fmf{phantom}{b1,vv2}
\fmf{phantom,label=$ $,l.s=right,l.d=0}{vv2,r2}
\fmf{plain,tension=1,right=0.5,label=$\q_{L}$,l.s=left,l.d=8}{r1,b1}
\fmf{plain,tension=1,right=0.5}{b1,r1}
\fmf{plain,tension=1,right=0.5,label=$\q_1$,l.s=left,l.d=11}{l1,b1}
\fmf{plain,tension=1,right=0.5}{b1,l1}
\fmfv{d.sh=circle,d.filled=full,label={\footnotesize $\bullet$},d.si=0.05thick,l.d=0}{b3}
\fmf{phantom,tension=2,label={\footnotesize $\bullet$},l.s=right}{r0,b3}
\fmf{phantom,tension=2,label={\footnotesize $\bullet$},l.s=left}{l0,b3}
\fmf{phantom}{b1,b2}
\end{fmfgraph*}
\end{fmffile}
\end{gathered}\\
=&~ \frac{1}{(n+2L)!}\cdot(2L+n)...(2L+1)\cdot (2L-1)!! \\
&\times \prod_{i=1}^{L}\left[\int_{q_i\leq k_S}[dq_i]\,
g^2\bar P_s(q_i)\right]g^{-2}\bar
\G'^w_{n+2L}(\k_1,...,\k_n,\q_1,-\q_1,...,\q_L,-\q_L)
\end{split}
\label{vertdress0}
\ee 
Using Eq.~(\ref{IRnm}) we obtain for the leading IR-enhanced part,   
\be
\label{vertdress1}
V_{w,n}^{L-loop,LO}=
\frac{(-g^2\S)^{L}}{L!}  g^{-2} \bar\G'^w_n(\k_1,...,\k_n) \,. 
\ee
Clearly, resummation of these diagrams results in the substitution of
the wiggly vertex, 
\be 
\label{vertdress2}
\bar \G_n'^w(\k_1,...,\k_n)\mapsto  \exp\left\{ -g^2 \S \right\} 
\bar \G_n'^w(\k_1,...,\k_n)\,.
\ee
Combining all terms together, one obtains the resummed bispectrum,
\be
\begin{split}
B^{IR\;res,LO}_{w,\delta \delta \delta}(\e;\k_1,\k_2,\k_3)=&
- g^4(\e)\delta^{(3)}\big(\sum_l \k_l\big)\bigg[
\prod_{i=1}^3 \bar P_s(k_i) \;e^{-g^2(\e)\S}\bar\G'^w_3(\k_1,\k_2,\k_3)\\
&+\sum_{j=1}^{3}e^{-g^2(\e)\S}\bar P_w(k_j)\prod_{\begin{smallmatrix}
i=1\\
i\neq j
\end{smallmatrix}}^3 \bar P_s(k_i) \;\bar\G'^s_3(\k_1,\k_2,\k_3)\\
 &+\sum_{i<j=1}^{3}
 e^{-g^2(\e)\S}\big(\bar P_w(k_j) \bar P_s(k_i)+\bar P_s(k_j) \bar
 P_w(k_i)\big) \;K_2(\k_i,\k_j)\bigg]\\
=&e^{-g^2(\e)\S}B_{w,\delta\delta\delta}^{tree}(\e;\k_1,\k_2,\k_3)
\,.
\end{split} 
\label{bispres}
\ee
Recall that in the last expression the operator $e^{-g^2\S}$ should be
understood as acting on every occurrence of $P_w$ in the tree-level
expression for the bispectrum.
In terms of the SPT kernels, Eq.~(\ref{bispres}) 
can be rearranged into a somewhat simpler form,
\be
\label{bispres1}
\begin{split}
B^{IR\;res,LO}_{w,\delta \delta \delta}(\e;\k_1,\k_2,\k_3)=2
\sum_{i<j=1}^3F_2(\k_i,\k_j)e^{-g^2(\e)\S}\left(P_w(\e;k_j) P_s(\e;k_i)+i \leftrightarrow j\right)\,.
\end{split} 
\ee
The total bispectrum is obtained by adding to this expression 
the smooth tree-level part.

The above result extends to any equal-time $n$-point correlation
functions $\mathfrak{C}_n(\k_1,...,\k_n)$ of $\delta$ or $\T$
with hard external legs. Namely, the IR resummation at LO 
amounts to simply substituting the wiggly part of the linear spectrum,
$P_w$, that enters via wiggly vertices and propagators within the TSPT
tree-level calculation, by the resummed expression $e^{-g^2\S}P_w$. 
This can be summarized in the following compact form,
\be
\label{Cncomp}
\mathfrak{C}_n^{IR\;res,LO}(\k_1,...,\k_n)=\mathfrak{C}_n^{tree}\big[P_s+e^{-g^2\S}P_w\big](\k_1,...,\k_n)\;,
\ee
where $\mathfrak{C}_n^{tree}$ is understood as a functional of the linear power
spectrum. 
Note that, since the tree-level $n$-point
correlation functions, when summed over all perturbative
contributions, 
coincide in SPT and in TSPT, one can equivalently  
use the replacement (\ref{replace}) in the usual
SPT computations. However, the clear diagrammatic 
representation as daisy resummation is only possible within TSPT. In
addition, TSPT allows to systematically 
compute corrections to the LO resummation presented above.

\section{Taking into account hard loops}
\label{sec:hard}
So far, we have considered and resummed the contributions that in the 
power-counting scheme of Sec.~\ref{sec:PCsub}
are of order 
$(\s_h^2)^0\times (\s_S^2\times 1/\ve^2)^{L_s}$. We now discuss corrections to this result. 
One can discriminate two types of corrections: 
\begin{enumerate}
\item[(1)] 
Soft diagrams
with non-maximal IR enhancement, characterized by $l=2L_s-1$ (see
Eq.~(\ref{eq:powerc})), as well as the subleading terms in the
diagrams considered in Sec.~\ref{sec:leading}.
These contributions
are suppressed by one power of $\varepsilon$ relative to the leading order. 
\item[(2)] 
Diagrams with
one hard loop, $L_h=1$, and otherwise maximal IR enhancement $l=2L_s$. These diagrams are 
suppressed by one factor of $\sigma_h^2$ relative to the leading order.
\end{enumerate}
We refer to these two types of contributions as NLO$_s$ and NLO$_h$, respectively.
When combined, they constitute the total NLO correction.  In this section we analyze the contributions of the second type, while NLO$_s$ corrections will be included in the next section.

We start from the `hard' 1-loop contribution to the wiggly matter power spectrum\footnote{For the $\T$ power spectrum one simply omits the diagrams containing the kernels $K_n$.}, 

\begin{align}
         P_{w,\delta\delta}^{1-loop}(\e;k)\Big|_{hard}= & ~
           \begin{gathered}
   \begin{fmffile}{w-vertex4_2}
        \begin{fmfgraph*}(60,20)
        \fmfpen{thick}
        \fmfkeep{1loop}
        \fmfleft{l1}
        \fmfright{r1}
        \fmfv{d.sh=square,d.filled=empty,d.si=.04w,label=$\bar \G^s_4$,l.a=-90,l.d=.1w}{b1}
		\fmf{wiggly,label=$ $}{l1,b1}
		\fmf{plain,label=$ $}{b1,r1}
	     \fmf{plain,right=0.5,tension=0.9,label=$ $}{b1,b1}
	    \fmfdotn{b}{1}      
	    \end{fmfgraph*}
	         \end{fmffile}
	         	     \end{gathered}  
	   ~
	     + 
	     ~        
	     \begin{fmffile}{w-vertex3_1}
	    \begin{gathered}
        \begin{fmfgraph*}(60,20)
        \fmfpen{thick}
        \fmfkeep{1loop_2}
         \fmfleft{l1}
        \fmfright{r1}
        \fmfv{d.sh=circle,d.filled=full,d.si=.01w,label=$\bar \G^s_3$,l.a=135,l.d=.06w}{v1}
        \fmfv{d.sh=circle,d.filled=full,d.si=.01w,label=$\bar \G^s_3$,l.a=45,l.d=.06w}{v2}
		\fmf{wiggly,label=$ $}{l1,v1}
		\fmf{plain,label=$ $}{v2,r1}
	    \fmf{plain,left=1.0,tension=0.5,label=$ $}{v1,v2}
	    \fmf{plain,left=1.0,tension=0.5,label=$ $,l.side=left}{v2,v1}
	    \fmfposition
	    \fmfdotn{v}{2}      
	    \end{fmfgraph*}
	    \end{gathered}    
  \end{fmffile}
  ~+~
  \begin{fmffile}{w-vertex3_k2_1}
	    \begin{gathered}
        \begin{fmfgraph*}(60,20)
        \fmfpen{thick}
        \fmfkeep{1loop_2}
         \fmfleft{l1}
        \fmfright{r1}
        \fmfv{d.sh=circle,d.filled=full,d.si=.07w,label=$\bar \G^s_3$,l.a=135,l.d=.09w}{v1}
        \fmfv{d.sh=circle,d.filled=full,d.si=7,label=$K_2$,l.a=45,l.d=.09w}{v2}
		\fmf{wiggly,label=$ $}{v1,l1}
		\fmf{phantom_arrow,label=$ $}{r1,v2}
	    \fmf{plain,left=1.0,tension=0.5,label=$ $}{v1,v2}
	    \fmf{plain,left=1.0,tension=0.5,label=$ $,l.side=left}{v2,v1}
	    \fmfposition
	    \end{fmfgraph*}
	    \end{gathered}    
  \end{fmffile}\notag
  \\
  \nonumber
    \\
  \nonumber 
   &+~   \begin{gathered}
   \begin{fmffile}{1lw-c2}
         \begin{fmfgraph*}(60,20)
        \fmfpen{thick}
        \fmfkeep{2loop}
        \fmfleft{l1}
        \fmfright{r1}
        \fmfv{d.sh=cross,d.si=.30w,label=$C_2$,l.a=90,l.d=.15w}{b1}      
		\fmf{wiggly,label=$ $}{l1,b1}
		\fmf{plain,label=$ $}{b1,r1}
	    \end{fmfgraph*}
	         \end{fmffile}
	         	     \end{gathered}
~
+ 	 
~ 
    \begin{gathered}
   \begin{fmffile}{k3w2}
        \begin{fmfgraph*}(60,20)
        \fmfpen{thick}
        \fmfkeep{1loop}
        \fmfleft{l1}
        \fmfright{r1}
        \fmfv{d.sh=circle,d.filled=full,d.si=7,label=$K_3$,l.a=-90,l.d=.1w}{b1}
		\fmf{phantom_arrow,tension=2,label=$ $}{l1,b1}
		\fmf{wiggly,label=$ $}{b1,r1}
	     \fmf{plain,right=0.5,tension=0.9,label=$ $}{b1,b1}
	    \end{fmfgraph*}
	         \end{fmffile}
	         	     \end{gathered}\\
  	     \nonumber
  	     \\
  &+~
      \begin{gathered}
   \begin{fmffile}{w-vertex4}
        \begin{fmfgraph*}(60,20)
        \fmfpen{thick}
        \fmfkeep{1loop}
        \fmfleft{l1}
        \fmfright{r1}
        \fmfv{d.sh=square,d.filled=empty,d.si=.07w,label=$\bar \G^s_4$,l.a=-90,l.d=.1w}{b1}
		\fmf{plain,label=$ $}{l1,b1}
		\fmf{plain,label=$ $}{b1,r1}
	     \fmf{wiggly,right=0.5,tension=0.9,label=$ $}{b1,b1}
	    \fmfdotn{b}{1}      
	    \end{fmfgraph*}
	         \end{fmffile}
	         	     \end{gathered}  
	     ~
	     + 	 
	     ~ 
	        \begin{fmffile}{w-vertex3_3}
	    \begin{gathered}
        \begin{fmfgraph*}(60,20)
        \fmfpen{thick}
        \fmfkeep{1loop_2}
         \fmfleft{l1}
        \fmfright{r1}
    \fmfv{d.sh=circle,d.filled=full,d.si=.01w,label=$\bar \G^s_3$,l.a=135,l.d=.06w}{v1}
        \fmfv{d.sh=circle,d.filled=full,d.si=.01w,label=$\bar \G^s_3$,l.a=45,l.d=.06w}{v2}
		\fmf{plain,label=$ $}{l1,v1}
		\fmf{plain,label=$ $}{v2,r1}
	    \fmf{wiggly,left=1.0,tension=0.5,label=$ $}{v1,v2}
	    \fmf{plain,left=1.0,tension=0.5,label=$ $,l.side=left}{v2,v1}
	    \fmfposition
	    \fmfdotn{v}{2}      
	    \end{fmfgraph*}
	    \end{gathered}    
 \end{fmffile}\notag
 \\
\nonumber\\	  
\nonumber 
  	  &+~          \begin{fmffile}{k22wig}
	    \begin{gathered}
        \begin{fmfgraph*}(60,20)
        \fmfpen{thick}
        \fmfkeep{1loop_2}
         \fmfleft{l1}
        \fmfright{r1}
            \fmfv{d.sh=circle,d.filled=full,d.si=7,label=$K_2$,l.a=135,l.d=.09w}{v1}
        \fmfv{d.sh=circle,d.filled=full,d.si=7,label=$K_2$,l.a=45,l.d=.09w}{v2}
		\fmf{phantom_arrow,label=$ $}{l1,v1}
		\fmf{phantom_arrow,label=$ $}{r1,v2}
	    \fmf{wiggly,left=1.0,tension=0.5,label=$ $}{v1,v2}
	    \fmf{plain,left=1.0,tension=0.5,label=$ $,l.side=left}{v2,v1}
	    \fmfposition
	    \end{fmfgraph*}
	    \end{gathered}    
  \end{fmffile} 
  ~+~
      \begin{gathered}
   \begin{fmffile}{k3w1}
        \begin{fmfgraph*}(60,20)
        \fmfpen{thick}
        \fmfkeep{1loop}
        \fmfleft{l1}
        \fmfright{r1}
        \fmfv{d.sh=circle,d.filled=full,d.si=7,label=$K_3$,l.a=-90,l.d=.1w}{b1}
		\fmf{phantom_arrow,tension=2,label=$ $}{l1,b1}
		\fmf{plain,label=$ $}{b1,r1}
	     \fmf{wiggly,right=0.5,tension=0.9,label=$ $}{b1,b1}
	    \end{fmfgraph*}
	         \end{fmffile}
	         	     \end{gathered}  
~+~
 	   \begin{fmffile}{w-vertex3_k2_2}
	    \begin{gathered}
        \begin{fmfgraph*}(60,20)
        \fmfpen{thick}
        \fmfkeep{1loop_2}
         \fmfleft{l1}
        \fmfright{r1}
    \fmfv{d.sh=circle,d.filled=full,d.si=.07w,label=$\bar \G^s_3$,l.a=135,l.d=.09w}{v1}
        \fmfv{d.sh=circle,d.filled=full,d.si=7,label=$K_2$,l.a=45,l.d=.09w}{v2}
		\fmf{plain,label=$ $}{l1,v1}
		\fmf{phantom_arrow,label=$ $}{r1,v2}
	    \fmf{wiggly,left=1.0,tension=0.5,label=$ $}{v1,v2}
	    \fmf{plain,left=1.0,tension=0.5,label=$ $,l.side=left}{v2,v1}
	    \fmfposition 
	    \end{fmfgraph*}
	    \end{gathered}    
 \end{fmffile}\\
 \notag
 \\
  \label{1lwigfull}  &+~
  	    \begin{gathered}
	   \begin{fmffile}{w-vertex4_3_a1}
          \begin{fmfgraph*}(60,20)
        \fmfpen{thick}
        \fmfkeep{1loop}
        \fmfleft{l1}
        \fmfright{r1}
        \fmfv{d.sh=circle,d.filled=shaded,d.si=.2w,label=$ $,l.a=-90,l.d=.0w}{b1}
		\fmf{plain,label=$\bar \G^w_4$,l.side=left}{l1,b1}
		\fmf{plain,label=$ $}{b1,r1}
	     \fmf{plain,right=0.4,tension=0.7,label=$ $}{b1,b1}
	    \end{fmfgraph*}
  \end{fmffile}
  	     \end{gathered}
 ~+~
  \begin{fmffile}{w-vertex3_5}
	    \begin{gathered}
        \begin{fmfgraph*}(60,20)
        \fmfpen{thick}
        \fmfkeep{1loop_2}
         \fmfleft{l1}
        \fmfright{r1}
            \fmfv{d.sh=circle,d.filled=shaded,d.si=.2w,label=$ $,l.a=135,l.d=.00w}{v1}
        \fmfv{d.sh=circle,d.filled=full,d.si=.04w,label=$\bar \G^s_3$,l.a=45,l.d=.06w}{v2}
		\fmf{plain,label=$\bar \G^w_3$,l.side=left}{l1,v1}
		\fmf{plain,label=$ $}{v2,r1}
	    \fmf{plain,left=1.2,tension=0.4,label=$ $}{v1,v2}
	    \fmf{plain,left=1.2,tension=0.4,label=$ $,l.side=left}{v2,v1}
	    \fmfposition
	    \end{fmfgraph*}
	    \end{gathered}    
  \end{fmffile} 
~+~
  \begin{fmffile}{w-vertex3_k2_3}
	    \begin{gathered}
        \begin{fmfgraph*}(60,20)
        \fmfpen{thick}
        \fmfkeep{1loop_2}
         \fmfleft{l1}
        \fmfright{r1}
            \fmfv{d.sh=circle,d.filled=shaded,d.si=.2w,label=,l.a=135,l.d=.00w}{v1}
        \fmfv{d.sh=circle,d.filled=full,d.si=7,label=$K_2$,l.a=45,l.d=.09w}{v2}
		\fmf{plain,label=$\bar \G^w_3$,l.side=left}{l1,v1}
		\fmf{phantom_arrow,label=$ $}{r1,v2}
	    \fmf{plain,left=1.2,tension=0.4,label=$ $}{v1,v2}
	    \fmf{plain,left=1.2,tension=0.4,label=$ $,l.side=left}{v2,v1}
	    \fmfposition
	    \end{fmfgraph*}
	    \end{gathered}    
  \end{fmffile}
  ~\,
\end{align}
where the wavenumber $\p$ running in the loop is taken to be above the separation scale $k_S$, $p>k_S$. 
Note the appearance of a diagram with the counterterm $C_2$ in the second line. Similarly to the case of the tree-level bispectrum, all wiggly elements in these graphs can be dressed with soft daisies producing contributions of order 
\be
\label{NLOhorder}
\s_h^2\times (\s_S^2\times 1/\ve^2)^{L_s}\;.
\ee
Resummation of these contributions proceeds in a straightforward manner using the general expressions (\ref{vertdress0}), (\ref{vertdress1}) and yields,
\be
\label{eq:nh=1}
P_{w,\delta\delta}^{IR\;res,NLO_h} (\e;k)=e^{-g^2(\e)\S}P_{w,\delta\delta}^{1-loop}(\e;k)\Big|_{hard}\,,
\ee 
where the r.h.s stands for the 1-loop diagrams (\ref{1lwigfull}) 
computed using the wiggly power spectrum $e^{-g^2\S} P_w$ instead of the
linear power spectrum $P_w$ as an input.

Two comments are in order. First, the diagrams in the third and fourth lines in (\ref{1lwigfull}), as well as their descendants obtained by dressing with daisies, contain a wiggly element inside the loop. This implies that the integrand of the corresponding loop integral is oscillating leading to cancellation between positive and negative contributions. As a result, the corresponding diagrams are further suppressed compared to the naive estimate (\ref{NLOhorder}). In fact, the suppression is formally exponential, as follows from the general formula for the Fourier transform of a smooth function,
\[
\int [dp] \sin(p/k_{osc}) f_{smooth}(p/k)\sim e^{-k/k_{osc}}\sim e^{-1/\ve}\;.
\] 
From the viewpoint of our power-counting scheme, these contributions
are `non-per\-tur\-ba\-tive' and can be included or neglected without
changing the accuracy of the perturbative calculation. We prefer to keep them as
it allows us to write the result of the IR resummation in the compact
form (\ref{eq:nh=1}). 

Second, an alert reader might have noticed that the contributions discussed so far, namely those obtained by the daisy dressing of (\ref{1lwigfull}), do not exhaust all possible diagrams that formally would be of order (\ref{NLOhorder}) by the power-counting rules of Sec.~\ref{sec:PCsub}. The remaining diagrams fall into two categories. First, there are diagrams where a hard line closes on a wiggly vertex, which is in its turn attached to the external legs via a soft loop; a two-loop example is given by
\vspace{1cm}
\begin{align}
  	       \label{fig:wiglcom}
         & 
      \begin{gathered}
   \begin{fmffile}{wiglcom}
        \begin{fmfgraph*}(80,20)
        \fmfpen{thick}
        \fmfkeep{1loop}
        \fmfleft{l1,l2}
        \fmfright{r1,r2}
        \fmfv{d.sh=circle,d.filled=full,d.si=.05w,label=$\bar \G^s_4$,l.a=-90,l.d=.1w}{b1}
         \fmfv{d.sh=circle,d.filled=shaded,d.si=.12w,label=$\bar \G^w_4$,l.a=0,l.d=.2w}{b2}
		\fmf{plain,label=$ $}{l1,b1}
		\fmf{plain,label=$ $}{b1,r1}
		\fmf{phantom,label=$ $}{l2,b2}
		\fmf{phantom,label=$ $}{b2,r2}
	    \fmf{plain,right=0.9,tension=0.001,label=$ $}{b1,b2}
	    \fmf{plain,right=0.9,tension=0.001,label=$\textbf{q}$}{b2,b1}
	    \fmf{plain,right=0.05,tension=0.8,label=$\textbf{p}$}{b2,b2} 
	    \end{fmfgraph*}
	         \end{fmffile}
	         	     \end{gathered} \,.
\\
\notag
\end{align}
However, these diagrams necessarily contain a wiggly vertex inside a hard loop and thus, according to the previous discussion, are exponentially suppressed. They can be safely neglected.
Another set of extra
diagrams is given by the graphs where a hard loop is attached to 
a smooth propagator that belongs itself to a soft loop. For example, at two-loop order these are
\vspace{0.9cm}	
\begin{align}
  	       \label{2lwig-prop3}
   &
      \begin{gathered}
   \begin{fmffile}{2lw-a1}
        \begin{fmfgraph*}(70,30)
        \fmfpen{thick}
        \fmfkeep{1loop}
        \fmfleft{l1,l2}
        \fmfright{r1,r2}
        \fmfv{d.sh=circle,d.filled=shaded,d.si=.14w,label=$\bar \G^w_4$,l.a=-90,l.d=.14w}{b1}
         \fmfv{d.sh=circle,d.filled=full,d.si=.04w,label=$\bar \G^s_4$,l.a=0,l.d=.2w}{b2}
		\fmf{plain,label=$ $}{l1,b1}
		\fmf{plain,label=$ $}{b1,r1}
		\fmf{phantom,label=$ $}{l2,b2}
		\fmf{phantom,label=$ $}{b2,r2}
	    \fmf{plain,right=0.9,tension=0.001,label=$ $}{b1,b2}
	    \fmf{plain,right=0.9,tension=0.001,label=$\textbf{q}$}{b2,b1}
	    \fmf{plain,right=0.5,tension=0.9,label=$\textbf{p}$}{b2,b2} 
	    \end{fmfgraph*}
	         \end{fmffile}
	         	     \end{gathered}  ~+~
	  \begin{fmffile}{2lw-a2}
         \begin{fmfgraph*}(70,30)
        \fmfpen{thick}
        \fmfkeep{1loop}
        \fmfleft{l1,l2}
        \fmfright{r1,r2}
        \fmfv{d.sh=cross,d.filled=empty,d.si=.15w,label=$C_2$,l.a=0,l.d=.2w}{b2}
         \fmfv{d.sh=circle,d.filled=shaded,d.si=.14w,label=$\bar \G^w_4$,l.a=-90,l.d=.14w}{b1}
		\fmf{plain,label=$ $}{l1,b1}
		\fmf{plain,label=$ $}{b1,r1}
		\fmf{phantom,label=$ $}{l2,b2}
		\fmf{phantom,label=$ $}{b2,r2}
	    \fmf{plain,right=1.1,tension=0.001,label=$ $}{b1,b2}
	    \fmf{plain,right=1.1,tension=0.001,label=$\textbf{q}$}{b2,b1}
	    \fmf{phantom,right=0.5,tension=0.9,label=$ $}{b2,b2} 
	    \end{fmfgraph*}
	         \end{fmffile}~+~
	         	        \begin{fmffile}{2lw-a4}
         \begin{fmfgraph*}(70,30)
        \fmfpen{thick}
        \fmfkeep{1loop}
        \fmfleft{l1,l2}
        \fmfright{r1,r2}
        \fmfv{d.sh=circle,d.filled=empty,d.si=0.25w,label=$\textbf{p}$,l.a=45,l.d=.2w}{b2}
         \fmfv{d.sh=circle,d.filled=shaded,d.si=.14w,label=$\bar \G^w_4$,l.a=-90,l.d=.14w}{b1}
		\fmf{plain,label=$ $}{l1,b1}
		\fmf{plain,label=$ $}{b1,r1}
		\fmf{phantom,label=$ $}{l2,b2}
		\fmf{phantom,label=$ $}{b2,r2}
	    \fmf{plain,right=1.2,tension=0.001,label=$ $}{b1,b2}
	    \fmf{plain,right=1.2,tension=0.001,label=$\textbf{q}$}{b2,b1}
	    \fmf{phantom,right=0.5,tension=0.9,label=$ $}{b2,b2} 
	    \end{fmfgraph*}
	         \end{fmffile} \\
	\nonumber         \\
	\nonumber
	         &= \frac{g^2}{2}\int_{q\leq k_S} [dq] P^{1-loop}_{s,\T\T}(q)\Big|_{hard} \D_{\q} \D_{-\q} P_w(\e;k)  \,.
\end{align}
The subdiagrams attached to the soft loop combine into 
the hard part of the one-loop correction to the smooth velocity divergence power spectrum
 $P_{s,\T\T}^{1-loop}(\e;q)$, see \cite{Blas:2015qsi}.
As before, the `hard part' means that the loop integration in $P^{1-loop}_{s,\T\T}(q)$ is restricted to $p>k_S$.
This correction could be viewed as a loop-correction to the operator $\S$ suggesting to substitute the smooth power spectrum in $\S$ by $P_s\mapsto P_s+ P^{1-loop}_{s,\T\T}\Big|_{hard}$.
While it is tempting to do this replacement, we refrain from this step for two reasons. First, this would correspond
to a partial resummation of \emph{hard} loops, and therefore goes against the rigorous expansion in our power counting
parameters. 
Second, and more importantly, the hard loop corrections to the power
spectrum are universally suppressed as $q^2/k_S^2$ at soft momenta
$q\ll k_S$ due to the well-known decoupling of long and short modes
\cite{Peebles:1980,Goroff:1986ep}
(see also \cite{Blas:2014hya}). 
This suppression essentially removes the IR enhancement of
the operator $\D_\q\D_{-\q}$ in (\ref{2lwig-prop3}), so that the term
(\ref{2lwig-prop3}) is suppressed by $\langle q\rangle^2/k_S^2$
compared to the contributions resummed in (\ref{eq:nh=1}). 
Here $\langle q\rangle$ is the characteristic IR scale which, as
discussed in Sec.~\ref{sec:PCsub}, is of order $k_{eq}$. Therefore,
the contribution (\ref{2lwig-prop3}) is small as long as $k_{eq}\ll
k_S$. In the real universe the hierarchy between $k_S$ and $k_{eq}$
cannot be too large. Nevertheless, we have checked numerically that
for the standard $\Lambda$CDM and the realistic range of values 
$k_S \sim O(0.1h/$Mpc$)$, the contribution (\ref{2lwig-prop3}) is
quantitatively unimportant. In what follows we neglect the diagrams
with hard loops inside soft ones.

The expression (\ref{eq:nh=1}) can be cast in a more convenient form
by the following steps. Let us add and subtract the soft part of
the one-loop diagrams,
\be
\begin{split}
P_{w,\delta\delta}^{IR\;res,NLO_h}=
&e^{-g^2\S}P_{w,\delta\delta}^{1-loop}\Big|_{hard}
+e^{-g^2\S}P_{w,\delta\delta}^{1-loop}\Big|_{soft}
-e^{-g^2\S}P_{w,\delta\delta}^{1-loop}\Big|_{soft}\\
=&e^{-g^2\S}P_{w,\delta\delta}^{1-loop}+e^{-g^2\S}g^2\S P_w\;,
\end{split}
\ee
where in the last line $P_{w,\delta\delta}^{1-loop}$ stands for the
{\em total} one-loop correction and we used Eq.~(\ref{1lpw}) to express
the leading soft-loop contribution. Combining this with the one-loop
correction to the smooth power spectrum and the LO expression
(\ref{eq:finalLO}) we obtain the final expression for the power
spectrum with NLO$_h$ corrections included,
\be
\label{finalNLOh}
P_{\delta\delta}^{IR\;res,LO+NLO_h}
=P_s+(1+g^2\S)e^{-g^2\S}P_w
+P_{\delta\delta}^{1-loop}\big[P_s+e^{-g^2\S}P_w\big]\;.
\ee
Here $P_{\delta\delta}^{1-loop}$ is understood as the functional of the
input linear power spectrum; the latter has been modified by the IR
resummation. Note the appearance of the term $g^2\S e^{-g^2\S}P_w$ in
the `tree-level' part of the expression (\ref{finalNLOh}) that is
important to avoid double counting of the soft contributions.

The formula (\ref{finalNLOh}) can be straightforwardly generalized to
other $n$-point functions, both of density and velocity divergence. We
give here the final result without derivation,
\be
\label{finalNLOhCn}
\mathfrak{C}_{n}^{IR\;res,LO+NLO_h}
=\mathfrak{C}_n^{tree}\big[P_s+(1+g^2\S)e^{-g^2\S}P_w\big]
+\mathfrak{C}_{n}^{1-loop}\big[P_s+e^{-g^2\S}P_w\big]\;.
\ee
Moreover, it is possible to include higher hard loops,
i.e. corrections of order $\s_h^{2L_h}$ with $L_h\geq 2$. Let us do it
for the power spectrum\footnote{The derivation does not depend on the
  type of the power spectrum ($\delta\delta$, $\T\T$ or $\T\delta$)
so we do
  not specify it explicitly.}. Repeating the arguments that led to
Eq.~(\ref{eq:nh=1}), one finds at the NNLO$_h$ order,
\be
\label{eq:nh=2}
P_w^{IR\;res,NNLO_h}=e^{-g^2\S}P_w^{2-loop}\Big|_{hh}\;,
\ee
where the 2-loop contribution on the r.h.s. is `double-hard', i.e. the
integration in both loops run over $p>k_S$. This is equivalently
written in the form,
\be
\label{eq:nh=2-1}
P_w^{IR\;res,NNLO_h}=e^{-g^2\S}P_w^{2-loop}
-e^{-g^2\S}P_w^{2-loop}\Big|_{hs}
-e^{-g^2\S}P_w^{2-loop}\Big|_{ss}\;,
\ee
where the first term contains the total 2-loop contribution, whereas the
second and third terms are `hard-soft' and `soft-soft'
respectively. Next, we use the relations valid at the leading soft
order,
\bseq
\begin{align}
\label{Pss}
&P_w^{2-loop}\Big|_{ss}=\frac{g^4\S^2}{2}P_w\;,\\
\label{Phs}
&P_w^{2-loop}\Big|_{hs}=-g^2\S P_w^{1-loop}\Big|_h
=-g^2\S P_w^{1-loop}-g^4\S^2P_w\;,
\end{align} 
\eseq 
where to obtain the last expression in (\ref{Phs}) we have again added
and subtracted the soft one-loop contribution. Combining everything
together and with the NLO$_h$ expression (\ref{finalNLOh}) we arrive
at a compact formula,
\be
\label{finalNNLOh}
\begin{split}
P^{IR\;res,LO+NLO_h+NNLO_h}
=&P_s+\bigg(1+g^2\S+\frac{g^4\S^2}{2}\bigg)e^{-g^2\S}P_w\\
+P^{1-loop}&\big[P_s+(1+g^2\S)e^{-g^2\S}P_w\big]
+P^{2-loop}\big[P_s+e^{-g^2\S}P_w\big]\;.
\end{split}
\ee
Its extension to other correlation functions and to higher orders in
hard loops can be done by proceeding analogously.

The expressions (\ref{finalNLOh}), (\ref{finalNLOhCn}),
(\ref{finalNNLOh}) do not take into account subleading soft
corrections. These are formally of order $\ve$ and thus can be
expected to be significant as in the real universe the latter
parameter is not small, $\ve\sim 0.2\div 0.3$. Therefore the next
section is devoted to a detailed computation of these
corrections. Notably, we will find that the dominant NLO$_s$ corrections
are already captured by the one-loop term in (\ref{finalNLOh}) for the
power spectrum, such that it works remarkably well. Still,
it is important to properly assess the NLO$_s$ contributions because
they are required for a reliable determination of the shift of the
BAO peak, as discussed in Sec.~\ref{sec:shift}.

\section{Resummation of infrared effects at next-to leading order}
\label{sec:NLO}

Here we discuss corrections to the power spectrum 
that are suppressed by one power of $\varepsilon$ relative to the
leading order IR resummed results discussed previously. 
There are two possibilities to obtain such corrections: 
($a$) On the one hand, one can consider the same daisy diagrams,
\emph{i.e.} with maximal IR enhancement $l=2L_s$, but take into
account the first sub-leading correction in the expansion of the
wiggly vertex \eqref{asympPs}. ($b$) On the other hand, one can
consider new diagrams with $l=2L_s-1$.  
All diagrams that contribute at $O(\varepsilon)$ to the velocity
divergence power spectrum at increasing number of loops  are given by 
 
\be
\begin{split}
  	       \label{1lwig-nlo}
          P_{w,\T\T}^{IR\;res,LO+NLO_s}(\e;k)= & ~
         \begin{fmffile}{w-ps-b}
	    \begin{gathered}
        \begin{fmfgraph*}(60,20)
        \fmfpen{thick}
        \fmfleft{l1}
        \fmfright{r1}
		\fmf{wiggly,label=$\textbf{k}$}{l1,r1}
	    \end{fmfgraph*}
	     \end{gathered}
	    \end{fmffile} ~+~ 
    \begin{gathered}
   \begin{fmffile}{1lw-1b}
         \begin{fmfgraph*}(60,20)
        \fmfpen{thick}
        \fmfkeep{2loop}
        \fmfleft{l1}
        \fmfright{r1}
        \fmfv{d.sh=circle,d.filled=shaded,d.si=.15w,label=$\bar \G^w_4$,l.a=-30,l.d=.2w}{b1}
		\fmf{plain,label=$ $}{l1,b1}
		\fmf{plain,label=$ $}{b1,r1}
	    \fmf{plain,right=0.5,tension=0.6,label=$ $,l.side=left}{b1,b1}
	    \end{fmfgraph*}
	         \end{fmffile}
	         	     \end{gathered}  
	         	     ~
	         	     +
	         	     ~
  	    \begin{fmffile}{1lw-2b}
	    \begin{gathered}
        \begin{fmfgraph*}(80,40)
        \fmfpen{thick}
         \fmfleft{l1,l2,l3}
        \fmfright{r1,r2,l3}
        \fmfv{d.sh=circle,d.filled=shaded,d.si=.12w,label=$\bar \G^w_3$,l.a=135,l.d=.06w}{v1}
        \fmfv{d.sh=circle,d.filled=full,d.si=.04w,label=$\bar \G^s_3$,l.a=45,l.d=.06w}{v2}
		\fmf{plain,tension=1.0,label=$ $}{l2,v1}
		\fmf{plain,tension=1.0,label=$ $}{v2,r2}
		\fmf{phantom,tension=1.0,label=$ $}{l1,b1}
		\fmf{phantom,tension=1.0,label=$ $}{b2,r1}
		\fmf{phantom,tension=0.05,label=$ $}{b1,b2}
		\fmf{phantom,left=0.01,tension=0.0001,label=$ $}{v2,b2}
	    \fmf{phantom,left=0.01,tension=0.0001,label=$ $}{b2,v2}
	    \fmf{phantom,right=0.65,tension=0.000001,label=$ $,l.side=left,l.d=0.03w}{v1,b1}
	    \fmf{phantom,right=0.5,tension=0.000001,label=$ $}{b1,v1}
	    \fmf{plain,left=1.0,tension=0.3,label=$ $}{v1,v2}
	    \fmf{plain,left=1.0,tension=0.3,label=$ $,l.side=left}{v2,v1}
	    \fmfposition
	    \end{fmfgraph*}
	    \end{gathered}    
  \end{fmffile}  \\ \\
~+~  
      \begin{gathered}
   \begin{fmffile}{2lw-3b}
         \begin{fmfgraph*}(60,20)
        \fmfpen{thick}
        \fmfkeep{2loop}
        \fmfleft{l1}
        \fmfright{r1}
        \fmfv{d.sh=circle,d.filled=shaded,d.si=.15w,label=$\bar \G^w_6$,l.a=-30,l.d=.2w}{b1}
		\fmf{plain,label=$ $}{l1,b1}
		\fmf{plain,label=$ $}{b1,r1}
	    \fmf{plain,right=0.5,tension=0.6,label=$ $,l.side=left}{b1,b1}
	    \fmf{plain,left=0.5,tension=0.6,label=$ $,l.side=left}{b1,b1}
	    \end{fmfgraph*}
	         \end{fmffile}
	         	     \end{gathered}  
	         	     ~
	         	     +
	         	     ~&
  	    \begin{fmffile}{2lw-4b}
	    \begin{gathered}
        \begin{fmfgraph*}(80,40)
        \fmfpen{thick}
         \fmfleft{l1,l2,l3}
        \fmfright{r1,r2,r3}
        \fmfv{d.sh=circle,d.filled=shaded,d.si=.12w,label=$\bar \G^w_5$,l.a=135,l.d=.06w}{v1}
        \fmfv{d.sh=circle,d.filled=full,d.si=.04w,label=$\bar \G^s_3$,l.a=45,l.d=.06w}{v2}
		\fmf{plain,tension=1.0,label=$ $}{l2,v1}
		\fmf{plain,tension=1.0,label=$ $}{v2,r2}
		\fmf{phantom,tension=1.0,label=$ $}{l1,b1}
		\fmf{phantom,tension=1.0,label=$ $}{b2,r1}
		\fmf{phantom,tension=0.05,label=$ $}{b1,b2}
		\fmf{phantom,left=0.01,tension=0.0001,label=$ $}{v2,b2}
	    \fmf{phantom,left=0.01,tension=0.0001,label=$ $}{b2,v2}
	    \fmf{plain,right=0.65,tension=0.000001,label=$ $,l.side=left,l.d=0.03w}{v1,b1}
	    \fmf{plain,right=0.5,tension=0.000001,label=$ $}{b1,v1}
	    \fmf{plain,left=1.0,tension=0.3,label=$ $}{v1,v2}
	    \fmf{plain,left=1.0,tension=0.3,label=$ $,l.side=left}{v2,v1}
	    \fmfposition
	    \end{fmfgraph*}
	    \end{gathered}    
  \end{fmffile} 
  ~+~ 
      \begin{gathered}
   \begin{fmffile}{2lw-5b}
         \begin{fmfgraph*}(80,40)
        \fmfpen{thick}
        \fmfkeep{2loop}
        \fmfleft{l3,l1,l2}
        \fmfright{r3,r1,r2}
\fmfv{d.sh=circle,d.filled=shaded,d.si=.12w,label=$\bar\G^w_5$,l.a=-90,l.d=.1w}{b1}
\fmfv{d.sh=circle,d.filled=full,d.si=.04w,label=$\bar\G^s_3$,l.a=90,l.d=.07w}{b2} 
		\fmf{plain,label=$ $}{l1,b1}
		\fmf{plain,label=$ $}{b1,r1}
		\fmf{phantom,label=$ $}{l2,b2}
		\fmf{phantom,label=$ $}{b2,r2}
		\fmf{plain,right=1.2,tension=0.001,label=$ $,l.side=left}{b1,b2}
		\fmf{plain,left=1.2,tension=0.001,label=$ $,l.side=left}{b1,b2}
		\fmf{plain,right=0.0,tension=0.001,label=$ $,l.side=left}{b1,b2}
	    \end{fmfgraph*}
	         \end{fmffile}
	         	     \end{gathered}   \\ \\
	       ~+~  	     
\begin{fmffile}{w-vertex8-b}
	    \begin{gathered}
        \begin{fmfgraph*}(80,64)
        \fmfpen{thick}
        \fmfkeep{3loop}
        \fmfleft{l1,l2,l3}
        \fmfright{r1,r2,r3}
        \fmfv{d.sh=circle,d.filled=shaded,d.si=.12w,label=$\bar \G^w_8$,l.a=90,l.d=.15w}{b1}
		\fmf{plain,l.side=left,label=$ $}{l2,b1}
		\fmf{plain,l.side=left,label=$ $}{b1,r2}
		\fmf{phantom}{l1,u1}
		\fmf{phantom}{u2,r1}
		\fmf{phantom,tension=2}{l3,v1}
		\fmf{phantom,tension=2}{v1,r3}
	    \fmf{plain,right=0.55,tension=0.01,l.side=left}{b1,u1}
	   \fmf{plain,left=0.55,tension=0.01,l.side=right,label=$ $}{b1,u1}
	   	    \fmf{plain,right=0.55,tension=0.01,l.side=left,label=$ $}{b1,u2}
	   \fmf{plain,left=0.55,tension=0.01}{b1,u2}
	     \fmf{plain,right=0.7,tension=0.00001,label=$ $}{b1,v1}
	   \fmf{plain,left=0.7,tension=0.00001}{b1,v1}
	    \end{fmfgraph*}
	    	    	    \end{gathered}
	    \end{fmffile}
 &~+~
	      	    \begin{fmffile}{3lw-1b}
	    \begin{gathered}
        \begin{fmfgraph*}(80,40)
        \fmfpen{thick}
         \fmfleft{l1,l2,l3}
        \fmfright{r1,r2,r3}
        \fmfv{d.sh=circle,d.filled=shaded,d.si=.12w,label=$\bar \G^w_7$,l.a=0,l.d=.12w}{v1}
        \fmfv{d.sh=circle,d.filled=full,d.si=.04w,label=$\bar \G^s_3$,l.a=45,l.d=.09w}{v2}
		\fmf{plain,tension=1.0,label=$ $}{l2,v1}
		\fmf{plain,tension=1.0,label=$ $}{v2,r2}
		\fmf{phantom,tension=1.0,label=$ $}{l1,b1}
		\fmf{phantom,tension=1.0,label=$ $}{b2,r1}
		\fmf{phantom,tension=0.05,label=$ $}{b1,b2}
		\fmf{phantom,left=0.01,tension=0.0001,label=$ $}{v2,b2}
	    \fmf{phantom,left=0.01,tension=0.0001,label=$ $}{b2,v2}
	    \fmf{plain,right=0.65,tension=0.000001,label=$ $,l.side=left,l.d=0.03w}{v1,b1}
	    \fmf{plain,right=0.5,tension=0.000001,label=$ $}{b1,v1}
	    \fmf{phantom,tension=1.0,label=$ $}{l3,a1}
		\fmf{phantom,tension=1.0,label=$ $}{a2,r3}
		\fmf{phantom,tension=0.05,label=$ $}{a1,a2}
		\fmf{phantom,left=0.01,tension=0.0001,label=$ $}{v2,a2}
	    \fmf{phantom,left=0.01,tension=0.0001,label=$ $}{a2,v2}
	    \fmf{plain,right=0.65,tension=0.000001,label=$ $,l.side=left,l.d=0.03w}{v1,a1}
	    \fmf{plain,right=0.5,tension=0.000001,label=$ $}{a1,v1}
	    \fmf{plain,left=1.0,tension=0.3,label=$ $}{v1,v2}
	    \fmf{plain,left=1.0,tension=0.3,label=$ $,l.side=left}{v2,v1}
	    \fmfposition
	    \end{fmfgraph*}
	    \end{gathered}    
  \end{fmffile}  ~+~ 
  \begin{gathered}
   \begin{fmffile}{3lw-2b}
         \begin{fmfgraph*}(80,40)
        \fmfpen{thick}
        \fmfkeep{2loop}
        \fmfleft{l3,l1,l2}
        \fmfright{r3,r1,r2}
        \fmfv{d.sh=circle,d.filled=shaded,d.si=.1w,label=$\bar\G^w_7$,l.a=-35,l.d=.15w}{b1}
\fmfv{d.sh=circle,d.filled=full,d.si=.04w,label=$\bar\G^s_3$,l.a=90,l.d=.07w}{b2}
		\fmf{plain,label=$ $}{l1,b1}
		\fmf{plain,label=$ $}{b1,r1}
		\fmf{phantom,label=$ $}{l2,b2}
		\fmf{phantom,label=$ $}{b2,r2}
		\fmf{plain,right=1.2,tension=0.001,label=$ $,l.side=left}{b1,b2}
		\fmf{plain,left=1.2,tension=0.001,label=$ $,l.side=left}{b1,b2}
		\fmf{plain,right=0.0,tension=0.001,label=$ $,l.side=left}{b1,b2}
	    \fmf{plain,tension=0.8,label=$ $,l.side=right}{b1,b1}
	    \end{fmfgraph*}
	         \end{fmffile}
\end{gathered} ~+~ ...
  ~ ~~~\,.
\end{split}
\ee

\noindent
Here all loops are soft, \emph{i.e.} integrated over wavenumbers below $k_S$.
The daisy diagrams containing the vertices 
$\bar\G_4^w, \bar\G_6^w, \bar\G_8^w\dots$ at 1, 2, 3-loop, and so on, belong
to the category ($a$). The diagrams containing $\bar\G_3^w, \bar\G_5^w, \bar\G_7^w\dots $ at 1, 2, 3-loop, and so on, 
have $l=1, 3, 5, \dots$, and therefore belong to category ($b$).
All diagrams in category ($b$) are related to two new diagrams:
the {\it fish} (last diagram in the first line) 
and {\it oyster} (last diagram in the second line).
All higher-loop diagrams of type ($b$) are obtained by dressing
the wiggly vertex contained
in these diagrams with daisy loops, 
such as \emph{e.g.} the middle diagram in the second
line and the last two diagrams in the last line.
The matter power spectrum  includes in addition at $O(\varepsilon)$ the
dressed fish diagram with the vertex $K_2$, 
\be
\label{cs-nlo}
\begin{split}
P^{IR\;res,LO+NLO_s}_{w,\delta \delta}(\e,k)= &~P^{IR\;res,LO+NLO_s}_{w,\T\T}(\e,k)  \\
  	 & + \; \begin{fmffile}{1lw-cs}
	    \begin{gathered}
        \begin{fmfgraph*}(80,40)
        \fmfpen{thick}
         \fmfleft{l1,l2,l3}
        \fmfright{r1,r2,l3}
        \fmfv{d.sh=circle,d.filled=shaded,d.si=.12w,label=$\bar \G^w_3$,l.a=135,l.d=.12w}{v1}
        \fmfv{decor.shape=circle,decor.filled=full,d.si=0.09w,label=$\quad K_2$,l.a=-45,l.d=.08w}{v2}
		\fmf{plain,tension=1.0,label=$ $}{l2,v1}
		\fmf{phantom_arrow,tension=1.0,label=$ $}{r2,v2}
	    \fmf{plain,left=1.0,tension=0.3,label=$ $}{v1,v2}
	    \fmf{plain,left=1.0,tension=0.3,label=$ $,l.side=left}{v2,v1}
	    \fmfposition
	    \end{fmfgraph*}
	    \end{gathered}    
  \end{fmffile} 
\end{split} 
  \raisebox{-4mm}{$  ~+~
  \begin{fmffile}{2lw-cs}
	    \begin{gathered}
        \begin{fmfgraph*}(80,40)
        \fmfpen{thick}
         \fmfleft{l1,l2,l3}
        \fmfright{r1,r2,l3}
        \fmfv{d.sh=circle,d.filled=shaded,d.si=.12w,label=$\bar \G^w_5$,l.a=135,l.d=.12w}{v1}
        \fmfv{decor.shape=circle,decor.filled=full,d.si=0.09w,label=$\quad K_2$,l.a=-45,l.d=.08w}{v2}
		\fmf{plain,tension=1.0,label=$ $}{l2,v1}
		\fmf{phantom_arrow,tension=1.0,label=$ $}{r2,v2}
		\fmf{phantom,tension=1.0,label=$ $}{l1,b1}
		\fmf{phantom,tension=1.0,label=$ $}{b2,r1}
		\fmf{phantom,tension=0.05,label=$ $}{b1,b2}
		\fmf{phantom,left=0.01,tension=0.0001,label=$ $}{v2,b2}
	    \fmf{phantom,left=0.01,tension=0.0001,label=$ $}{b2,v2}
	    \fmf{plain,right=0.65,tension=0.000001,label=$ $,l.side=left,l.d=0.03w}{v1,b1}
	    \fmf{plain,right=0.5,tension=0.000001,label=$ $}{b1,v1}
	    \fmf{plain,left=1.0,tension=0.3,label=$ $}{v1,v2}
	    \fmf{plain,left=1.0,tension=0.3,label=$ $,l.side=left}{v2,v1}
	    \fmfposition
	    \end{fmfgraph*}
	    \end{gathered}    
  \end{fmffile} ~+~~~...\,.
  $}
\ee

We first consider the diagrams of type ({\it b}). Dressing of the fish
diagrams simply leads to the already familiar replacement (\ref{replace}).
We do not need these contributions explicitly, as our eventual
goal is to combine them with the hard corrections (\ref{eq:nh=1}) to
form the total fish diagrams evaluated using the modified wiggly power
spectrum (\ref{replace}).  

For the oyster diagram, a straightforward evaluation using the
expansion (\ref{IRnm}) for $\bar\G_5^w$ and the exact expression for
$\bar\G_3^s$ yields,
\vspace{0.3cm}
\be
\label{oyster}
\delta P_{w,oyster}^{1-loop}\Big|_{soft}
\equiv
\begin{gathered}
\begin{fmffile}{2lw-5b-1}
         \begin{fmfgraph*}(80,40)
        \fmfpen{thick}
        \fmfkeep{2loop}
        \fmfleft{l3,l1,l2}
        \fmfright{r3,r1,r2}
\fmfv{d.sh=circle,d.filled=shaded,d.si=.12w,label=$\bar\G^w_5$,l.a=-90,l.d=.1w}{b1}
\fmfv{d.sh=circle,d.filled=full,d.si=.04w,label=$\bar\G^s_3$,l.a=90,l.d=.07w}{b2} 
		\fmf{plain,label=$ $}{l1,b1}
		\fmf{plain,label=$ $}{b1,r1}
		\fmf{phantom,label=$ $}{l2,b2}
		\fmf{phantom,label=$ $}{b2,r2}
		\fmf{plain,right=1.2,tension=0.001,label=$ $,l.side=left}{b1,b2}
		\fmf{plain,left=1.2,tension=0.001,label=$ $,l.side=left}{b1,b2}
		\fmf{plain,right=0.0,tension=0.001,label=$ $,l.side=left}{b1,b2}
	    \end{fmfgraph*}
\end{fmffile}
\end{gathered} ~\Bigg|_{soft}
=g^6(\S^a+2\varkappa \S^b)\bar P_w(k)\;,
\ee
where we introduced the operators,
\bseq
\begin{align}
\label{SSa}
\S^a\bar P_w(k)=&
2\!\!\!\!\!\int\limits_{q,q'\leq k_S}\!\!\!\![dq][dq']\bar P_s(q)\bar P_s(q')
\frac{(\k\cdot\q)^2(\k\cdot\q')(\q\cdot\q')}{q^4q'^4}
\sinh(\q\!\cdot\!\nabla)\big(1-\cosh(\q'\!\!\cdot\!\nabla)\big)\bar
P_w(k)\,,\\
\S^b\bar P_w(k)=&\frac{6}{7}
\int\limits_{q,q'\leq k_S}\!\!\!\![dq][dq']\bar P_s(q)\bar P_s(q')
\sin^2(\q,\q')\notag\\
&\times\frac{(\k\cdot\q)(\k\cdot\q')(\k\cdot(\q+\q'))}{q^2q'^2(\q+\q')^2}
\sinh(\q\!\cdot\!\nabla)\big(1-\cosh(\q'\!\!\cdot\!\nabla)\big)\bar
P_w(k)\,,
\label{SSb}
\end{align}
\eseq
and $\varkappa=1$ ($\varkappa=0$) in ED (ZA). The dressing of the
wiggly vertex by daisies again leads to the replacement
(\ref{replace}) in this formula,
\be
\label{oysterres}
\delta
P_{w,oyster}^{IR\;res,NLO_s}=g^6\big(\S^a+2\varkappa\S^b\big)
e^{-g^2\S}\bar P_w(k)\;.
\ee 

Next, we turn to the contributions of type ({\it a}), i.e. daisy
diagrams expanded to the NLO$_s$ order. Remarkably, these also can be
resummed. To see this, we need the first corrections in $\ve$ to
Eq.~(\ref{asympPs}). The latter are derived in 
Appendix~\ref{app:sublead} with the result,
\be
\label{NLOG}
 \begin{split}
&\bar\G'^{w,LO+NLO}_{n}\Big(\k,-\k-\sum^{n-2}_{i=1}\q_i,\q_1,..,\q_{n-2}\Big)\\
&\qquad=(-1)^{n-1}\left[\prod_{i=1}^{n-2}\D_{\q_i}+\sum_{j=1}^{n-2}\E_{\q_j}
\prod_{\begin{smallmatrix}
i=1\\
i\neq j
\end{smallmatrix}
}^{n-2}\mathcal{D}_{\q_i} +
\sum_{\begin{smallmatrix}j_1,j_2=1\\
j_1<j_2\end{smallmatrix}}^{n-2}\F_{\q_{j_1}\q_{j_2}}
\prod_{\begin{smallmatrix}i=1\\i\neq j_1,j_2\end{smallmatrix}}^{n-2}
\D_{\q_i}\right]\frac{\bar P_w(k)}{\bar P^2_s(k)}\,,
\end{split} 
\ee
where
$\E_\q$ and $\F_{\q_1\q_2}$ are new finite-difference operators 
acting on $\bar P_w$ that are defined in 
\eqref{eq:sublop1} and \eqref{eq:sublop2}.
Inserting this expression into Eq.~(\ref{leadlooplim}) and summing
over the number of loops $L$ we obtain,
\be
\label{dasiessub}
\delta P_{w,daisy}^{IR\;res,LO+NLO_s}
=\big(g^2+g^4\S^c-g^6\S^a-g^6\varkappa\S^b\big)e^{-g^2\S}\bar P_w(k)\;,
\ee
where
\be
\label{SSc}
\S^c\bar P_w(k)=\frac{1}{2}\int_{q\leq k_S}
[dq]\bar P_s(q)\big(\E_\q\D_{-\q}+\E_{-\q}\D_\q+\F_{\q,-\q}\big)\bar P_w(k)\;,
\ee
and we used the relation
\be
\frac{1}{2}\int_{q,q'\leq k_S}
[dq][dq']\bar P_s(q)\bar P_s(q')\F_{\q,\q'}\D_{-\q}\D_{-\q'}
=-(\S^a+\varkappa\S^b)\;.
\ee

It is desirable to connect the expression (\ref{dasiessub}) to the 
one-loop daisy graph computed using the modified linear power spectrum,
\be
\delta P_{w,daisy}^{1-loop,LO+NLO_s}[P_s+e^{-g^2\S} P_w]\;.
\ee 
One may be tempted to compute this by simply making the replacement
(\ref{replace}) in the NLO expression (\ref{eq:G4nlo}) for the vertex
function $\bar \G_4'^w$. However, this would produce a mistake at the
NLO$_s$ order. The reason is that the operator $\S$ depends on the
hard wavenumber $\k$ (see Eq.~(\ref{Slead})). When $\bar \G_4'^w$ is
evaluated with the new wiggly power spectrum $e^{-g^2\S}\bar P_w$, the
operator $\S$ gets shifted,
\[
\S\big|_{\k}\mapsto \S\big|_{\k\pm\q}=\S\big|_{\k}\pm\Delta \S\big|_\q\;.
\]
The correct expression  for the vertex 
taking this shift into account is derived in
Appendix~\ref{app:sublead}, see Eq.~(\ref{eq:expdressed}).
Using it to evaluate the loop we obtain,
\be
\label{daisynew}
\delta P_{w,daisy}^{1-loop,LO+NLO_s}[P_s+e^{-g^2\S}P_w]\Big|_{soft}
=\big(-g^4\S+g^4\S^c-g^6\S^a\big)e^{-g^2\S}\bar P_w(k)\;.
\ee
This should be compared with (\ref{dasiessub}). We have 
\be
\label{nlodais}
\begin{split}
\delta P_{w,daisy}^{IR\;res,LO+NLO_s}
=&g^2(1+g^2\S)e^{-g^2\S}\bar P_w
+\delta P_{w,daisy}^{1-loop,LO+NLO_s}[P_s+e^{-g^2\S}P_w]\Big|_{soft}\\
&-g^6\varkappa\S^be^{-g^2\S}\bar P_w\,.
\end{split}
\ee
Notice that the operators $\S^a$, $\S^c$ have dropped out of
this relation.

We are now ready to combine all NLO contributions together. These
include the resummed hard loops (\ref{eq:nh=1}), soft fish diagrams,
the contribution of the soft oyster diagram (\ref{oysterres}), and the
resummed daisies (\ref{nlodais}). Adding to them the smooth part we
obtain,
\be
\label{irres3}
\begin{split}
P^{IR\;res,LO+NLO}_w=&P_s+(1+g^2\S)e^{-g^2\S}P_w
+P^{1-loop}[P_s+e^{-g^2\S} P_w]\\
&+g^4(\S^a+\varkappa\S^b)e^{-g^2\S} P_w\,.
\end{split} 
\ee
This is our final result for the full NLO power spectrum.
At face value, it differs from the 
NLO$_h$ formula (\ref{finalNLOh}) only 
by the last term. However, 
we emphasize that now all subleading IR corrections have been
consistently included.  
In particular, the one-loop contribution differs for ZA and ED as well
as for density and velocity correlators at the NLO$_s$ order. 
The last term scales as $g^4$ compared to the leading piece. This
reflects 
that it receives contributions only starting from
2-loops. Nevertheless, this term is of order $O(\ve)$ in our power
counting and therefore must be retained. 
Still, we will find below that it happens to be numerically small.

\section{Practical implementation and comparison with other methods}\label{sec:pract}

In this section we first discuss how the IR resummed power spectrum obtained
in TSPT can be evaluated in practice,
 and then compare to other analytical approaches as well as to $N$-body data. 
 In the last part  we discuss the predictions for the shift of the BAO peak. 

\subsection{Evaluation of IR resummed power spectrum}

After decomposing the linear power spectrum $P^{lin}(z,k)=D(z)^2\big(P_s(k)+ P_w(k)\big)$
into smooth and oscillating (wiggly) contributions\footnote{
In this section we adopt the conventional notations and denote the growth factor 
$g \mapsto D(z)$ and the linear power spectra at $z=0$ simply by 
$\bar P_{s(w)}(k) \mapsto P_{s(w)}(k)$.}, we need to evaluate the
derivative operator $\S$ defined in \eqref{Slead}, that describes the IR enhancement.
This is done using 
\[
\nabla_{\a_1}\cdots\nabla_{\a_{2n}} P_w(k) = (-1)^n \frac{\hat k_{\a_1}\cdots \hat k_{\a_{2n}}}{k_{osc}^{2n}} 
P_w(k)\big(1+O(\varepsilon)\big)\;,
\] 
where $\varepsilon$ is the small expansion parameter related to IR enhancement defined in \eqref{defeps} and
$\hat \k=\k/k$. Recall that $k_{osc}=h/(110\,$Mpc) is the scale setting the period of the 
BAO oscillations. A straightforward computation yields
\be\label{Sleading}
 \S   P_w(k) = k^2\Sigma^2   P_w(k) \times \big(1+O(\varepsilon)\big)\,,
\ee
with
\be
  \Sigma^2 \equiv \frac{4\pi}{3} \int_{0}^{k_S}dq  P_s(q)\bigg[1-j_0\left(\frac{q}{k_{osc}}\right)+2j_2\left(\frac{q}{k_{osc}}\right) \bigg]\,, \label{Sigmadefk} \\
\ee
where $j_n$ are spherical Bessel functions and $k_S$ is the (a priori arbitrary) separation scale of long and short modes, that has been introduced in
order to treat the perturbative expansion in the two regimes separately. Since the exact result for the power spectrum and other observables
is independent of $k_S$, any residual dependence on it can be taken as an estimate of the perturbative uncertainty.  
The IR resummed power spectrum at leading order  following from \eqref{eq:finalLO} is given by
\be
\label{eq:LOfinal}
P^{IR\;res, LO}(z, k)=  D(z)^2  \left(  P_s(k) + e^{-k^2D(z)^2\Sigma^2} P_w(z)\right)\,,
\ee
where the first term corresponds to the smooth part of the linear spectrum.
The leading effect of IR enhanced loop contributions is an exponential damping of the oscillatory
part of the spectrum.

\begin{figure}
\begin{center}
\includegraphics[width=.6\textwidth]{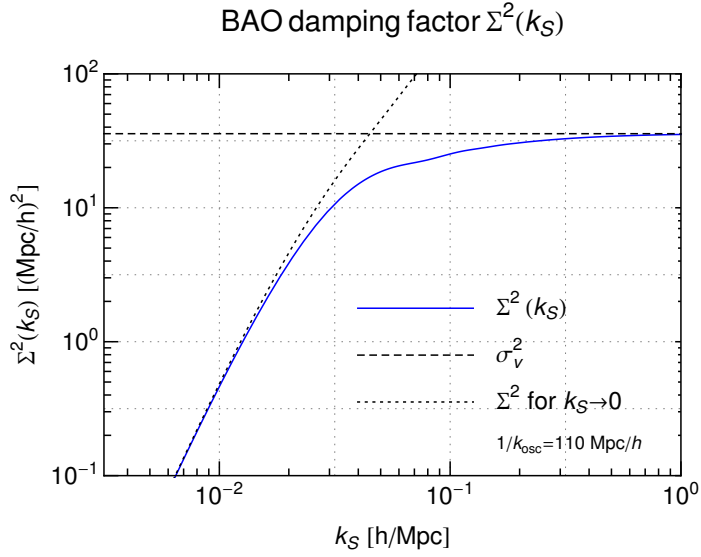}
\caption{\label{fig:SigmaL} Dependence of the BAO damping factor $\Sigma^2$ on the separation scale $k_S$. Dashed curves show the limiting cases discussed in the text. 
}
\end{center}
\end{figure}

Let us now discuss the choice of $k_S$. By inspection of the integrand in
(\ref{Sigmadefk}) we find that it peaks at $q\sim 0.03 h/$Mpc, but gives significant contribution
into the integral up to wavenumbers $q\sim 0.2 h/$Mpc. This is corroborated by the numerical evaluation of
the damping factor $\Sigma^2$ as a function of $k_S$; the result is shown in Fig.~\ref{fig:SigmaL}
for a realistic $\Lambda$CDM model.
For very small values of the IR separation scale $k_S\lesssim k_{osc}$ it approaches the
limiting form $\Sigma^2\to 2\pi/5 \int_0^{k_S} dq\, q^2  P_s(q)/k_{osc}^2$, while for
large $k_S\gg k_{osc}$ it asymptotes to a constant 
$\Sigma^2\to \sigma_v^2 \equiv 4\pi/3 \int_0^{\infty} dq\,   P_s(q)$. It is desirable to take $k_S$ as large as possible to include more IR contributions and minimize the dependence of the damping factor on $k_S$. On the other hand, $k_S$ cannot be taken too large as the previous analysis relies on the IR expansions which are valid for $q\ll k$. As a compromise, we consider several values of $k_S$ around the BAO scale
$k_{BAO}\sim 0.1 h/$Mpc. We are going to see that the dependence of our results on the precise choice of $k_S$ in this range is very mild.

At NLO, the IR resummed power spectrum \eqref{irres3} can be written in the 
form\footnote{
Since we are keeping NLO terms, one should in principle keep also the first sub-leading
corrections in the evaluation of the derivative operator in \eqref{Sleading}. However,
it turns out that this correction cancels in \eqref{eq:NLOfinal} at NLO precision.
The simplest way to see this is to go back to Eqs.~(\ref{dasiessub}), (\ref{daisynew}) and substitute in them the expansion (\ref{Sleading}) keeping track of the $O(\ve)$ terms. By comparing the resulting expressions one finds that the relation (\ref{nlodais}) holds with NLO precision if $\S$ is everywhere replaced by $k^2\Sigma^2$. As all other contributions comprising (\ref{irres3}) do not contain an $O(1)$ part, the replacement 
$\S\mapsto k^2\Sigma^2$ in them is also justified leading to (\ref{eq:NLOfinal}).  
}
\be
\label{eq:NLOfinal}
\begin{split}
P^{IR\;res,LO+NLO}&(z, k) =D(z)^2 \left(P_s(k)+ \big(1+ k^2D(z)^2\Sigma^2 \big)e^{-k^2D(z)^2\Sigma^2}  P_w(k) \right)\\
&+\!D(z)^4 P^{1-loop}[ P_s+e^{-k^2D(z)^2\Sigma^2}  P_w]
   +D(z)^6 e^{-k^2D(z)^2\Sigma^2}\!(\S^a\!+\!\varkappa\S^b)P_w(k)\,,
\end{split}
\ee
where $\varkappa=1$ in ED ($\varkappa=0$ in ZA).
The first term in the second line of \eqref{eq:NLOfinal} corresponds to the standard one-loop result,
but computed with the LO IR resummed power spectrum, instead of
the linear one. 
As was demonstrated in \cite{Blas:2015qsi},
the sum over all one-loop diagrams in TSPT agrees with the SPT result. Therefore,
in practice, one can use the usual expression $P^{1-loop}=P_{22}+2P_{13}$,
however evaluating the loop integrals $P_{22}$ and $P_{13}$ with the input
spectrum \eqref{eq:LOfinal} instead of the linear spectrum.

Finally, the finite-difference operators in the last term  can be evaluated similarly to
$\S$.  After somewhat lengthy but straightforward calculation, 
we obtain
\begin{subequations} \label{eq:practical}
\begin{align}
\label{eq:Siga}
 \S^a  P_w(k) =&  \frac{8\pi}{5}  \,  k_{osc} k^3 \Sigma^2
\int_0^{k_S} dq P_s(q) q\,  \left[3j_1\left(\frac{q}{k_{osc}}\right)-2j_3\left(\frac{q}{k_{osc}}\right)\right]
\frac{d  P_{w}(k)}{dk}
\equiv k^3\Sigma_a^2\frac{d  P_{w}(k)}{dk} ,\\
 \S^b  P_w(k) =& - (4\pi)^2 \,  k^2_{osc}k^3  
   \int_0^{k_S} dq dq' P_s(q) P_s(q') \,  h\left(\frac{q'}{k_{osc}},\frac{q}{k_{osc}}\right)
   \frac{d  P_{w}(k)}{dk}
\equiv k^3\Sigma^2_{b}\frac{d  P_{w}(k)}{dk},
\label{eq:cb}
\end{align}
\end{subequations}
with 
\be
\label{hxy}
  h(x,y) = \frac37 \big( h_1(x,y) + h_2(x,y) + h_2(y,x) \big).
\ee
The functions $h_{1,2}(x,y)$ are given in  App.~\ref{app:NLOpol}.
The result \eqref{eq:NLOfinal} is valid for both density, velocity and
cross power spectra when using the appropriate expressions for the
one-loop correction. In addition, one can obtain the result in ZA by
using the corresponding one-loop expression with kernels computed in ZA
and setting $\varkappa=0$.

The IR resummed result for the bispectrum 
is given at leading order in $\varepsilon$ and $\sigma_h^2$ by (see
Eq.~(\ref{bispres1})) 
\be
\begin{split}
B^{IR\;res;LO}_{\delta \delta \delta}(z;\k_1,\k_2,\k_3)&=B^{tree}_{s,\delta \delta \delta}(z;\k_1,\k_2,\k_3)\\
&+2D(z)^4\sum_{i<j=1}^3F_2(\k_i,\k_j)\left(e^{-k_j^2D(z)^2\Sigma^2}  P_w(k_j)   P_s(k_i)+i \leftrightarrow j\right)\,,
\end{split} 
\ee
where $B^{tree}_{s,\delta \delta \delta}(z;\k_1,\k_2,\k_3)$ is the tree-level result for the smooth part, as obtained also in SPT, and 
$F_2(\k_i,\k_j)$ is the usual SPT kernel \cite{Bernardeau:2001qr}. The IR resummation again corresponds to a damping of the
oscillating contributions.

\subsection{Comparison with other approaches}
\label{sec:compother}

Let us now compare our results to other approaches existing
in the literature.
From a phenomenological viewpoint, it is well-known that an exponential damping factor
applied to the oscillating component of the power spectrum gives a reasonable description of
the BAO peak in the measured two- and also three-point correlation functions, see e.g. \cite{Slepian:2015hca} and references therein.
Therefore, the aim of perturbative descriptions is to derive this behavior from first principles,
identify effects that go beyond a simple damping, and give a definite quantitative prediction as
well as an estimate of the theoretical error.

There exist many schemes to derive non-linear corrections to the BAO
peak within cosmological perturbation theory, 
here we focus on a few of them.
In \cite{Crocce:2007dt}, the RPT formalism  \cite{CrSc1} was used to obtain a formula of the form.
\be
\label{PRPT}
P(z,k) = G^2(z,k)P_{lin}(k) + P_{MC}(z,k)\;,
\ee 
where $G(z,k)$ is the propagator and $P_{MC}$ is the part due to the mode coupling. 
The propagator describes
how a perturbation evolves over time and is not a Galilean invariant
quantity. As such, it contains IR enhanced contributions corresponding
to the translation 
of inhomogeneities by large-scale flows. 
When resummed at the leading order, these contributions produce an exponential damping factor at high $k$, 
\be\label{eq:RPT}
G^2(z, k)= D(z)^2 \exp[- k^2 D(z)^2 \sigma_v^2], 
\ee 
with $\sigma_v^2\equiv 4\pi/3 \int_0^\Lambda dq P_{lin}(q)$,
 and $\Lambda$ being a UV cut-off of the theory ($k\ll \Lambda$).
 In the RPT-based approach this form of the propagator is 
substituted into (\ref{PRPT}). 
A similar result is derived in the 
Lagrangian picture in \cite{Matsubara:2007wj,Noh:2009bb}. 
Notice that the exponential damping in this case applies to the whole linear power spectrum, including both wiggly and smooth parts.
Further developments of this idea have been proposed in \cite{Anselmi:2012cn}.

While being successful on a phenomenological level, this approach is quite different
from ours. In RPT, there is no clear parametric dependence that would
single out the resummed set of contributions. In particular, for a
{\em smooth} power spectrum the contributions resummed in (\ref{eq:RPT}) are of the same order as those comprising $P_{MC}$, and actually cancel with them 
\cite{Jain:1995kx, Scoccimarro:1995if, Peloso:2013zw, Kehagias:2013yd, Blas:2013bpa, Sugiyama:2013gza}
as required by the equivalence principle.
On the other hand, our approach is based on well-defined power-counting rules formulated directly for the perturbative expansion of equal-time correlation functions. As a result, we obtain 
a damping only of the oscillating part
of the power spectrum, in line with the expected cancellation of IR enhancement
for the smooth part. Furthermore, the damping factor given by \eqref{Sigmadefk} 
has a different structure from (\ref{eq:RPT}). Although for the real universe their numerical values happen to be very close to each other, the situation would be different for a universe with more power in very soft modes with wavenumbers $q\ll k_{osc}$.  
Finally, when
taking NLO corrections into account our result \eqref{eq:NLOfinal} cannot be described anymore
by a simple exponential damping of the overall power spectrum or its
wiggly part. 

Ref.~\cite{Baldauf:2015xfa} proposed a description of the IR enhanced
effects on the BAO peak 
motivated by consistency relations between the bispectrum and the
power spectrum based on 
the equivalence principle. This approach is related to the
earlier perturbative framework developed in \cite{Senatore:2014via}.
At leading order, our result \eqref{eq:LOfinal} coincides with the
results of \cite{Baldauf:2015xfa} when choosing $k_S=k/2$. The
agreement essentially extends to the corrections produced by hard
loops (see Sec.~\ref{sec:hard}). Subleading soft corrections were not
considered in \cite{Baldauf:2015xfa}. 
TSPT gives a simple diagrammatic description of IR enhancement and
provides a tool to systematically derive, scrutinize and extend the 
results found in \cite{Baldauf:2015xfa}. 
In particular, the subleading soft corrections computed in the present
work and entering in \eqref{eq:NLOfinal} capture the shift of the BAO
peak, as we will see 
below, and the power counting allows in principle to go beyond NLO in
a systematic way. 
Furthermore, the IR resummation in TSPT readily generalizes 
beyond the power spectra and applies to any $n$-point
correlation functions. 

Finally, it is worth mentioning that we have checked our results by
comparing with known exact expressions in the Zel'dovich
approximation, see Appendix~\,\ref{app:ZA}. Additionally, in
Appendix~\ref{app:SPT_TSPT} we rederive the TSPT IR resummed power spectra within
the SPT framework.

\subsection{Comparison with $N$-body data}\label{sec:Nbody}

We consider a $\Lambda$CDM model with cosmological parameters matching those of
the Horizon simulation \cite{Kim:2011ab}. 
The linear power spectrum is
obtained from the CLASS code \cite{Blas:2011rf} and decomposed into smooth and oscillating components
as described in Sec.\,\ref{sec:BAO}.

\begin{figure}[th]
  \includegraphics[width=0.5\textwidth]{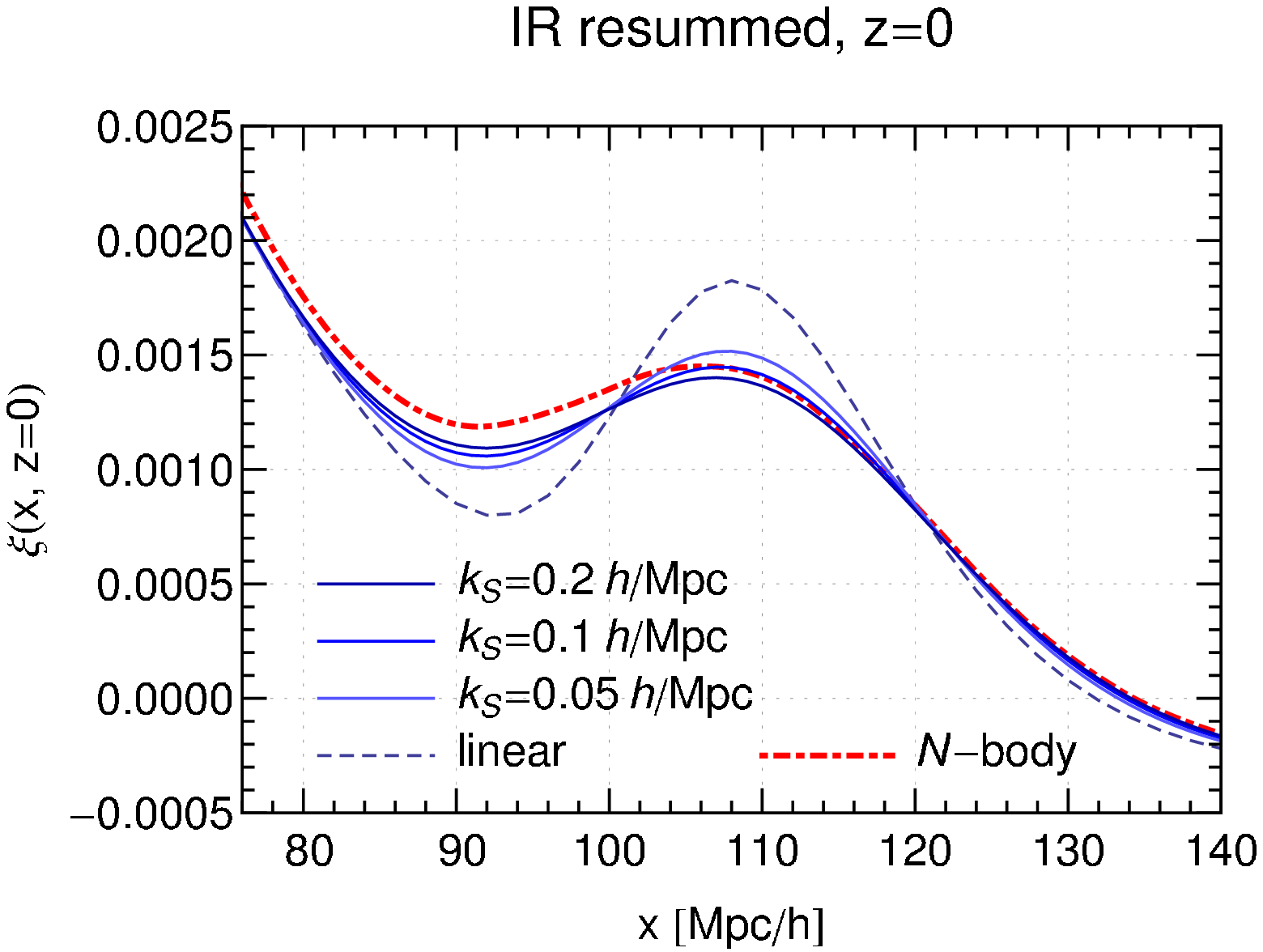}
  \includegraphics[width=0.5\textwidth]{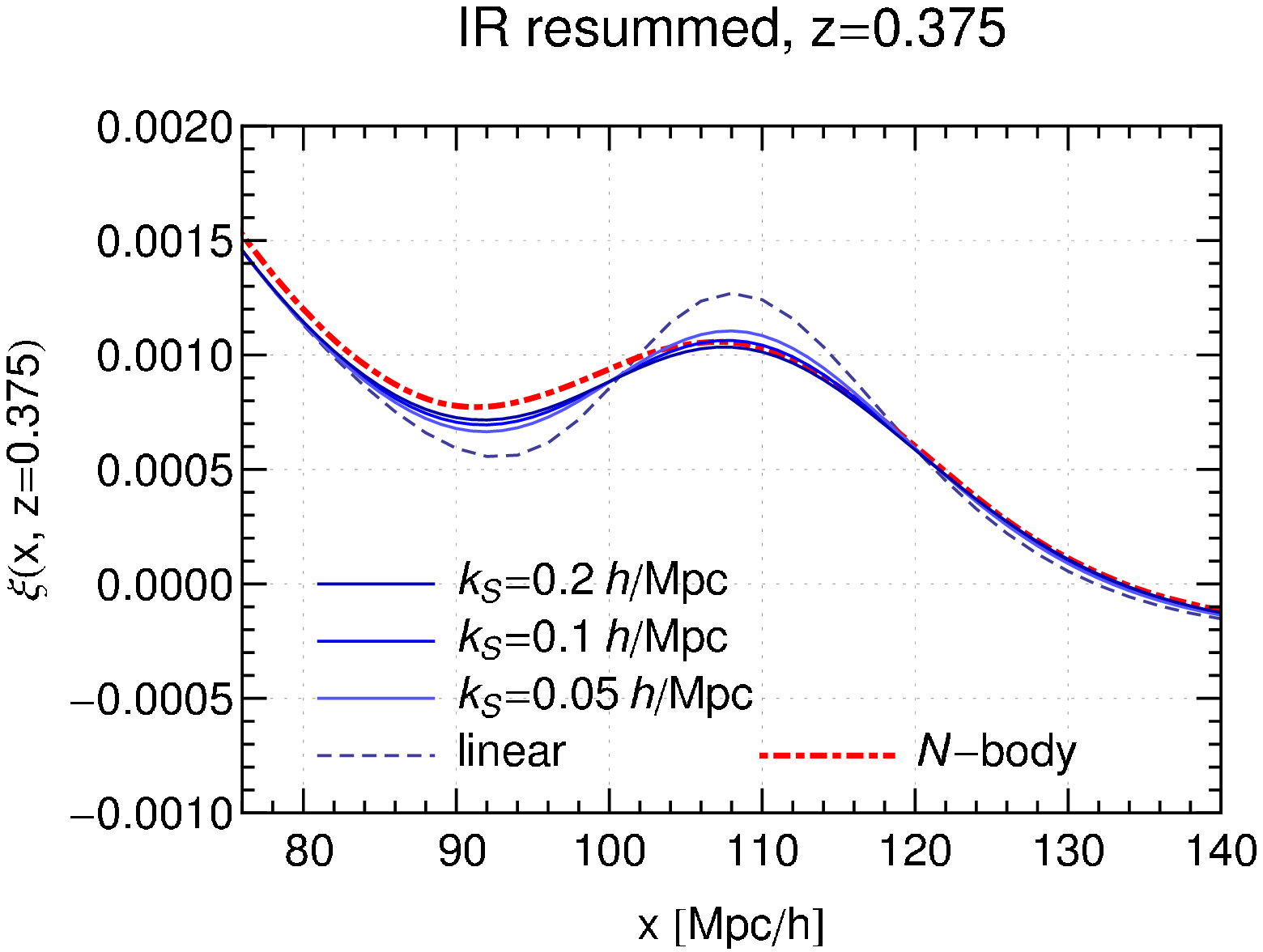}
  \caption{\label{fig:Pirres} Infra-red resummed matter correlation function at LO obtained in TSPT for three different
  values of the IR separation scale $k_S$, and two different redshifts (left: $z=0$, right: $z=0.375$). Also shown is the linear result (dashed) and the result of the Horizon Run 2 large-scale $N$-body simulation \cite{Kim:2011ab}. We use $1/k_{osc}=110\,$Mpc$/h$.}
\end{figure}

In the following we show results for the correlation function, because it exhibits a clear
separation between the BAO peak and the small-distance part of the correlations, and allows to
visualize the effects on the BAO feature in a transparent way.
The matter correlation function is related to the power spectrum as
\be
\label{eq:correlN}
  \xi(x,z) = \frac{4\pi }{x}\int_0^\infty dk \, k P(k,z)\, \sin(k x)\,.
\ee
In Fig.\,\ref{fig:Pirres}, we show the leading-order IR resummed result for three different choices
of $k_S$ (blue solid lines). The damping of the BAO oscillations described by $\Sigma$ corresponds to
a broadening of the BAO peak in real space and gives
already a relatively good description of the $N$-body result shown by the 
red dashed line \cite{Kim:2011ab}, especially when compared
to the linear prediction (thin dashed line). Nevertheless,  there are some differences, and the dependence
on $k_S$ is not negligible.

\begin{figure}[t]
  \includegraphics[width=0.5\textwidth]{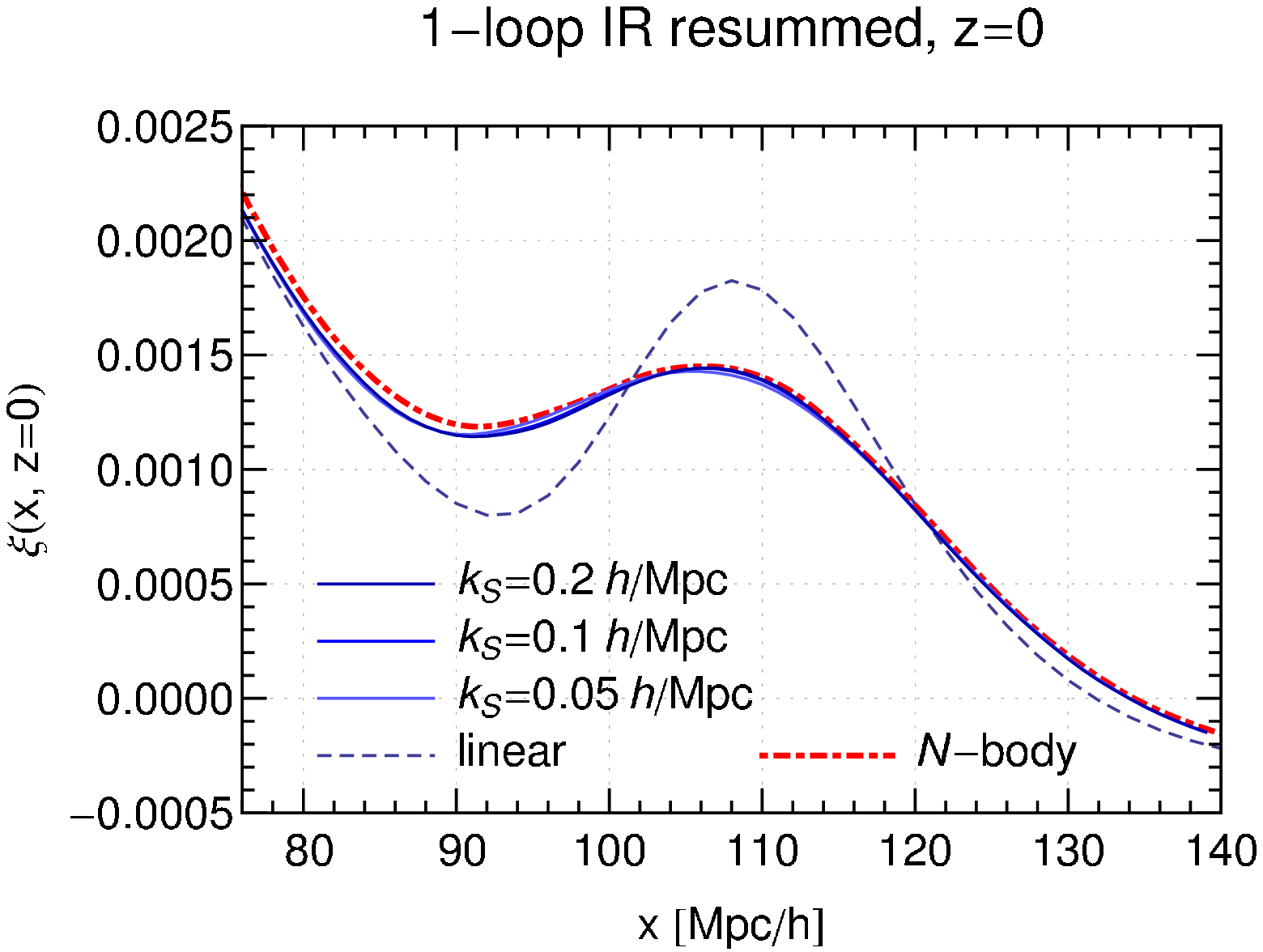}
  \includegraphics[width=0.5\textwidth]{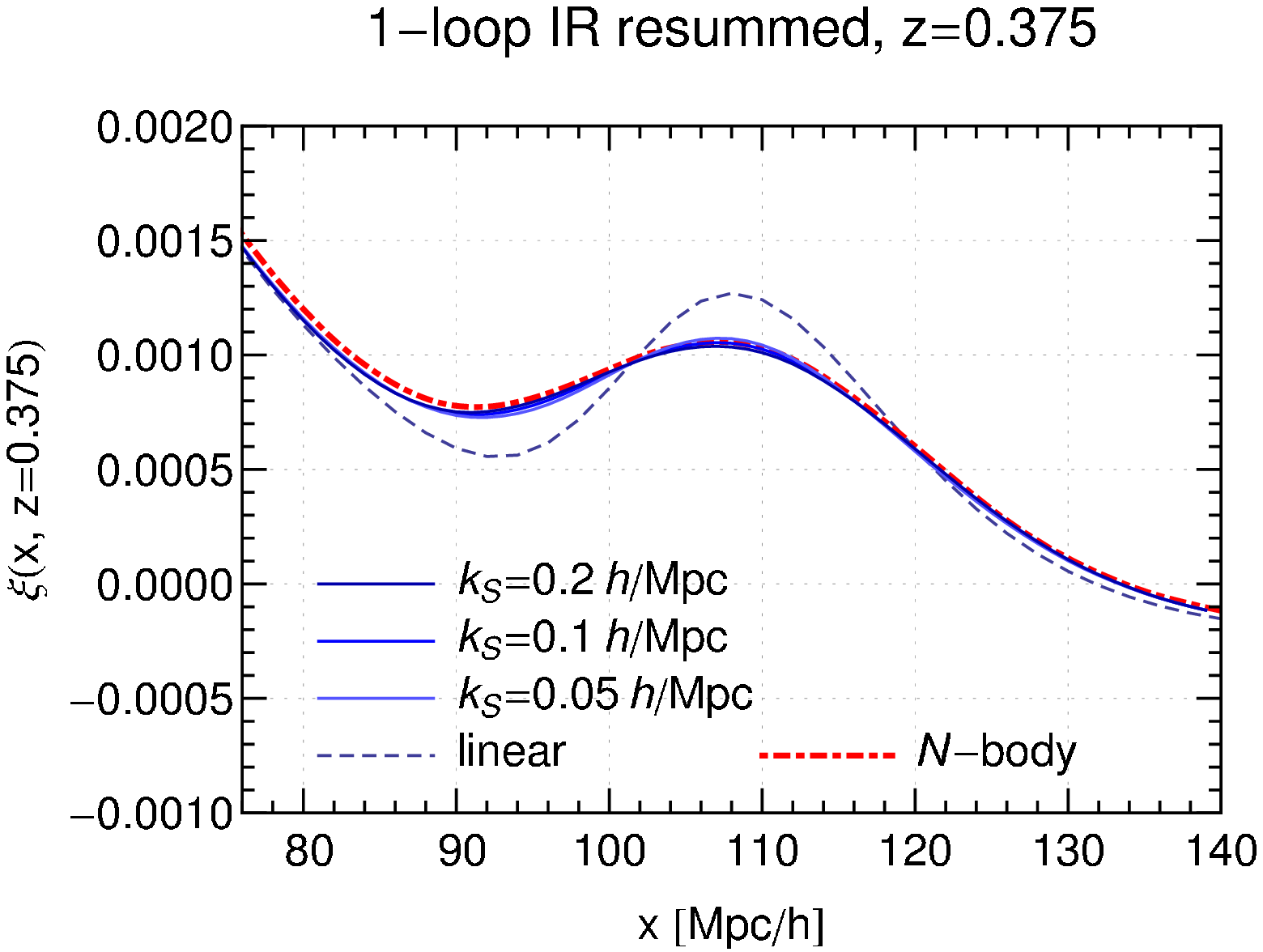}
  \caption{\label{fig:Pirres1L} Same as Fig.\,\ref{fig:Pirres}, but showing the infra-red resummed matter correlation function obtained in TSPT at next-to-leading order (blue lines)
compared to the Horizon Run 2 $N$-body data (red line). Note that the three lines for the three values of $k_S$ are almost indistinguishable.}
\end{figure}

We now turn to the NLO result. The comparison of the matter correlation function obtained using 
\eqref{eq:NLOfinal} with the $N$-body data is shown in Fig.\,\ref{fig:Pirres1L}.
One observes that the agreement is considerably improved compared to the LO. Furthermore,
the dependence on the separation scale ${k_S}$ is reduced. This is an important consistency check, because
the dependence on $k_S$ vanishes in principle in the exact result. Thus, any residual dependence on $k_S$ can be
taken as an estimate of the perturbative uncertainty, and it is reassuring that this uncertainty is reduced
when going from LO to NLO. 

We conclude that the systematic IR resummation gives a very accurate description of the correlation function
at BAO scales. The residual discrepancies at shorter distances visible in Fig.~\ref{fig:Pirres1L} 
are expected due to several effects. The variance due to the finite boxsize, and the finite resolution of the $N$-body data leads to an uncertainty
of several percent\footnote{Ref.~\cite{Kim:2011ab} does not give error bars for the simulation data points.
An estimate of the statistical variance using the number of available modes in the simulation as well as the finite resolution suggests
that the uncertainty is at the few percent level in the range of scales relevant for BAO. This level of accuracy is also consistent with
the difference between the correlation function extracted from Horizon Run 2 ($L=7.2$Gpc$/h$, $N=6000^3$) versus Horizon Run 3 ($L=10.8$Gpc$/h$, $N=7210^3$) data presented in~\cite{Kim:2011ab}.}.
In addition, the correlation function is sensitive to the UV physics which has been left
beyond the scope of our present study. 

\begin{figure}[t]
\begin{center}
  \includegraphics[width=0.5\textwidth]{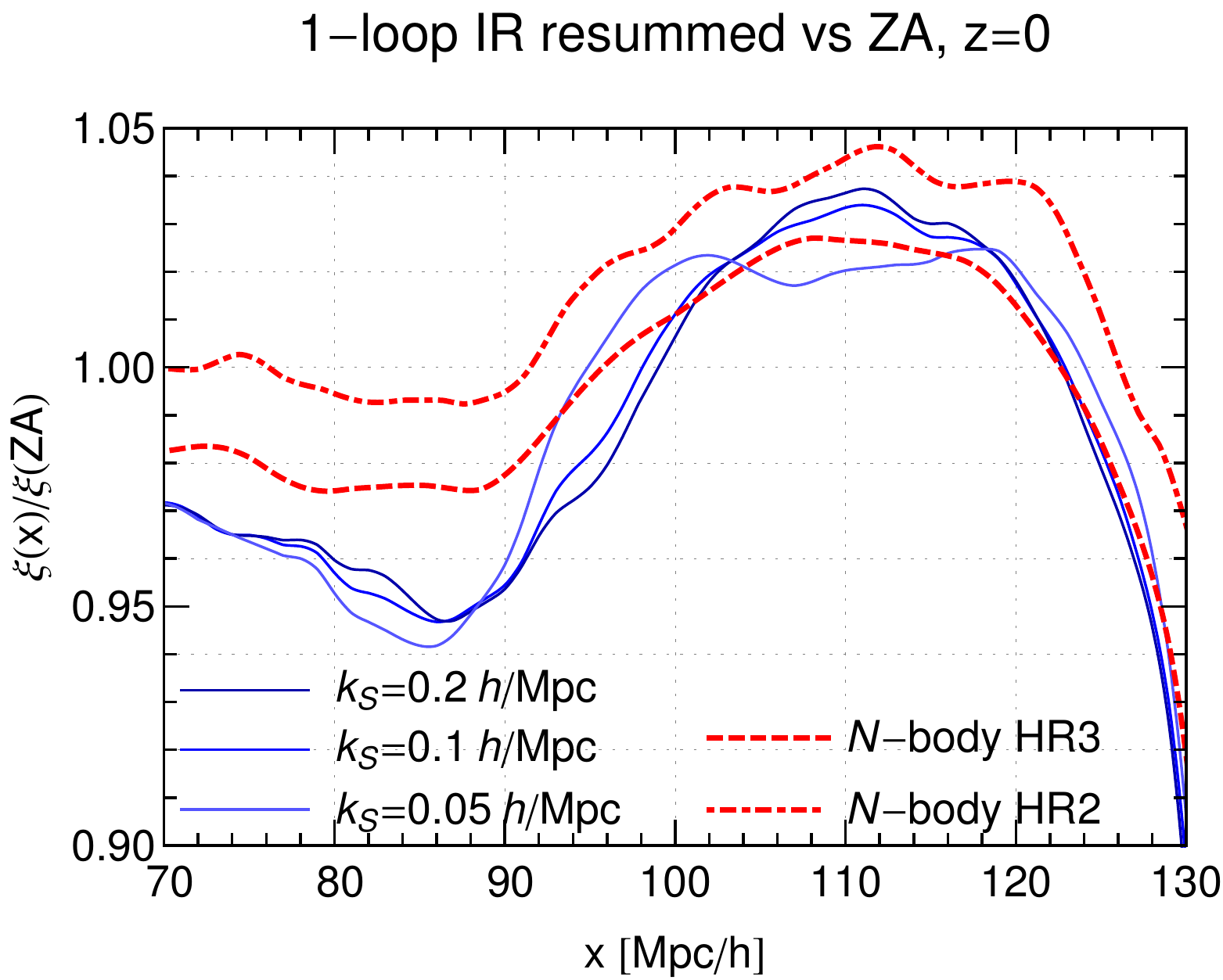}
\end{center}
  \caption{\label{fig:xiVsZA} 
The correlation function computed in TSPT at NLO normalized to the correlation function in the Zel'dovich approximation (blue curves).
We also show the correlation functions of the Horizon Run 2 and 3~\cite{Kim:2011ab} divided by the Zel'dovich approximation (red curves). 
 }
\end{figure}

In Fig.\,\ref{fig:xiVsZA} we show the ratio of the NLO result to the correlation function obtained in the Zel'dovich
approximation\footnote{Here by the Zel'dovich approximation we mean the leading order of Lagrangian perturbation theory.
The 2-point correlation function  in ZA was computed with the publicly available code ZelCa \cite{Tassev:2013zua}.}.
The differences are around $5\%$ in the BAO range, and therefore our
results are broadly consistent with ZA, as expected. 
Nevertheless, the
differences are larger than the ultimate precision that is desired to
match future surveys. The ratio between the N-body correlation
function and the one obtained in ZA is also shown on the same plot by
the red line.
The TSPT result agrees with the N-body data somewhat better than ZA in the
BAO peak region, though the error range of the N-body data does not
allow at the moment to clearly discriminate between the two.
As discussed before, the TSPT framework can be
systematically extended to NNLO, and further corrections from UV modes can be incorporated, which is left for future work.

Finally, we have compared the results for the 
correlation function computed using the
full NLO formula (\ref{eq:NLOfinal}) and its reduced version without
the last term containing the operators $\S^a$, $\S^b$. The relative
difference $\Delta\xi/\xi$ at $z=0$ does not exceed
$0.5\%$.
Given the strong dependence of the omitted term on the growth factor
and hence its quick decrease 
with redshift, one concludes
that this term is negligible for all practical purposes.

\subsection{Shift of the BAO peak}
\label{sec:shift}

A valuable piece of information provided by the BAO
peak is its position as a function of redshift that can be
used as a standard ruler to infer cosmological parameters and probe
possible alternatives to $\Lambda$CDM (see \cite{Bellini:2015oua} for
a recent discussion in the context of modified gravity). The upcoming
surveys aim at measuring this quantity with sub-percent accuracy
\cite{Weinberg:2012es}. 
Therefore, it is important to assess how non-linear dynamics 
offsets the BAO peak as 
compared to the linear prediction.

For concreteness we focus on the position of the maximum of the
BAO peak which we denote by $x_{BAO}$.
There are two effects that contribute to its shift with respect to 
the linear result
$x_{BAO}^{lin}$. First, the damping of the wiggly component in
the power spectrum, which occurs already at leading order of the
IR resummation, shifts the maximum because the correlation function
$\xi(x)$ is not symmetric. Second, at NLO the interactions of the
modes in the BAO region with soft modes shift the phase of the
BAO. This, in turn, translates into an additional shift of the peak in
position space. Let us discuss these two contributions one by one.

It is convenient to decompose the correlation function into
a smooth component and a component that describes the BAO peak,
\be\label{eq:wCF}
 \xi(x) = \xi_s(x) + \xi_w(x)\,,
\ee
that are inherited from the decomposition of the power spectrum into
smooth and wiggly parts. In the region of the peak these two
contributions are of the same order with $\xi_w$ being a factor of a
few larger than $\xi_s$. At the linear level, the condition for the
maximum of the peak reads,
\be
\label{peaklin}
0=\int dk\, k^2 P_w(k)\cos\big(k x_{BAO}^{lin}\big)-
\frac{1}{x_{BAO}^{lin}}\int dk\, kP_w(k)\sin\big(k x_{BAO}^{lin}\big)
+\frac{x_{BAO}^{lin}}{4\pi}\xi_s'(x_{BAO}^{lin})\;.
\ee
To obtain analytic estimates we represent the wiggly power spectrum as
a product of the oscillating part and a smooth envelope
(cf. Eq.~(\ref{pwfit})),
\be
\label{envelop}
P_w(k)=f_{env}(k)\sin(k/k_{osc})\;.
\ee
This implies that the position of the peak is close to
$x_{BAO}^{lin}\approx k_{osc}^{-1}$. In what follows we seek
the corrections to this
relation. Writing 
\[
x_{BAO}^{lin}=k_{osc}^{-1}+\delta x^{lin}
\]
and treating the product $k\,\delta x^{lin}$ as small we obtain from
(\ref{peaklin}) 
\be
\label{deltaxlin}
\delta x^{lin}=-k_{osc}\frac{\int dk\, k f_{env}(k)
-(2\pi k_{osc}^2)^{-1}\xi_s'(k_{osc}^{-1})}{\int dk\, k^3f_{env}(k)}\;,
\ee
where we have neglected the integrals of rapidly oscillating
functions. In particular, we have used the relation 
\[
\int dk\,k f_{env}\sin^2(k/k_{osc})=\int dk\,k f_{env}\frac{1-\cos(k/k_{osc})}{2}
\approx \frac{1}{2} \int dk\, k f_{env}(k)\;.
\]

If instead of the linear power spectrum, we consider the LO expression
(\ref{eq:LOfinal}) with damped wiggly component, we obtain the
position of the corresponding peak as 
\[
x_{BAO}^{LO}=k_{osc}^{-1}+\delta x^{LO}\;,
\]
where $\delta x^{LO}$ is given by the expression (\ref{deltaxlin}),
but with $f_{env}$ replaced by $f_{env}(k)e^{-k^2D(z)^2\Sigma^2}$. One
concludes that the shift of the LO peak relative to the linear one is
\be
\begin{split}
\label{shiftdamp}
\frac{\Delta x^{LO}}{x_{BAO}}\equiv\frac{\delta x^{LO}-\delta
  x^{lin}}{x_{BAO}}
=&k_{osc}^2\bigg[\frac{\int dk\,kf_{env}(k)}{\int dk\,k^3f_{env}(k)}
-\frac{\int dk\,kf_{env}(k)e^{-k^2D(z)^2\Sigma^2}}{
\int dk\,k^3f_{env}(k)e^{-k^2D(z)^2\Sigma^2}}\bigg]\\
&-\frac{\xi_s'(k_{osc}^{-1})}{2\pi}
\bigg[\frac{1}{\int dk\,k^3f_{env}(k)}
-\frac{1}{
\int dk\,k^3f_{env}(k)e^{-k^2D(z)^2\Sigma^2}}\bigg]\;.
\end{split}
\ee
The two contributions in this formula are of the same order 
$(k_{osc}/k)^2$, where $k\sim 0.1 h/$Mpc is the characteristic
range of wavenumbers corresponding to BAO. At $z=0$ we have
\be
\label{DxLO}
\frac{\Delta x^{LO}}{x_{BAO}}\sim -1\%.
\ee
We see that this LO shift is quite significant. It is worth
emphasizing that it is exclusively due to the damping of BAO by
large IR effects. The damping is the same in ED and ZA. Therefore
$\Delta x^{LO}$ is expected to be removed by the BAO reconstruction
procedure which essentially uses the ZA to evolve the density field
backward in time. To understand if this procedure can leave any
residual shift we have to go to the next-to-leading order.

Writing the correlation function as $\xi=\xi^{LO}+\xi^{NLO}$ one
easily derives the additional shift of the peak induced by the second
term,
\be
\label{eq:shiftder}
\frac{\Delta x^{NLO}}{x_{BAO}}=
-\frac{(\xi^{NLO})'}{x(\xi^{LO})''}\bigg|_{x=x_{BAO}}\;.
\ee
One can check that the contribution of the smooth correlation function
into this expression is negligible, so that one can safely replace in
it $\xi\mapsto\xi_w$.

We evaluate the LO and NLO contributions into the correlation function
numerically using Eqs.~(\ref{eq:LOfinal}), (\ref{eq:NLOfinal}) for the
power spectrum. The shift of the peak is then computed either directly
by comparing the full correlation function to that at LO, 
or using (\ref{eq:shiftder}). The
results of this evaluation are presented in Table~\ref{tableshift0}
(second and third columns). We consider three choices of the
separation scale $k_S$ bounding the IR region. 
\begin{table*}[h]
\centering
  \begin{tabular}{|l|ccc|}
     \hline
\multirow{2}{*}{$k_S\,,\; h/$Mpc }
&\multicolumn{3}{|c|}{$\Delta x^{NLO}/x_{BAO}$}\\\cline{2-4}
       &  ~Full~  & ~~Eq.~\eqref{eq:shiftder}~~ &  ~Eq.~\eqref{xNLOfinal}  \\ \hline 
   $ 0.05$ & $-0.38\%$ & $-0.43\%$ &  $-0.46\%$  \\ \hline
   $ 0.1$ & $-0.41\%$ & $-0.45\%$    &$-0.40\%$   \\ \hline
   $0.2$  & $-0.45\%$ & $-0.50\%$   &  $-0.32\%$ \\ \hline
   \end{tabular}
\caption{
\label{tableshift0}
Shift of the BAO peak at redshift $z=0$ for three values of the separation scale $k_S$. First column: full NLO result.
Second column: evaluation using Eq. \eqref{eq:shiftder}. 
Third column: analytic estimate Eq.~\eqref{eq:shiftfin0}.}
\end{table*}
Modulo some scatter introduced by the $k_S$ dependence, our estimate
for the NLO shift is around $0.4\%$. This lies in the ballpark of the
estimates obtained using different approaches 
\cite{Smith:2007gi,Crocce:2007dt,Padmanabhan:2009yr,Sherwin:2012nh,Prada:2014bra}
and agrees well with the value of the so-called `physical' shift
\cite{Smith:2007gi,Crocce:2007dt}
measured in the simulations
\cite{Xu:2010tf,Seo:2009fp,Prada:2014bra}. While we expect that the
NLO shift considered here agrees essentially with the `physical' shift,
the precise relation is not completely clear to us and we leave the task
of understanding it for future work.

It is instructive to derive an analytic estimate for $\Delta
x^{NLO}$. It is shown in Appendix~\ref{app:shift} that the NLO wiggly
power spectrum has the form, 
\be
\label{eq:Psnlo}
P_w^{IR\;res,NLO}(z, k) = D(z)^4 e^{-k^2 D(z)^2 \Sigma^2}
\bigg(H(k) P_w(k)+S(z,k)\frac{d P_w(k)}{dk} 
\bigg) \,.
\ee
The first term in brackets receives contributions both from hard and
soft modes, whereas the second term is exclusively due to soft modes
with wavenumbers $q\lesssim k_{osc}$. It describes a phase shift of
the wiggly component of the power spectrum.
The precise form of the function
$H(k)$ is not relevant to us; it is only important that it depends
smoothly on its argument. 
For $S(z,k)$ we find,
\be
\label{eq:Szk}
S(z,k)=s\,k+\big(\Sigma_{Silk}^2+\varkappa D(z)^2\Sigma_b^2\big)k^3\;,
\ee
where $\Sigma_b^2$ is defined in (\ref{eq:cb}). The other two
coefficients are related to the density variance at the scale
$k_{osc}$,
\be
s\sim \sigma^2(k_{osc})\equiv\int_0^{k_{osc}} dq\,q^2P_s(q)~,~~~~~~
\Sigma_{Silk}^2\sim \sigma^2(k_{osc})/k^2_{Silk}\;,
\ee
where in the last formula $k_{Silk}\sim 0.2\, h/$Mpc is the Silk damping scale.
The detailed expressions are given in Appendix~\ref{app:shift}. It is
worth to point out that the formula for $s$ is different in ED and ZA,
as well as for the density and velocity divergence power
spectra. Notice also the presence of the coefficient $\varkappa$ in
(\ref{eq:Szk}) that discriminates between ZA ($\varkappa=0$) and ED
($\varkappa=1$).  

Next, we substitute (\ref{eq:Psnlo}) into (\ref{eq:shiftder}) which
yields,
\be
\frac{\Delta x^{NLO}}{x_{BAO}}=
k_{osc}D(z)^2\frac{\int dk\,k^2\cos(k/k_{osc})\big[H(k)P_w(k)
+S(z,k)\frac{dP_w}{dk}\big]e^{-k^2D(z)^2\Sigma^2}}{
\int dk\,k^3\sin(k/k_{osc})P_w(k)e^{-k^2D(z)^2\Sigma^2}}\;.
\ee
Recalling the form (\ref{envelop}) of $P_w$ one observes that the
integral involving the first term in the numerator contains a rapidly
oscillating function and thus gives a negligible contribution. In the
second term we integrate by parts. Neglecting again integrals of
rapidly oscillating functions and using (\ref{eq:Szk}) we arrive at 
\be
\label{xNLOfinal}
\frac{\Delta x^{NLO}}{x_{BAO}}=
D(z)^2\,s
+D(z)^2\big(\Sigma_{Silk}^2+\varkappa D(z)^2\Sigma_b^2\big)
\frac{\int dk\,k^5P_w e^{-k^2D(z)^2\Sigma^2}\sin(k/k_{osc})}{
\int dk\,k^3P_w e^{-k^2D(z)^2\Sigma^2}\sin(k/k_{osc})}\;.
\ee
For realistic power spectra the ratio of integrals in the second term
is of order $0.02\,[h/\mathrm{Mpc}]^2$ at $z=0$. It is worth noting
that numerically the
second term gives a subdominant contribution, so approximately one can
write,
\be
\label{eq:shiftfin0}
\frac{\Delta x^{NLO}}{x_{BAO}}\approx  D(z)^2\, s\;.
\ee
Still, we prefer to use the complete expression
(\ref{xNLOfinal}). Evaluating various contributions entering into it
using the expressions from Appendix~\ref{app:shift} and Eq.~(\ref{eq:cb})
we obtain the estimates for the shift listed in the fourth column of
Table~\ref{tableshift0}. They are in reasonable agreement with the
values obtained by the direct numerical evaluation of the correlation
function.   

As already mentioned before, the values of $s$ and $\varkappa$ are
different in ED and ZA. Consequently, the BAO shift computed in ZA is
somewhat lower than in ED: 
$$(\Delta x^{NLO}/x_{BAO})^{ZA}\simeq
-(0.28\div 0.22)\%\;,$$ 
depending on the
choice of $k_S$.
Thus, while the ZA gives a rather accurate description of the BAO broadening, it underestimates the
BAO shift. The difference is expected since the terms responsible for the BAO
shift originate 
from the non-dipole parts of the interaction vertices, which are
different in ZA and in ED. One concludes that, in principle, the BAO
reconstruction based on ZA is expected to leave a small residual shift
of order $0.1\%$. However, this discrepancy is likely to be too small
to have any significant effect on the determination of the BAO peak
position. 

Let us make a cautious remark. Although the above analysis provides a
qualitative understanding of the origin of the BAO shift, as well as a
trustable estimate of its order of magnitude, the concrete numbers
listed in Table~\ref{tableshift0} should be taken with a grain of
salt. They are smaller than the typical percent accuracy of our
calculations, which calls for a re-assessment of various
approximations made in their derivation. Also a realistic calculation
of the BAO shift must include the effect of the bias
\cite{Padmanabhan:2009yr,Sherwin:2012nh,Matsubara:2016wth}. 
We leave the study of these issues for future work. 

Before closing this section, let us mention that the term proportional
to $k^3$ in \eqref{eq:Szk} generates also a \emph{distortion} of the
BAO peak that tends to make it 
more asymmetric. However, in $\Lambda$CDM this effect is subdominant
compared to the initial asymmetry of the peak present already at the linear level
and amplified by various other terms in the NLO power spectrum.
Still, the different contributions are not completely degenerate and
it would be interesting to understand the impact of non-linear
distortion 
on an accurate description of the BAO data.

\section{Conclusions and outlook}\label{sec:conclusions}

In this work we have developed a systematic approach to describe the
non-linear evolution 
of the feature imprinted in the matter correlation functions by baryon
acoustic oscillations. 
We have provided a theoretical framework to efficiently resum
 corrections arising from non-linear interactions with
long-wavelength modes that are 
particularly enhanced for the baryon acoustic feature.

Our approach is based on the framework of TSPT, that provides a
perturbative description of 
structure formation manifestly free from spurious infrared
divergences. Besides, it is based on an Eulerian description
and therefore its practical implementation does not suffer from the
complications arising in Lagrangian perturbation
theory. These features make TSPT a convenient framework to discuss the
effect of bulk 
flows on the BAO feature. We have first developed a formalism to
isolate IR enhanced effects 
by splitting the TSPT propagators and vertices into smooth and
oscillating (wiggly) contributions. 
Next, we identified the IR enhanced loop 
contributions, taking modes below an (a
priori arbitrary) 
separation scale $k_S$ into account. These have a simple diagrammatic
representation, with the dominant diagrams
corresponding to daisy graphs. 
Finally, we have shown that within TSPT one can develop a
modified perturbative expansion in which the large IR effects are
resummed to all orders, and we computed 
next-to leading corrections including loops with hard
wavenumbers 
as well as
 subleading contributions of the soft loops. 
Our leading IR resummed result agrees with
that obtained in \cite{Baldauf:2015xfa} using the symmetry
arguments. TSPT provides a 
useful framework to systematically extend this result to higher
$n$-point functions and compute relevant corrections
in a controlled way.

Our analysis provides a simple prescription for practical evaluation of the
resummed correlation functions. At the leading order, it amounts to
replacing the linear power spectrum in all calculations by the
spectrum with damped wiggly component. This essentially remains true
upon inclusion of hard loops, whereas the subleading soft loops
introduce new terms. Our result for the IR resummed power spectrum
with inclusion of all next-to-leading corrections is given in
\eqref{eq:NLOfinal}. 
It describes the non-linear evolution of the BAO peak with sub-percent
accuracy, when compared to large-scale $N$-body simulations.
Although we found that the soft NLO corrections are rather small, they
are important to capture the shift in the position of the peak
maximum.  

The
residual dependence of our results on the artificial separation scale
$k_S$ provides 
an estimate of the theoretical error, similar to analogous scale-dependencies in
quantum field theory 
computations. At LO our result for the two-point correlation function close
to the BAO peak exhibits a dependence on this scale 
at the level of several percent,
when varying $k_S$ in the plausible range $(0.05\div 0.2)h/$Mpc. 
As expected, the scale-dependence is reduced in the NLO result and is
well below the 
percent level close to the BAO peak. The theoretical error estimated in this way is
consistent with the agreement with $N$-body data, except for short
scales sensitive to the UV effects that were not considered in
this paper. 

Our results suggest several directions for future research.
First, one can use
the systematic TSPT approach to investigate the effects on the 
BAO peak in cosmological models beyond $\Lambda$CDM. 
As examples we mention inclusion of neutrino masses or modifications
of gravity. 
In particular, the NLO corrections affecting the BAO shift 
are sensitive to non-dipole corrections to the
non-linear evolution that are not protected by the equivalence
principle, and therefore can be 
particularly sensitive to modifications of the dynamics.
Second, it will be very interesting to study in detail 
the BAO feature
in the three-point function, as well as to extend the analysis to biased
tracers. 
Finally, the TSPT framework can also be used to address 
the contributions of UV modes that  
influence correlation functions at shorter distances.

\section*{Acknowledments}\label{sec:ackn}
We thank Stefano Anselmi, Tobias Baldauf, Mart\'in Crocce, Guido D'Amico, Vincent Desjacques,
Sergei Dubovsky, Mehrdad Mirbabayi, David Pirtskhalava,
Francisco Prada, Roman Scoccimarro, Gabriele Trevisan, Fillippo
Vernizzi, Matias Zaldarriaga and Miguel Zumalac\'arregui 
for helpful comments and discussions.
We are grateful to Valery Rubakov for encouraging interest.
S.S. acknowledges the hospitality of CCPP at NYU while this work was
being finalized. 
D.B. and S.S. 
thank the Galileo Galilei Institute for Theoretical Physics for the
hospitality and the INFN for partial support during the completion of
this work. 
This work was supported in part by the Swiss National Science Foundation
(M.I. and S.S.)
and the RFBR grant 14-02-00894 (M.I.). 

\appendix 

\section{Recursion relations for TSPT vertices}\label{app:rec}

To define the building blocks of TSPT one starts with the non-linear 
SPT kernels,
\be
\label{alphabeta}
\alpha(\k_1,\k_2)\equiv\frac{(\k_1+\k_2)\cdot \k_1}{k_1^2}\,, \quad \quad 
\b(\k_1,\k_2)\equiv\frac{(\k_1+\k_2)^2(\k_1\cdot \k_2)}{2k_1^2k_2^2}\,. 
\ee
These are used to write down the recursion relations for the vertices
$K_n$ and $\bar \G_n$ \cite{Blas:2015qsi}. The seeds for these
relation are $K_1=1$ and $\bar \G_2$ given by
Eq.~(\ref{eq:initial}). Higher vertices are different for ZA and
ED. We have 
\begin{align}
\text{ZA:}~~~~&K_2(\k_1,\k_2)=\sin^2(\k_1,\k_2)\equiv
1-\frac{(\k_1\cdot \k_2)^2}{k_1^2k_2^2}\;,\\
\text{ED:}~~~~&K_2(\k_1,\k_2)=\frac{4}{7}\sin^2(\k_1,\k_2)\;.
\label{K2ED}
\end{align}
For $n\geq 3$ the recursion relations read,
\bseq
\begin{align}
&\text{ZA:}\notag\\
&K_n(\k_1,...,\k_n)=\frac{1}{n}\bigg[\sum_{i=1}^n
\a\Big(\k_i,\sum_{1\leq j\leq n,j\neq i}\k_j\Big)
K_{n-1}(\k_1,...,\check\k_i,...,\k_n)\notag\\
&\qquad\qquad\qquad\qquad-\sum_{1\leq i <j\leq n}I_2(\k_i,\k_j)
K_{n-1}(\k_i+\k_j,\k_1,...,\check\k_i,...,\check \k_j,...,\k_n)\bigg],\label{recurKZA}
\\
&\bar\G_{n}(\k_1,...,\k_{n})=- \frac{1}{n-2}
\sum_{1\leq i<j\leq n}I_2(\k_{i},\k_{j})\,
\bar\G_{n-1}(\k_{i}+\k_j,\k_1,...,\check\k_i,...,\check\k_j,...,\k_n)\,,
\label{recurGZA}
\end{align}
\eseq
and 
\bseq
\label{eq:recGED}
\begin{align}
&\text{ED:}\notag\\
&K_n(\k_1,...,\k_n)=
\frac{2}{2n+3}\bigg[ \sum_{i=1}^{n}\a\Big(\k_i,\!\sum_{1\leq j\leq n,j\neq i} \!\!\k_j\Big)\,
K_{n-1}(\k_1,...,\check{\k}_i,...,\k_{n})\notag\\
&-\sum_{1\leq i<j \leq n}I_2(\k_i,\k_j)\, 
K_{n-1}(\k_i+\k_j,\k_1,...,\check\k_i,...,\check\k_j,...,\k_n)\notag\\
&-\frac{3}{2} \sum_{p=3}^{n-1}\frac{1}{p!(n-p)!} \sum_{\s}
K_{p}\big(\k_{\s(1)},...,\k_{\s(p)}\big)\,K_{n-p+1}
\Big(\sum_{l=1}^p\k_{\s(l)},\k_{\s(p+1)},...,\k_{\s(n)}\Big)
\Bigg] \,,\label{Kned}\\
&\bar\G_n(\k_1,...,\k_n)=-\frac{1}{n-2} 
\!\sum_{1\leq i<j\leq n}\!\!
I_2(\k_i,\k_j)
\bar\G_{n-1}(\k_i\!+\!\k_j,\k_1,...,\check\k_i,...,\check\k_j,...,\k_n)
\notag\\
&-\frac{3}{2(n-2)} \sum_{p=3}^{n-1}
\frac{1}{p!(n-p)!}
\sum_{\s}
K_{p}\big(\k_{\s(1)},...,\k_{\s(p)}\big)
\bar\G_{n-p+1}\Big(\sum_{l=1}^p \k_{\s(l)},\k_{\s(p+1)},...,\k_{\s(n)}\Big)\,.
\label{Gned}
\end{align}
\eseq
where
\be
\label{I2}
I_2(\k_1,\k_2)=\begin{cases}
2\b(\k_1,\k_2)& \text{for ZA}\\
2\b(\k_1,\k_2)
+\frac{3}{2}K_2(\k_1,\k_2)& \text{for ED}
\end{cases}
\ee
The notation $\check \k_i$ above means that the momentum $\k_i$ is
absent from the arguments of the corresponding function, and in the
last lines of (\ref{Kned}), (\ref{Gned}) the summation is performed over all
permutations $\sigma$ of $n$ indices.

\section{Decomposition of smooth and wiggly components}
\label{app:wiggly}

In this appendix we describe our algorithms to separate
the smooth and oscillating contributions to the matter power
spectrum. In the first case, 
in order to obtain a smooth power spectrum we perform
a fit to the linear power spectrum obtained
with the CLASS code \cite{Blas:2011rf} that is inspired by the Eisenstein-Hu formula \cite{Eisenstein:1997jh}.
Specifically, we find that the smooth part can be well described by
the parametric form
\be
 \bar P_{s}(k) = A k^{n_s} T(k)^2 \times (1+ \Delta(k))\,,
\ee
where the transfer function is parameterized as in \cite{Eisenstein:1997jh},
\bea
 T(k) &=& \frac{L(k)}{L(k)+C(k) k_{eff}^2} \nn\\
 L(k) &=& \ln(e+ c_1 k_{eff}) \nn\\
 C(k) &=& 14.4 + \frac{325}{1+60.5 k_{eff}^{1.08}} \nn\\
 k_{eff} &=& c_2 k \left[1+\frac{c_3}{1+(c_4 k)^4}\right]^{-1}\,.
\eea
Here we treat the coefficients $A$ and $c_i$ as free parameters that
are determined by fitting to the linear power spectrum, see Tab.\,\ref{tableSmooth}.
In addition, we introduce a correction of the Eisenstein-Hu fitting
formula to account for a slight residual offset at large $k$ given by
\be
 \Delta(k) = c_5 \left[ \tanh\left(\ln(c_6 k)/c_7\right)+1\right]\,.
\ee
The values of $c_5-c_7$ we used are also given in Tab.\,\ref{tableSmooth}.
The oscillating component is then given by
$P_w(k)=P_{lin}(k)-P_s(k)$. It is plotted in Fig.~\ref{fig:Pwiggly}
by the solid line.

\begin{table*}[t]
\centering
  \begin{tabular}{|cccccccc|}
     \hline
     $A$ &$c_1$&$c_2$&$c_3$&$c_4$&$c_5$&$c_6$&$c_7$ \\ \hline 
     $1.496\cdot 10^4$ & $1.967$ & $6.831$ & $0.4043$ & $55.53$ & $0.00425$ & $2.5$ & $0.35$ \\
     \hline
   \end{tabular}
\caption{The values of the coefficients $c_i$ used to describe the smooth part of
the power spectrum for a $\Lambda$CDM model with parameters given as in \cite{Kim:2011ab}, in particular $n_s=0.96$. The dimensionful
coefficients $c_2, c_4, c_6$ are given in units of Mpc$/h$, and $A$ in units $($Mpc$/h)^3$.}
\label{tableSmooth}
\end{table*}

The second algorithm is different in that the smooth part is
determined by a spline approximation using several ($\sim 20$) 
points to the right and
left of the scales of BAO 
and a pivot point at $k=0.03\, h/$Mpc \cite{ZaldaICTP}. This leads to the wiggly power
spectrum shown by the dashed curve in Fig.~\ref{fig:Pwiggly}.

\section{Asymptotic behavior of the $\G_n$ vertices in the soft limit}
\label{app:TSPT}

In this Appendix we study in detail the form of the TSPT vertices with
soft legs. First, we derive the leading order expression
\eqref{IRnm}. 
Next, we extend the analysis to include the subleading corrections and
obtain Eq.~\eqref{NLOG}, as well as the expression for the
four-point vertex evaluated on the modified linear power spectrum used
in the derivation of Eq.~(\ref{daisynew}).

\hiddensubsection{$\G_n$ vertices in the soft limit: leading order}
\label{app:LO}

We split the arguments of the vertices into `hard' momenta
$\k_i$, $1\leq i\leq m$, $\k_m=-\sum_{i=1}^{m-1}\k_i$, 
that are are fixed and `soft' momenta  $\q_j$, which are 
sent uniformly to zero,
\be
\label{eq:uniform}
\q_j=\varepsilon \,\tilde \q_j\,,\quad \varepsilon  \to 0\,.
\ee
To prove (\ref{IRnm}) we proceed by induction. Let us fix $m\geq 2$
and assume that Eq.~(\ref{IRnm}) has been already proved for all
$m'< m$. Equation (\ref{IRnm}) holds trivially for $n=m$. Now,
suppose that it is valid for all $n'$ such that $m\leq n' < n$. Our
task is to show that it also holds for $n$.

We focus on  ED; the derivation for ZA can be recovered by simply 
ignoring all contributions due to the $K_n$ kernels. We will use the
shorthand 
$
\q\equiv \sum_{j=1}^{n-m}\q_j\,. 
$
It is convenient to decompose the $\bar\G'^w_n$ into two pieces,
\be
\label{G2AB}
\bar \G^w_{n}=\bar \G^{w}_{n,A}+ \bar \G^{w}_{n,B}\,,
\ee
which correspond respectively to the first and second lines in the
recursion relation (\ref{Gned}).
Consider first $\bar \G^{w}_{n,A}$. We have,
\be
\begin{split}
\bar \G'^{w}_{n,A}(\k_1,...,\k_m-\q,\q_1,...,&\q_{n-m})
=\frac{-1}{n-2}\bigg[\sum_{1\leq i<j< m}
\!\!I_2(\k_i,\k_j)
\bar \G'^w_{n-1}(\k_i+\k_j,...,\check{\k}_i,...,\check\k_i,...)\\
&+\sum_{1\leq i<m} I_2(\k_i, \k_m-\q)
\bar
\G'^w_{n-1}(\k_i+\k_m-\q,...,\check\k_i,...,\check{\k}_m-\check{\q},...)\\
&+\sum_{i=1}^{m-1}\sum_{j=1}^{n-m}  
I_2(\k_i,\q_j)
\bar \G'^w_{n-1}(...,\k_i+\q_j,...,\check\q_j,...)\\
&+\sum_{j=1}^{n-m} I_2(\k_m-\q,\q_j)
\bar \G'^w_{n-1}(...,\k_m-\q+\q_j,...,\check\q_j,...)\\
&+\sum_{1\leq i<j\leq n-m}I_2(\q_i,\q_j)
\bar \G'^w_{n-1}(...,\q_i+\q_j,...,\check\q_i,...,\check\q_j,...)
\bigg]\,.
\end{split}
\label{GA1}
\ee
Let us analyze the soft enhancement of various terms in this
expression. The vertex functions in the first two lines have $n-m$
soft arguments, and hence, by the induction assumption, are of order 
$O(\varepsilon^{-n+m})$. On the other hand, the vertices
$\bar\G_{n-1}'^w$ in the last three lines have one soft argument less
and thus are only $O(\varepsilon^{-n+m+1})$. In the third and fourth
lines this is compensated by the enhancement of the kernels
$I_2$. Indeed,
\be
\label{I2soft}
I_2(\k_i,\q_j)\approx\frac{\k_i\cdot\q_j}{q_j^2}=O(1/\ve)\;.
\ee
Finally, the kernel $I_2$ in the last line of (\ref{GA1})
is of order $O(\ve^0)$ and
thus this term can be neglected in the leading approximation.

Keeping only contributions of order $O(\varepsilon^{-n+m})$ we obtain,
\be
\begin{split}
\label{leadproof}
\bar \G'^{w,LO}_{n,A}= \frac{-1}{n-2}\bigg[&
\sum_{1\leq i<j\leq m}
I_2(\textbf{k}_i,\textbf{k}_j)\bar 
\G'^{w}_{n-1}(\k_i+\k_j,...\check\k_i,...,\check\k_j,...)\\
&+\sum_{j=1}^{n-m}\bigg(\sum_{i=1}^{m-1}  
\frac{(\k_i\cdot \q_j)}{q_j^2}
\bar \G'^{w}_{n-1}(...,\k_i+\q_j,...,\check{\q}_j,...)\\
&\qquad\qquad+\frac{(\k_m\cdot\q_j)}{q_j^2}
\bar
\G'^{w}_{n-1}\Big(...,\k_m-\sum_{l\neq j}\q_l,...,\check{\q}_j,...\Big)\bigg)
\bigg]\\
=\frac{-1}{n-2}\bigg[&\sum_{1\leq i<j\leq m}
I_2(\textbf{k}_i,\textbf{k}_j)\bar 
\G'^{w}_{n-1}(\k_i+\k_j,...\check\k_i,...,\check\k_j,...)\\
&+\sum_{j=1}^{n-m}\D_{\q_j}
\bar \G'^{w}_{n-1}(\k_1...,\k_m-\sum_{l\neq j}\q_l,...,\check{\q}_j,...)
\bigg]\;,
\end{split}
\ee
where in passing to the second equality we substituted 
$\k_m=-\sum_{i=1}^{m-1}\k_i$ and used
\be
\begin{split}
\sum_{i=1}^{m-1}\frac{(\k_i\cdot\q_j)}{q_j^2}
\Big[&\bar\G_{n-1}'^w\big(\k_1,...,\k_i+\q_j,...,
\k_m-\q,...,\check\q_j,...\big)\\
&-\bar\G_{n-1}'^w\Big(\k_1,...,\k_i,...,
\k_m-\sum_{l\neq j}\q_l,...,\check\q_j,...\Big)
\Big]\\
&\qquad\qquad=
\D_{\q_j}\bar\G_{n-1}'^w\Big(\k_1,...,\k_i,...,
\k_m-\sum_{l\neq j}\q_l,...,\check\q_j,...\Big)\;.
\end{split}
\ee
Finally, inserting the expansion (\ref{IRnm}) 
for the vertices in (\ref{leadproof}) and using that the operators
$\D_{\q_j}$ commute with each other, we arrive at
\be
\begin{split}
\label{leadproof1}
\bar \G'^{w,LO}_{n,A}= \frac{(-1)^{n-m}}{n-2}
\prod_{l=1}^{n-m}\D_{\q_l}
\bigg[&-
\sum_{1\leq i<j\leq m}
I_2(\textbf{k}_i,\textbf{k}_j)\bar 
\G'^{w}_{m-1}(\k_i+\k_j,...\check\k_i,...,\check\k_j,...,\k_m)\\
&+(n-m)
\bar \G'^{w}_{m}(\k_1,...,\k_m)
\bigg]\;.
\end{split}
\ee

We now turn to the second piece in (\ref{G2AB}). By inspection of the
second line of (\ref{Gned}) one concludes that the terms of order
$O(\ve^{-n+m})$ can arise only if all $(n-m)$ soft wavenumbers 
$\q_j$ appear as the arguments of the vertex function
$\bar\G_{n-p+1}^w$ (recall that the kernels $K_p$ do not depend on the
wiggly power spectrum and hence are not enhanced). 
The vertex must also depend on at
least two hard momenta, which implies that $p$ cannot exceed
$m-1$. This yields for the LO contributions,
\be
\label{Krec}
\begin{split}
\bar \G'^{w,LO}_{n,B}(\k_1,...,\k_m-\q,\q_1,...,\q_{n-m})
=-\frac{3}{2(n-2)}\sum_{p=3}^{m-1}\frac{1}{p!(n-p)!}
\times(n-m)!\,C_{n-p}^{n-m}
\\
\times\sum_{\sigma}
K_p(\k_{\sigma(1)},...,\k_{\sigma(p)})
\,
\bar\G'^w_{n-p+1}\left(\sum_{l=1}^{p}\k_{\sigma(l)},\k_{\s(p+1)},...,
\k_{\s(m)}-\q,\q_{1},...,\q_{n-m}\right)\;,
\end{split} 
\ee
where, in contrast to (\ref{Gned}),
we have replaced the summation over all permutations of the
arguments by the sum only over permutations of 
the hard wavenumbers and accounted for the multiplicity of the retained
terms with the appropriate
symmetry factor. Using Eq.~(\ref{IRnm}) we obtain,
\be
\label{Krec1}
\begin{split}
\bar \G'^{w,LO}_{n,B}=\frac{3(-1)^{n-m}}{2(n-2)}
\prod_{l=1}^{n-m}\D_{\q_l}
\bigg[-&\sum_{p=3}^{m-1}\frac{1}{p!(m-p)!}
\sum_{\sigma}
K_p(\k_{\sigma(1)},...,\k_{\sigma(p)})\\
&\times\bar\G'^w_{m-p+1}\left(\sum_{l=1}^{p}\k_{\sigma(l)},\k_{\s(p+1)},...,
\k_{\s(m)}\right)\bigg]\;.
\end{split}
\ee
One notices that this expression, when combined with the first line of
(\ref{leadproof1}), gives precisely the recursive formula for 
\be
\frac{(-1)^{n-m}}{n-2}\prod_{l=1}^{n-m}\D_{\q_l}
\big[(m-2)\bar\G_m'^w(\k_1,...,\k_m)\big]\;.
\ee
Adding to this the second line of (\ref{leadproof1}) yields
(\ref{IRnm}). This is our final result. It has the same form 
in ED and ZA.

\hiddensubsection{$\G_n$ vertices in the soft limit: NLO}
\label{app:sublead}

In this section we study the subleading corrections of order
$O(\ve^{-n+3})$ to the expression (\ref{asympPs}).
We carry out the derivation for ZA and ED in parallel. 
Whenever there are differences between these two cases
we will use indices $(ZA)$ or $(ED)$.

Let us start with $\bar \Gamma'^w_3$. Expanding Eq.~(\ref{preG3}) in
the soft wavenumber $\q$ and keeping term up to 
$O\big((q/k)^0\big)$ we obtain,
\be
\label{eq:G3nlo}
\bar \G'^{w,LO+NLO}_3(\k,-\k-\q,\q)=
\left[\D_{\q}+\E_\q \right]\frac{\bar P_w(k)}{\bar P^2_s(k)}\,,
\ee
where we introduced the new operator $\E_\q$ defined as,
\be
\label{eq:sublop1}
\begin{split}
 \E^{(ZA)}_\q \bar P_w= &
 \Big(2\cos^2(\q,\k)-1+2\cos^2(\q,\k)\big(1-n_s(k)\big)e^{\q\cdot
   \nabla}\Big)\bar P_w\,,\\
 \E^{(ED)}_\q \bar P_w =&\bigg(\E_\q^{(ZA)}+
 \frac{6}{7}\sin^2(\q,\k)\left(e^{\q\cdot \nabla}+1\right)\bigg) \bar
 P_w\,, 
\quad\qquad 
 n_s\equiv  \frac{d \ln \bar P_s(k)}{d \ln k}\,.
\end{split} 
\ee
As in the case of $\D_\q$, the operator $\E_\q$ acts only on the wiggly
power spectrum, leaving all smoothly varying functions intact. Its
action is of order $O(1)$,
\be
\label{Eorder}
\E_\q \bar P_w\sim O(1)\bar P_w\;.
\ee  
A new structure appears when we consider the subleading expansion of 
$\bar \G'^w_4$. After a somewhat lengthy, but straightforward
calculation using the recursion relations (\ref{recurGZA}),
(\ref{Gned}) 
one obtains,
\be
\label{eq:G4nlo}
\begin{split}
\bar \G'^{w,LO+NLO}_4(\k,-\k-\q_1-\q_2,\q_1,\q_2)=-\left[\D_{\q_1} \D_{\q_2}+\D_{\q_1}\E_{\q_2} + \D_{\q_2} \E_{\q_1}+\F_{\q_1\q_2}\right]\frac{\bar P_w(k)}{\bar P^2_s(k)}\,,
\end{split} 
\ee
where 
\be
\label{eq:sublop2}
\begin{split}
\F^{(ZA)}_{\q_1\q_2}\bar P_w= & \frac{(\q_1\cdot \q_2)}{q_1^2q_2^2}\Big[(\k
\cdot \q_1)e^{\q_1\cdot \nabla}(e^{\q_2\cdot \nabla}-1)+(\k \cdot
\q_2)e^{\q_2\cdot \nabla}(e^{\q_1\cdot \nabla}-1)\Big]\bar P_w\,,\\
\F^{(ED)}_{\q_1\q_2}\bar P_w= &\bigg[\F^{(ZA)}_{\q_1\q_2}
+ \frac{3}{7}\sin^2(\q_1,\q_2)\D_{\q_1+\q_2}\bigg]\bar P_w\,.
\end{split} 
\ee
This operator is of order $O(\ve^{-1})$. The expressions
(\ref{eq:G3nlo}), (\ref{eq:G4nlo}) are special cases of
Eq.~(\ref{NLOG}) from the main text for $n=3$ and $4$. We now prove
Eq.~(\ref{NLOG}) for general $n$ by induction.

Suppose that \eqref{NLOG} holds for $\bar \G'^{w,NLO}_{n'}$ for any $n'< n$.
Assuming that all $\q_i$ go uniformly to zero as in \eqref{eq:uniform},
we will expand the vertex 
$\bar \G'^{w}_{n}\big(\k,-\k-\sum^{ n-2}_{i=1}\q_i,\q_1,..,\q_{n-2}\big)$
in powers of soft momenta and focus on the terms scaling 
like $\varepsilon^{-n+2}$ (leading order)
and $\varepsilon^{-n+3}$ (next-to-leading order).
The terms of the form
$\G_{n-p+1} K_{p}$ in the recursion relation \eqref{Gned}
scale at most as $\varepsilon^{-n+4}$ because the $K_p$
kernels are infrared safe and cannot produce new poles. 
Hence, we concentrate on the part of the recursion relation 
\eqref{Gned} with the $\bar\G_{n-1}$ vertex,
\be 
\label{asympsub1}
\begin{split}
\bar \G'^{w}_{n}(\k,-\k-\q,&\q_1,..,\q_{n-2})
=\frac{-1}{n-2}
\Bigg[ \sum_{j=1}^{n-2}I_2(\k,\q_j)\,
\bar \G'^{w}_{n-1}(\k+\q_j,-\k-\q,...,\check{\q}_j,...)\\
&+\sum_{j=1}^{n-2}I_2(-\k-\q,\q_j)\,
\bar \G'^{w}_{n-1}(\k,-\k-\q+\q_j,...,\check{\q}_j,...)
\\
&+\sum_{\bsm i,j=1\\
i<j\esm}^{n-2}I_2(\q_i,\q_j)
\bar \G'^{w}_{n-1}(\k,-\k-\q,\q_i+\q_j,...,\check \q_i,...,\check\q_j,...)
\Bigg]+O(\varepsilon^{-n+4})\,,
\end{split}
\ee
where we used the shorthand notation $\q=\sum_{i=1}^{n-2}\q_i$.
The hard argument of the vertex in the first term in (\ref{asympsub1})
is shifted by $\q_j$ with respect to $\k$. Thus, we have to shift the
argument in the expansion \eqref{NLOG} for $\bar\G_{n-1}'^w$ and keep
the subleading terms in $\q_j$. Multiplying by $I_2(\k,\q_j)$
Taylor-expanded up to zeroth order in $q_j$ we obtain,
\be
\label{monster1}
\begin{split}
A_j&\equiv I_2(\k,\q_j)\,
\bar \G'^{w}_{n-1}(\k+\q_j,-\k-\q,...,\check{\q}_j,...)
\approx (-1)^{n-2} \bigg[\frac{(\k\cdot\q_j)}{q_j^2}\prod_{l\neq
  j}^{n-2} \D_{\q_l}\\
&+\frac{(\k\cdot\q_j)}{q_j^2}\sum_{m\neq j}^{n-2}\!\!
\bigg(\E_{\q_m}+\frac{(\q_j\cdot\q_m)}{q_m^2}(e^{\q_m\cdot\nabla}-1)\bigg)
\!\!\prod_{l\neq j,m}^{n-2}\!\!\D_{\q_l}
+\frac{(\k\cdot\q_j)}{q_j^2}\!\!\!\!\!\sum_{\bsm m_1<m_2\\m_1,m_2\neq j\esm}^{n-2}
\!\!\!\!\!\F_{\q_{m_1}\q_{m_2}}
\!\!\!\prod_{l\neq j,m_1,m_2}^{n-2}\!\!\!\D_{\q_l}\\
&+\bigg(2(1-n_s)\cos^2(\k,\q_j)+\frac{6}{7}\sin^2(\k,\q_j)\bigg)
\prod_{l\neq j}^{n-2}\D_{\q_l}\bigg]
\frac{e^{\q_j\cdot\nabla}\bar P_w(k)}{\bar P_s^2(k)}\;.
\end{split} 
\ee
In the second term in (\ref{asympsub1}) we can directly substitute
Eq.~(\ref{NLOG}) for $\bar \G_{n-1}'^w$ which yields,
\be
\label{monster2}
\begin{split}
B_j\equiv &I_2(-\k-\q,\q_j)\,
\bar \G'^{w}_{n-1}(\k,-\k-\q+\q_j,...,\check{\q}_j,...)
\approx (-1)^{n-2}\bigg[-\frac{(\k\cdot\q_j)}{q_j^2}
\prod_{l\neq j}^{n-2} \D_{\q_l}\\
&-\frac{(\k\cdot\q_j)}{q_j^2}\sum_{m\neq j}^{n-2}\E_{\q_m}
\prod_{l\neq j,m}^{n-2}\D_{\q_l}
-\frac{(\k\cdot\q_j)}{q_j^2}
\!\!\!\!\!\sum_{\bsm m_1<m_2\\m_1,m_2\neq j\esm}^{n-2}
\!\!\!\!\!\F_{\q_{m_1}\q_{m_2}}
\!\!\!\prod_{l\neq j,m_1,m_2}^{n-2}\!\!\!\D_{\q_l}\\
&+\bigg(2\cos^2(\k,\q_j)-1+\frac{6}{7}\sin^2(\k,\q_j)\bigg)
\prod_{l\neq j}^{n-2}\D_{\q_l}
-\sum_{m\neq j}^{n-2}\frac{(\q_m\cdot\q_j)}{q_j^2}\prod_{l\neq j}^{n-2}\D_{\q_l}\bigg]
\frac{\bar P_w(k)}{\bar P_s^2(k)}\,.
\end{split} 
\ee
Together the above contributions sum up to
\be
\label{monster12}
\begin{split}
&\frac{-1}{n-2}\sum_{j=1}^{n-2}(A_j+B_j)=\frac{(-1)^{n-1}}{n-2}\bigg[
(n-2)\prod_{l=1}^{n-2}\D_{\q_l}
+(n-2)\sum_{m=1}^{n-2}\E_{\q_m}\prod_{l\neq m}^{n-2}\D_{\q_l}\\
&\quad+(n-4) \sum_{m_1<m_2}^{n-2}\F_{\q_{m_1}\q_{m_2}}
\prod_{l\neq m_1,m_2}^{n-2}\D_{\q_l}
+\sum_{m<j}^{n-2}\frac{(\q_m\cdot\q_j)}{q_m^2q_j^2}
\Big(\big(\k\cdot(\q_m+\q_j)\big)(e^{(\q_m+\q_j)\cdot\nabla}-1)\\
&\quad-2(\k\cdot\q_m)e^{\q_m\cdot\nabla}-2(\k\cdot\q_j)e^{\q_j\cdot\nabla}\Big)
\prod_{l\neq m,j}^{n-2}\D_{\q_l}\bigg]\frac{\bar P_w(k)}{\bar P_s^2(k)}\;.
\end{split}
\ee
It remains to include the third term in (\ref{asympsub1}). It is
sufficient to consider only the leading behaviour of the vertex, as
$I_2(\q_i,\q_j)$ is an order-one function. We obtain,
\be
\label{monster3}
\begin{split}
&\frac{-1}{n-2}\sum_{i<j}^{n-2}I_2(\q_i,\q_j)
\bar \G'^{w}_{n-1}(\k,-\k-\q,\q_i+\q_j,...,\check \q_i,...,\check\q_j,...)
\\
&\approx \frac{(-1)^{n-1}}{n-2}\sum_{i<j}^{n-2}
\bigg[\frac{\big(\k\cdot(\q_i+\q_j)\big)}{q_i^2 q_j^2}
(e^{(\q_i+\q_j)\cdot\nabla}-1)
+\frac{6}{7}\sin^2(\q_i,\q_j)\D_{\q_i+\q_j}\bigg]
\prod_{l\neq i,j}^{n-2}\D_{\q_l}
\frac{\bar P_w(k)}{\bar P_s^2(k)}\,.
\end{split} 
\ee
Combining this with (\ref{monster12}) we obtain the representation 
(\ref{NLOG})
for $\bar \G_n'^w$. QED.\\

In Sec.~\ref{sec:NLO} we used the NLO expression for the four-point vertex $\bar\G_n'^w(\k,-\k,\q,-\q)$
evaluated using the modified power spectrum $e^{-g^2\S}\bar P_w$. Let us derive this expression. 
As emphasized in the main text, it would be incorrect to just substitute 
$\bar P_w\mapsto e^{-g^2\S}\bar P_w$ in the formula (\ref{eq:G4nlo}) because the 
combination $\D_{-\q}\D_\q[e^{-g^2\S}\bar P_w]$
involves terms containing the operator $\S$ with shifted argument. This shift generates additional NLO contributions that must be properly taken into account. Indeed, we have,
\be
\label{DS}
\begin{split}
\D_\q[e^{-g^2\S}\bar P_w](\k)&=\frac{(\k\cdot\q)}{q^2}\Big(e^{-g^2\S\big|_{\k+\q}}
\bar P_w(|\k+\q|)-e^{-g^2\S\big|_{\k}}\bar P_w(k)\Big)\\
&=\D_\q e^{-g^2\S\big|_{\k}}\bar P_w(k)
-g^2\frac{(\k\cdot\q)}{q^2}\Delta \S\big|_{\q}e^{-g^2\S\big|_{\k}}\;e^{\q\cdot\nabla}\bar P_w(k)+O(\ve)\;,
\end{split}
\ee
where
\be
\Delta \S\big|_{\q}=2\int_{q'\leq k_S} [dq']\bar P_s(q')\frac{(\k\cdot\q')(\q\cdot\q')}{q'^4}
\big(1-\cosh(\q'\cdot\nabla)\big)\;.
\ee
Acting on (\ref{DS}) with $\D_{-\q}$ and again taking into account the shift in the argument of $\S$ in the first term we obtain the desired expression,
\be
\label{eq:expdressed}
\begin{split}
&\bar \G'^w_4(\k,-\k,\q,-\q)[e^{-g^2\S}\bar P_w]=
-\big[\D_{\q}\D_{-\q}+\E_{\q}\D_{-\q}+\D_{\q}\E_{-\q}
+\F_{\q,-\q}\big]\frac{e^{-g^2\S}\bar P_w(k)}{\bar P_s^2(k)}\\
&~~~~+4\frac{(\k \cdot \q)^2}{q^4}
\int_{q'\leq k_S}[dq']
\bar P_s(q')\frac{(\k\cdot \q')(\q\cdot\q')}{q'^4} 
\sinh(\q\cdot\nabla) \big(1-\cosh(\q'\cdot \nabla)\big)
\frac{e^{-g^2\S}\bar P_w(k)}{\bar P_s^2(k)}\,.
\end{split} 
\ee
This formula was used in the derivation of Eq.~(\ref{daisynew}).

\section{NLO polynomials}\label{app:NLOpol}
The functions appearing Eq.\,\eqref{hxy} are given by
\begin{align}
   h_1(x&,y) =\frac{1}{k_{osc}}\int \frac{d\Omega_q d\Omega_{q'}}{(4\pi)^2} \, 
  \frac{c_{kq}c_{kq'}(qc_{kq}+q'c_{kq'})s_{qq'}^2qq'}{(q^2+2qq'c_{qq'}+{q'}^2)}
  \sin\left((q c_{kq}+q' c_{kq'})/k_{osc}\right)\Big|_{\bsm q=x k_{osc}\\ q'=y k_{osc}\esm}  \nn\\
 =& -\frac{1}{60 x^2 y^2 } \left(720+x^4-12 y^2+y^4-2 x^2 \left(6+y^2\right)\right) \cos[x] \cos[y] \nn\\
 & -\frac{1}{240 x^3 y^3 }\left(x^2-y^2\right)^4 (\text{Ci}[|x-y|]-\text{Ci}[x+y])\nn\\
&-\frac{1}{240 x^3 (x^2-y^2) y^3 } \Bigg\{-2
y \Bigg(x^8-2 x^6 \left(3+2 y^2\right)+2 x^4 \left(-300+5 y^2+3 y^4\right) \nn\\ 
 & +y^2 \left(-1440+24 y^2-2 y^4+y^6\right)-2 x^2 \left(-720-240 y^2+y^4+2
y^6\right)\Bigg) \cos[y] \sin[x] \nn\\
 & +2 \Bigg(x \Big(x^8-2 x^6 \left(1+2 y^2\right)+x^4 \left(24-2 y^2+6 y^4\right)+2 x^2 \left(-720+240
y^2+5 y^4-2 y^6\right) \nn\\
 & +y^2 \left(1440-600 y^2-6 y^4+y^6\right)\Big) \cos[x]+(x^2-y^2) \Big(1440+x^6-600 y^2-6 y^4 \nn\\
 & +y^6-x^4 \left(6+y^2\right)-x^2
\left(600-204 y^2+y^4\right)\Big) \sin[x]\Bigg) \sin[y]\Bigg\}\,,
\end{align}
and
\bea
   h_2(x,y) &=& -\frac{1}{k_{osc}}\int \frac{d\Omega_q d\Omega_{q'}}{(4\pi)^2} \, 
  \frac{c_{kq}c_{kq'}(qc_{kq}+q'c_{kq'})s_{qq'}^2qq'}{(q^2+2qq'c_{qq'}+{q'}^2)}
  \sin\left(q c_{kq}/k_{osc}\right)\Big|_{q=x k_{osc}, q'=y k_{osc}}  \nn\\
 &=& -\frac{1}{192 x^8 y^3}\Bigg\{x \cos[x] \Bigg(4 x y \Big(3 x^8+27 y^6-3 x^2 y^4 \left(21+y^2\right)+x^4 y^2 \left(-3+5 y^2\right) \nn\\
 &&-x^6 \left(9+5 y^2\right)\Big)  +3
(x-y)^3 (x+y)^3 \left(x^4-9 y^2+x^2 \left(-3+y^2\right)\right) \log\left[\frac{(x-y)^2}{(x+y)^2}\right]\Bigg) \nn\\
 && {} +\Bigg(4 x y \left(9
x^6-6 x^8+x^4 \left(3+4 x^2\right) y^2+x^2 \left(63-26 x^2\right) y^4+3 \left(-9+4 x^2\right) y^6\right) \nn\\
 && -3 (x-y)^3 (x+y)^3 \left(2 x^4-9 y^2+x^2
\left(-3+4 y^2\right)\right) \log\left[\frac{(x-y)^2}{(x+y)^2}\right]\Bigg) \sin[x]\Bigg\}\,,
\eea
where $c_{kq}=\k\cdot\q/(kq)$, $c_{kq'}=\k\cdot\q'/(kq')$, $c_{qq'}=\q\cdot\q'/(qq')$, $s_{qq'}^2=1-c_{qq'}^2$.
Note that instead of averaging over the directions of $\q$ and $\q'$ one can equivalently
average over the direction of the external wavenumber $\k$ and either $\q$ or $\q'$.
In particular, $h_2$ can be evaluated setting $\q'=q'(0,0,1)$, $\k=k(0,s_k,c_k)$,
$\q=q(s_q s_\phi, s_q c_\phi, c_q)$ and integrating first over $\phi$, then $c_k$ and then $c_q$.
The integral $h_1$ can be evaluated by first averaging over the direction of $\k$ in a frame
where the $z$-axis is chosen along $\q+\q'$, and $\q=q(0,s_q,c_q)$. Then one uses that
\be
  c_q = \frac{(\q+\q')^2+q^2-{q'}^2}{2q |\q+\q'|},\quad c_{qq'} = \frac{(\q+\q')^2-q^2-{q'}^2}{2q q'}
\ee
and performs the average over $c_{qq'}$. For the last integral it is helpful to do the
substitution $dc_{qq'}=|\q+\q'|/(qq')\, d|\q+\q'|$, with integration boundaries from $|q-q'|$ to $q+q'$.

\section{Shift  of the BAO in momentum space}
\label{app:shift}

As discussed in the main text, at LO the IR resummation leads to a
simple damping of the wiggly component in the power spectrum. In this
appendix we evaluate the NLO contribution and show that it contains a
term describing the shift of the phase of BAO. Namely, we derive
Eq.~(\ref{eq:Psnlo}). Our calculation is similar to the one performed
in \cite{Sherwin:2012nh} with the main difference that the IR resummation
procedure allows us to consistently take
into account the BAO damping. Besides, we retain certain NLO terms that were
omitted in~\cite{Sherwin:2012nh}.

According to Eq.~(\ref{eq:NLOfinal}) we need to compute the 1-loop
correction to the power spectrum using the damped wiggly spectrum as
an input. It is convenient to use the standard SPT expression (see
\cite{Blas:2015qsi} for the derivation using TSPT),
\be
\label{SPT1loop}
P^{1-loop}(z,k)=D(z)^4\Big[6P(k)\!\!\int\! [dq] E_3(\k,\q,-\q)P(q)
+2\!\!\int\! [dq]\big(E_2(\k-\q,\q)\big)^2P(q)P(|\k-\q|)\Big],
\ee
where $E_n$ are the SPT kernels $F_n$ ($G_n$) for the density
(velocity divergence) power spectrum. The first and second terms in
brackets are identified as $P_{13}$ and $P_{22}$
respectively. Performing the smooth + wiggly decomposition we see that
the term $P_{13}$ contributes only to the part proportional to
$P_w(k)$ and thus does not affect the BAO phase. In the $P_{22}$ term
we split the integration into soft and hard parts,
\be
\label{Pw22}
\begin{split}
P_{w,22}\big[e^{-k^2D(z)^2\Sigma^2}P_w\big]
=4D(z)^4\bigg[\int_{q\leq k_S}[dq]+\int_{q\geq k_S}[dq]\bigg]
&\big(E_2(\k-\q,\q)\big)^2\\
\times &P_s(q)P_w(|\k-\q|)
e^{-(\k-\q)^2D(z)^2\Sigma^2}\;.
\end{split}
\ee
The integral over the hard modes vanishes as it involves a rapidly
oscillating function. In the soft part we use the expression
(cf. Eqs.~(\ref{eq:Fnexpansion}), (\ref{eq:FnSub})),
\be
\label{F2exp}
E_2(\k-\q,\q)=\frac{(\k\cdot\q)}{2q^2}\bigg(1+\frac{\k\cdot\q}{k^2}
+\Upsilon\frac{q^2}{(\k\cdot\q)}\sin^2(\k,\q)+O(q^2/k^2)\bigg)\;,
\ee
where the coefficient $\Upsilon$ is different in ED and ZA and depends
on the type of the power spectrum under consideration; its values are
given in Table~\ref{table} (Sec. \ref{sec:longsc}). Next, using the representation
(\ref{envelop}) we write,
\be
\label{Pwshifted}
\begin{split}
P_w(|\k-\q|)=P_w(k)\bigg[&\cos\frac{(\hat\k\cdot\q)}{k_{osc}}
+\frac{d\log f_{env}}{d\log k}
\bigg(\frac{k_{osc}}{k}\sin\frac{(\hat\k\cdot\q)}{k_{osc}}
-\frac{(\hat\k\cdot\q)}{k}\cos\frac{(\hat\k\cdot\q)}{k_{osc}}\bigg)\\
&+\frac{q^2}{2kk_{osc}}\sin^2(\k,\q)\sin\frac{(\hat\k\cdot\q)}{k_{osc}}\bigg]
+k_{osc}\frac{dP_w(k)}{dk}
\bigg[-\sin\frac{(\hat\k\cdot\q)}{k_{osc}}\\
&+\frac{d\log f_{env}}{d\log k}\frac{(\hat\k\cdot\q)}{k}
\sin\frac{(\hat\k\cdot\q)}{k_{osc}}
+\frac{q^2}{2kk_{osc}}\sin^2(\k,\q)\cos\frac{(\hat\k\cdot\q)}{k_{osc}}\bigg]\;.
\end{split}
\ee
Expanding also the exponent in (\ref{Pw22}) and integrating over
angles one obtains,
\be
\label{Pw22-2}
P_{w,22}\big[e^{-k^2D(z)^2\Sigma^2}P_w\big]
=D(z)^4e^{-k^2D(z)^2\Sigma^2}\bigg[\tilde H(z,k)P_w(k)
+\tilde S(z,k)\frac{dP_w(k)}{dk}\bigg]\;, 
\ee
where the form of $\tilde H(z,k)$ is unimportant for our purposes.
For the second term we have
\be
\label{tildeS}
\tilde S(z,k)=
k\bigg(1-\frac{1}{2}\frac{d\log f_{env}}{d\log k}\bigg) s_1
+k (s_2+s_3)-k^3 D(z)^2\Sigma_a^2
\ee
where
\bseq
\label{eq:subleading}
\begin{align}
s_1= &  -8\pi k_{osc}\int_0^{k_S} dq q P_s(q) 
\int_{0}^1 d\mu \, \mu^3\sin({q\mu}/{k_{osc}})\\
s_2= &-8\pi\, \Upsilon k_{osc}\int_0^{k_S} dq q P_s(q)
\int_{0}^1 d\mu \, \mu (1-\mu^2) \sin(\mu q/k_{osc})\,,\\
s_3=& 2\pi\int_0^{k_S} dq\, q^2 P_s(q) 
\int_0^1 d\mu\,\mu^2(1-\mu^2)\cos({q\mu}{k_{osc}})\;,
\end{align} 
\eseq
and $\Sigma_a^2$ has been defined in (\ref{eq:Siga}). In the region of
BAO oscillations the enveloping function is well approximated by the
form (cf. Eq.~(\ref{pwfit}))
\be
f_{env}(k)\propto k^{-m} e^{-(k/k_{Silk})^2}
\ee
with $m\approx 3/2$. Finally, combining (\ref{Pw22-2}) with the
contribution of $P_{w,13}$ and the last term in (\ref{eq:NLOfinal}) we
obtain Eqs.~(\ref{eq:Psnlo}), (\ref{eq:Szk}) with
\be
\label{eq:finsub}
s=(1+m/2)s_1+s_2+s_3~,~~~~~
\Sigma_{Silk}^2=s_1/k^2_{Silk}\;.
\ee

\section{IR resummation of power spectra in SPT}\label{app:SPT_TSPT}

In this appendix we derive the IR resummed result for
the power spectrum within SPT for an EdS background
cosmology. The derivation is less intuitive than in TSPT,
for example there is no simple diagrammatic interpretation of the
terms that need to be resummed, but it serves as a cross check.
We follow the same strategy as in TSPT, \emph{i.e.} introduce a separation scale
$k_S$ in order to treat IR and UV contributions to loop integrals separately.
We derive first the leading IR resummed result, as well as the subleading (NLO) terms in the
expansion in $\varepsilon\sim\langle q\rangle/k\lesssim k_S/k$. 
Then we discuss subleading terms in $\sigma_{h}^2$,
\emph{i.e.} taking hard loop corrections into account.

\hiddensubsection{Power spectrum in SPT}

The matter density power spectrum in SPT, for an EdS background, is given by \cite{Blas:2013aba}
\be
  P(\eta;k) = \sum_{l,r\geq 0}\sum_{m\geq 1} P_{(l,r,m)}(\eta;k)
\ee
where 
\bea
\label{eq:Plrm}
P_{(l,r,m)} (\eta;k) &=& \frac{(2l+m)! (2r+m)!}{2^{(l+r)} m! l! r!} g^{2(l+r+m)}(\e) \int \, [dQ] \, \nn \\
&& F_{2l+m}(\k_1,\dots, \k_m, \q_1, -\q_1, \dots, \q_l, - \q_l) \, \nn \\
&& F_{2r+m}(\k_1,\dots, \k_m, \p_1, -\p_1, \dots, \p_r, - \p_r) \, , \, 
\eea
where $F_n$ are the standard symmetrized SPT kernels \cite{Bernardeau:2001qr}. The velocity divergence power spectrum
can be obtained from the same expression by replacing $F_n\to G_n$,
the velocity kernels, and the cross spectrum by doing this  
replacement only in one of the two factors.
The integration measure is given by 
\bea
\label{eq:measure}
\int [dQ] &=& \left( \prod_{i=1}^{l+r+m} \int  \, [dQ_i] \, \bar P(Q_i) \right) \, \delta^{(3)}\left(\k - \sum_{i=1}^m \k_i\right) \, .
\eea
Here $Q_i$ run over $k_1 \dots k_m$, $q_1 \dots q_l$ and $p_1 \dots p_r$.
The summation over $l,r,m$ corresponds to the sum over all perturbative contributions.
The `loop' order of each term is given by $L=l+r+m-1$. Within the conventional notation, $P_{n_1n_2}$ denotes the sum over all $P_{(l,r,m)}$
with $n_1=2l+m$ and $n_2=2r+m$. For example, $P_{1-loop}=P_{22}+2P_{13}$ with $P_{22}=P_{(0,0,2)}$, $P_{13}=P_{(0,1,1)}$ and
$P_{2-loop}=P_{33}+2P_{24}+2P_{15}$ with $P_{33}=P_{(0,0,3)}+P_{(1,1,1)}$, $P_{24}=P_{(0,1,2)}$, $P_{15}=P_{(0,2,1)}$. The advantage of the
three-index notation is that each summand corresponds to a \emph{single} `diagram' with two vertices, $m$ lines connecting them, $l$
lines starting and ending at the left vertex, and $r$ at the right one. Note that this diagrammatic representation is very different
from the one arising in TSPT, although the full result summed over all diagrams must agree at a given loop order \cite{Blas:2015qsi}. 
Therefore, we also expect the IR resummed expressions to agree, when applying the same
power counting rules. This will be checked in what follows.

\hiddensubsection{Contribution from long scales}
\label{sec:longsc}

We are interested in obtaining an approximate result for the contribution from
long wavelength modes $q\ll k_S$ to the matter power spectrum at scale $k\gtrsim k_S$ (for some IR scale $k_S$, \emph{e.g.} slightly below the BAO scale).
To extract this contribution, we first rewrite the integrand of $P_{(l,r,m)}$ in  (\ref{eq:Plrm}) in the IR-safe form  described in \cite{Blas:2013aba}:
Due to the symmetry of the integrand w.r.t. permutations of $k_i$, we can restrict the
integration to the region where $|\k_i|\leq|\k_m|$ (for $i=1,\dots,m-1$), and compensate by multiplying by a factor $m$.
Then we use the delta function to integrate over $k_m$. This gives the equivalent expression\footnote{We use the notation $\t(x)$ for the Heaviside theta-function.}
\bea
\label{eq:Plrm1}
P_{(l,r,m)} (k,\eta) &=& m \frac{(2l+m)! (2r+m)!}{2^{(l+r)} m! l! r!} 
g^{2(l+r+m)}(\e)\int \, [d'Q] \, \nn \\
&& \bar P(|\k - \k_1 - \dots - \k_{m-1}|)\theta(|\k_m|-|\k_1|)\cdots \theta(|\k_m|-|\k_{m-1}|) \nn\\
&& F_{2l+m} (\k_1,\dots, \k_m, \q_1, -\q_1, \dots, \q_l, - \q_l) \, \nn \\
&& F_{2r+m} (\k_1,\dots, \k_m, \p_1, -\p_1, \dots, \p_r, - \p_r)\Bigg|_{\k_m = \k - \k_1 - \dots - \k_{m-1}} \, , \, 
\eea
where
\bea
\label{eq:measuren}
\int [d'Q] &=& \left( \prod_{i=1}^{l+r+m-1} \int  \, [dQ_i ]\, \bar P(Q_i) \right) \,,
\eea
and the $Q_i$ now denote the momenta $k_1,\dots,k_{m-1},p_1,\dots,p_r,q_1,\dots,q_l$.
Since we are interested in the contribution from long modes
we split the integration,
\be
  P_{(l,r,m)} (k,\eta)= P^{IR}_{(l,r,m)} (\eta;k,k_S)+P^{rest}_{(l,r,m)} (\eta;k,k_S)
\ee
where in the first term the loop momenta satisfy $|{\bf Q}_i|\leq k_S$, and the second one contains the integration over the remaining
parts (\emph{i.e.} at least of order $\sigma^2_{h}$ or higher in our power counting). For $P^{IR}$, in the limit $k\gg k_S$, we can  drop
the $\theta$ functions in (\ref{eq:Plrm1}), and write the power-spectrum as
\be
  \bar P(|\k - \k_1 - \dots - \k_{m-1}|) 
  = e^{-\sum_{i=1}^{m-1} \k_i \cdot \nabla_k} \bar P(k)\;.
\ee
This gives
\bea
\label{eq:Plrm2}
P^{IR}_{(l,r,m)} (\eta;k,k_S) &= & m \frac{(2l+m)! (2l+m)!}{2^{(l+r)} m! l! r!} g^{2(l+r+m)}(\e) \int_{|{\bf Q_i}|<k_S} \, [d'Q] \, \nn \\
&& \Big[ F_{2l+m} (\k_1,\dots, \k_m, \q_1, -\q_1, \dots, \q_l, - \q_l) \, \nn \\
&& F_{2r+m} (\k_1,\dots, \k_m, \p_1, -\p_1, \dots, \p_r, - \p_r) \Big]_{\k_m = \k - \k_1 - \dots - \k_{m-1}}\nn\\ 
&& e^{-\sum_{i=1}^{m-1} \k_i \cdot \nabla_k} \bar P(k) \,.
\eea

The expression (\ref{eq:Plrm2}) still contains all orders in an expansion in the scale of the loop wavenumbers $Q_i$ (bounded by $k_S$ by our definition)
and the external wavenmuber $k$. To obtain IR resummed expressions as an expansion in $\varepsilon$ we need to expand the SPT kernels in this parameter. Using the recursion relations for the SPT kernels (see \emph{e.g.} \cite{Bernardeau:2001qr})
one obtains the following results:
\be\label{eq:Fnexpansion}
  {E}_n(\k-\sum_{i=1}^{n-1}\q_i,\q_1,\dots,\q_{n-1}) = \frac{1}{n!}\frac{(\k\cdot \q_1)}{q_1^2}\cdots\frac{(\k\cdot \q_{n-1})}{q_{n-1}^2}\left(1 + {\cal K}^a + {\cal K}^b + {\cal K}^d + {O}(q_i^2/k^2)\right),
\ee
where ${E}=(F, G)$ can stand either for density or velocity divergence kernels. The leading term is given by the prefactor, and it is identical for density and velocity divergence kernels. Furthermore it is also identical for ED and ZA.

The terms in the bracket denote subleading terms at ${ O}(q_i/k)$, 
\bea\label{eq:FnSub}
{\cal K}^a &=& \sum_{i=1}^{n-1}\frac{(\k\cdot \q_i)}{k^2}\,, \quad \quad
{\cal K}^b =\Upsilon \sum_{i=1}^{n-1}\frac{q_i^2}{(\k\cdot \q_i)} \sin^2(\k,\q_i)\,, \nn\\
 {\cal K}^d &=& \varkappa\cdot \frac{3}{14}\sum_{i,j=1}^{n-1}\frac{q_i^2}{(\k\cdot \q_i)}\frac{q_j^2}{(\k\cdot \q_j)}\frac{(\k\cdot(\q_i+\q_j))}{(\q_i+\q_j)^2} \sin^2(\q_i,\q_j) \,.
\eea
The values of $\Upsilon$ and $\varkappa$ for the density/velocity PS in
ED/ZA are given in Table~\ref{table}.
 Note that, together with the prefactor, ${E}_n$ is regular for $(\k\cdot \q_i)\to 0$
and also for $|\q_i+\q_j|\to 0$. 

\begin{table*}[t]
\centering
  \begin{tabular}{|lccccc|}
     \hline
    $ $      & ED $P_{\delta \delta}$& ED $P_{\T \T}$ & ZA $P_{\delta \delta}$& ZA $P_{\T \T}$&  \\ \hline 
   $\Upsilon$  & $\frac{3}{7}$& $-\frac{1}{7}$ & $0$ &$-1$ &  \\ \hline
   $\varkappa$  & $1$ & $1$ & $0$ & $0$ &   \\ \hline
   \end{tabular}
\caption{The values of $\Upsilon, \varkappa$  for different power
  spectra in ED and ZA.} 
\label{table}
\end{table*}

\hiddensubsection{Cancellation of IR divergences for a smooth power spectrum}

Before discussing IR resummation for a wiggly power spectrum, we briefly review the case of a smooth spectrum
(\emph{cf.} \emph{e.g.} \cite{Jain:1995kx, Scoccimarro:1995if, Peloso:2013zw, Kehagias:2013yd, Blas:2013bpa, Sugiyama:2013gza}).
Operationally, this means we count derivatives acting on the power spectrum as $\nabla = { O}(1/k)$, or more
precisely 
\be
\q_i\cdot \nabla \bar P_s(k) = {O}(q_i/k) \bar P_s(k) \qquad \mbox{(smooth spectrum)}\;.
\ee
In this case we expect that all IR-enhanced contributions (IR `divergences') cancel. 
Note that in SPT the cancellation of
IR `divergences' only takes place when summing over all contributions at a fixed loop order $L=l+r+m-1$.
This can be checked explicitly by inserting the expansion of the kernels Eq.\,(\ref{eq:Fnexpansion}) into Eq.\,(\ref{eq:Plrm2}).

Let us first discuss the leading-order case, in which the computation is equivalent to \cite{Jain:1995kx},
and can also be considered as a special case of the general result discussed in \cite{Blas:2013bpa}.
At this order we may replace the kernels in Eq.\,(\ref{eq:Plrm2}) by the leading-order limit, \emph{i.e.} keeping
only the prefactor in Eq.\,(\ref{eq:Fnexpansion}), and, according to the
power counting for a smooth spectrum, omit the derivative operators acting on $\bar P(k)$. Then, summing over all
terms that contribute at $L$-loop order, one obtains ($\delta_{I,J}$ is the Kronecker delta)
\bea
\label{eq:cancel}
  P_{L-loop} &=& \sum_{l,r\geq 0}\sum_{m\geq 1} \delta_{l+r+m,L+1} P_{(l,r,m)}(\eta;k) \nn\\
  & \to & g^{2(L+1)}(\e) \bar P(k) \sum_{l,r\geq 0}\sum_{m\geq 1} \delta_{l+r+m,L+1} \frac{(k^2\sigma^2_d)^{m-1}}{(m-1)!}\frac{(-k^2\sigma^2_d)^{l}}{2^l l!}\frac{(-k^2\sigma^2_d)^{r}}{2^r r!} \nn\\
 & = & g^{2(L+1)}(\e) \bar P(k) \left( k^2\sigma^2_d - \frac{k^2\sigma^2_d}{2} - \frac{k^2\sigma^2_d}{2}\right)^{L-1} = 0\;,
\eea
\emph{i.e.} the leading IR `divergences' cancel at each loop order, as expected\footnote{Formally,
the result \eqref{eq:cancel} is valid for 
$L\geq 1$. 
In the degenerate case $L=0$, \emph{i.e.} at the linear order, one gets $P_{0-loop}=g^2(\e)\bar P$.}. Note that here we defined
\be
  k^2 \sigma^2_d \equiv \int_{|\q|<k_S} [dq] \, \frac{(\k\cdot \q)^2}{q^4} \bar P(q) \;.
\ee

Let us now look at the first sub-leading terms. For the power counting rule for
a smooth spectrum, there are two sources for subleading terms in Eq.\,(\ref{eq:Plrm2}): (i) contributions where one derivative acts on the power spectrum,
and (ii)  the contribution from the subleading terms in the expansion of the SPT kernels in Eq.\,(\ref{eq:FnSub}). Both vanish identically because
the integrand is odd under ${\bf Q}_i\to -{\bf Q}_i$. Non-trivial contributions arise for the first time at NNLO in $\varepsilon$, which is beyond the order we
are interested in here (this conclusion will change when looking at the wiggly part of the spectrum). Nevertheless, note that the cancellation of \emph{all} sub-leading IR `divergences' has been demonstrated analytically
up to 2-loop (and numerically up to 4-loop) in \cite{Blas:2013bpa}, along with arguments that support a cancellation of all sub-leading IR-enhanced terms at
any loop order. Similar conclusions have also been reached in \cite{Scoccimarro:1995if, Peloso:2013zw, Kehagias:2013yd} based on symmetry arguments.

\hiddensubsection{Leading IR resummed result for the wiggly spectrum}

We now consider the case where the power spectrum can be decomposed in a smooth
and a wiggly contribution, $\bar P=\bar P_s+\bar P_w$ (\emph{cf.} \eqref{wigsm}). For the purpose of the discussion here
the essential property of the wiggly part is that the power counting of derivatives
is different,
\be
\q_i\cdot \nabla \bar P_w(k) = {O}(q_i/k_{osc}) \bar P_w(k) \qquad \mbox{(wiggly spectrum)}\;,
\ee
where $k_{osc}$ denotes the typical wavenumber of the oscillations imprinted in
the wiggly component. We directly consider the general case where we assume no hierarchy between $k_S$ and $k_{osc}$, \emph{i.e.}
 $k_S/k\gtrsim k_{osc}/k\sim \varepsilon$.

The wiggly power spectrum at leading order in $\sigma_{h}^2$  (recall \eqref{eq:sigmah})
is obtained by replacing $\bar P(k) \to \bar P_w(k)$
in the last line of Eq.\,(\ref{eq:Plrm2}), \emph{i.e.} keeping only
terms where the derivative operators act on the wiggly part, 
\be\label{eq:PIRw}
  P^{IR}_w(\eta;k) = \sum_{l,r\geq 0}\sum_{m\geq 1} P^{IR}_{(l,r,m)}(\eta;k)\big|_{\bar P(k)\to \bar P_w(k)}\,.
\ee

The leading order in $\varepsilon$ is obtained by inserting  the prefactor in Eq.\,(\ref{eq:Fnexpansion}), into Eq.\,(\ref{eq:PIRw}) and
using Eq.\,(\ref{eq:Plrm2}) we get,
\bea\label{eq:PIRwleading}
P^{IR\;res,LO}_w(\e;k) &=& \sum_{l,r\geq 0}\sum_{m\geq 1} \frac{g^{2(l+r+m)}(\e) }{2^{(l+r)} (m-1)! l! r!} \prod_{i=1}^{m-1} \Bigg[ \int_{|\k_i|<k_S} \, [dk_i] \, \frac{(\k\cdot \k_i)^2}{k_i^4} \bar P_s(k_i) e^{-\k_i\cdot \nabla_{k'}}\Bigg]\nn \\
&& {} \times \prod_{j=1}^{l} \Bigg[ - \int_{|\q_j|<k_S} \, [dq_j] \, \frac{(\k\cdot \q_j)^2}{q_j^4} \bar P_s(q_j) \Bigg] 
\nn\\
&& {} 
\times \prod_{j=1}^{r} \Bigg[ - \int_{|\p_j|<k_S} \, [dp_j] \, \frac{(\k\cdot {\mathbf p}_j)^2}{p_j^4}\bar P_s(p_j) \Bigg] \bar P_w(k')\Bigg|_{k'=k} \nn\\
&=& \exp\left\{ \int_{|\q|<k_S} \, [dq] \, \left(\frac{\k\cdot \q}{q^2}\right)^2 g^2\bar P_s(q) \, e^{-\q\cdot \nabla_{k'}} \right\} 
\nn\\
&& {} 
\times \exp\left\{ - \int_{|q|<k_S} \, [dq] \, \left(\frac{\k\cdot \q}{q^2}\right)^2 g^2\bar P_s(q) \right\}
 g^2\bar P_w(k')\Bigg|_{\k'=\k} \nn\\
&=& \exp(- g^2\S)  P_w(\eta;k')|_{\k'=\k} \,,
\eea
which coincides with the result \eqref{Irressim}. As discussed before, this also coincides with the result of \cite{Baldauf:2015xfa} when assuming that the derivative
operators can be evaluated using $e^{-\q\cdot \nabla_{k'}} \to e^{-\q\cdot \hat k/k_{osc}}$, where $\hat k=\k/|k|$.
On the other hand, one can formally also recover the result for the smooth spectrum, by replacing $e^{-\q\cdot \nabla_{k'}} \to 1$, which confirms that all the IR-enhanced terms cancel in this case.
Finally, we stress again that the leading IR-resummed power spectrum is identical for the density and velocity fields both in ZA and ED.
This follows from the fact that the leading term in the expansion 
Eq.\,(\ref{eq:Fnexpansion}) of the kernels is identical for ZA and ED, as well as for density and velocity
components.

\hiddensubsection{NLO IR resummed results for the wiggly spectrum}

The first sub-leading correction to Eq.\,(\ref{eq:PIRwleading}) with
respect to the expansion in powers of $\varepsilon$ 
can be obtained by inserting the first sub-leading corrections  in
Eq.\,(\ref{eq:Fnexpansion}) into Eq.\,(\ref{eq:PIRw}), 
using  Eq.\,(\ref{eq:Plrm2}) and Eq.\,(\ref{eq:FnSub}). 
After somewhat lengthy but
straightforward calculation one finds 
\be
P^{IR\;res,LO+NLO_s}_w(\e;k) = \exp(- g^2\S)  \{ 1 + \delta_a + \delta_b + \delta_d \} P_w(\eta;k')|_{\k'=\k} \,,
\ee
where the contributions that arise from the sub-leading terms are given by
\bea
 \delta_a &=& - 2 \int_{|\q|<k_S} [dq]  P_s(\eta;q) \frac{\k\cdot \q}{q^2} \cos^2(\k,\q) \sinh(\q\cdot \nabla_{k'}) \,,\nn\\
 \delta_b &=& - 2\Upsilon \int_{|\q|<k_S} [dq]  P_s(\eta;q) \frac{\k\cdot \q}{q^2} \sin^2(\k,\q) \sinh(\q\cdot \nabla_{k'})\,, \nn\\
 \delta_d &=& - \varkappa \times \frac67 \int_{|\q|<k_S} [dq]  P_s(\eta;q) \int_{|\p|<k_S} [dp]  P_s(\eta;p) \frac{\k\cdot \q}{q^2}\frac{\k\cdot \p}{p^2} \nn\\
 && \times \frac{\k\cdot(\p+\q)}{(\p+\q)^2} \sin^2(\p,\q) \sinh(\q\cdot \nabla_{k'}) \left( \cosh(\p\cdot \nabla_{k'})-1\right)\,.
\eea
This result matches precisely the soft NLO
terms\footnote{To evaluate the soft part of the loop entering in
  \eqref{eq:NLOfinal}, one can use the standard SPT expression and
  expand the corresponding kernels using once more
  Eq.~(\ref{eq:Fnexpansion}).}
 in  
\eqref{eq:NLOfinal}. 

Let us now discuss the corrections in the hard expansion parameter $\sigma_{h}$ .
We separate the integration over wavenumbers in (\ref{eq:Plrm}) into a hard domain $Q_i>k_S$
and a soft domain $Q_i<k_S$. For the sets of wavenumbers labeled by $l$, we split $l=l_s+l_h$
where $l_h$ labels the number of hard momenta, 
and $l_s$ labels the  number of soft ones. 
Similarly, we split
$r=r_s+r_h$ and $m=m_s+m_h$. The case in which all loop momenta are soft, that has been discussed before,
corresponds to the case $l_h=r_h=0$ and $m_h=1$. Note that $m_h=1$ because the external wavenumber $k$
is also considered as hard, such that at least one of the wavenumbers $k_1,\dots,k_m$ has to be hard.
Next, there are three possibilities that lead to a contribution with one hard loop, given by
$(l_h,r_h,m_h)=(1,0,1), (0,1,1), (0,0,2)$, and $l_s\geq 0$, $r_s\geq 0$, $m_s\geq 0$. 

Since the integrand is symmetric in permutations among the momenta $q_i$, we can select that the hard momenta
have indices $q_1,\dots, q_{l_h}$, and the soft ones $q_{l_h+1},\dots,q_l$.
The number of possibilities to select the hard momenta is $l!/l_h!/l_s!$,
and we have to multiply the integrand by this factor to account for the fact that we have chosen one
particular possibility to select the hard momenta.
A similar rearrangement can be done for the momenta $p_i$ and $k_i$, respectively.
The corresponding contribution to the power spectrum with $(l_h,r_h,m_h)$ hard wavenumbers
is given by
\be
\begin{split}
\label{eq:Plrmhard}
P_{(l,r,m)} (k,\eta)&|_{l_h,r_h,m_h\ hard} = \frac{(2l+m)! (2r+m)!}{2^{(l+r)} m! l! r!} g^{2(l+r+m)}(\e) \int \, [d'Q_h] \int \, [d'Q_s]\,  \\
& \frac{l!}{l_h!l_s!}\frac{r!}{r_h!r_s!}\frac{m!}{m_h!m_s!} \,  
F_{2l+m} (\k_1,\dots, \k_m, \q_1, -\q_1, \dots, \q_l, - \q_l) \, \\
& F_{2r+m} (\k_1,\dots, \k_m, \p_1, -\p_1, \dots, \p_r, - \p_r) \,\bar P(k_{m_h}) \bigg|_{\k_{m_h}=\k-\sum\limits_{i\not=m_h}^m \k_i} \,,
\end{split}
\ee
where we defined
\bea
\label{eq:measurehs}
\int [d'Q_h] &=& \left( \prod_{i=1}^{l_h+r_h+m_h-1} \int_{|Q^h_i|>k_S}  \, d^3Q^h_i \, \bar P(Q_i^h) \right) \,, \nn\\
\int [d'Q_s] &=& \left( \prod_{i=1}^{l_s+r_s+m_s} \int_{|Q^s_i|<k_S}  \, d^3Q^s_i \, \bar P(Q_i^s) \right)  \,.
\eea
Here $Q_i^{h}$ labels the hard momenta $k_1,\dots,k_{m_h-1},q_1,\dots,q_{l_h},p_1,\dots,p_{r_h}$ and
$Q_i^s$ the soft momenta $k_{m_h+1},\cdots,k_m,p_{l_h+1},\cdots,q_l,p_{r_h+1},\dots,p_r$. Furthermore, we used the momentum conservation delta function
to eliminate the integration over one of the hard momenta, chosen to be $k_{m_h}$. This is consistent
because we treat the external momentum as a hard scale compared to the IR cutoff $k_S$. For $l_h=r_h=0$
and $m_h=1$ it coincides with (\ref{eq:Plrm2}) up to a renaming of $k_1$ and $k_m$.
The measures are defined as in \eqref{eq:measuren}, for hard and soft modes. 

As a next step, we expand the SPT kernels in the soft momenta. For this it is helpful to use the general
relation \cite{Sugiyama:2013pwa}
\be
  F_n(\k_1,\dots,\k_n) = \frac{m!}{n!} F_{m}(\k_1,\dots,\k_m) \frac{(\k\cdot \k_{m+1})}{k_{m+1}^2}\cdots\frac{(\k\cdot \k_n)}{k_n^2}\big(1+{O}(k_{soft}/k_{hard})\big)\;,
\ee
in the limit where $k_{m+1},\dots,k_n$ are soft, \emph{i.e.} much smaller than the arguments $k_1,\dots,k_m$, and $\k\equiv \sum \k_i$. 
Inserting this expansion into (\ref{eq:Plrmhard}) and summing over the number of soft loops gives the IR resummed power spectrum at order $(l_h,r_h,m_h)$,
and at leading order in~$\varepsilon$,
\bea
\label{eq:PlrmIRhard2}
P_{(l_h,r_h,m_h)}^{IR} (\eta;k) &\equiv& \sum_{l_s\geq 0}\sum_{r_s\geq 0}\sum_{m_s\geq 0} P_{(l,r,m)}(\eta;k)|_{l_h,r_h,m_h\ hard,\ LO\ in\ \varepsilon} \nn\\
&&  \hspace{-2.5cm}=\sum_{l_s\geq 0}\sum_{r_s\geq 0}\sum_{m_s\geq 0}  \frac{(2l+m)! (2r+m)!}{2^{(l+r)} m! l! r!} g^{2(l+r+m)} \int \, [d'Q_h] \int \, [d'Q_s]\, \nn \\
&&\hspace{-2.5cm}\times \frac{l!}{l_h!l_s!}\frac{r!}{r_h!r_s!}\frac{m!}{m_h!m_s!} \, \frac{(2l_h+m_h)!}{(2l+m)!}  
F_{2l_h+m_h} (\k_1,\dots, \k_{m_h}, \q_1, -\q_1, \dots, \q_{l_h}, - \q_{l_h}) \, \nn \\
&&\hspace{-2.5cm} \frac{(2r_h+m_h)!}{(2r+m)!} F_{2r_h+m_h} (\k_1,\dots, \k_{m_h}, \p_1, -\p_1, \dots, \p_{r_h}, - \p_{r_h}) \, (-1)^{l_s+r_s}\\
&& \hspace{-2.5cm}\prod_{i=m_h+1}^m 
\frac{(\k\cdot \k_i)^2}{k_i^4}
\prod_{i=l_h+1}^l 
\frac{(\k\cdot \q_i)^2}{q_i^4} 
\prod_{i=r_h+1}^r 
\frac{(\k\cdot \p_i)^2}{p_i^4}
\bar P(k_{m_h}) \bigg|_{\k_{m_h}=\k-\sum\limits_{i\not=m_h}^m \k_i} \,. \nn
\eea
 The power spectrum $\bar P(k_{m_h})$
can be rewritten as 
\be
  \bar P(k_{m_h}) \bigg|_{\k_{m_h}=\k-\sum\limits_{i\not=m_h}^m \k_i} = e^{-\sum_{i=m_h+1}^m \k_i \cdot \nabla_{k'}} \bar P(|\k'-\sum_{i=1}^{m_h-1}\k_i|)\Big|_{\k'=\k}\;,
\ee
where the soft momenta appear in the derivative operator, and the hard ones remain in the argument of the power spectrum. Furthermore, at leading order
in $\ve$ one can replace $\k_{m_h}\to \k-\sum_{i=1}^{m_h-1}\k_i$ inside the SPT kernels $F_{2l_h+m_h}$ and $F_{2r_r+m_h}$. The soft loops can now
be resummed into exponential derivative operators similar to the case without hard loops, and one obtains
\bea
\label{eq:PlrmIRhard3}
P_{(l_h,r_h,m_h)}^{IR} (\eta;k) &=& \exp\left\{ \int_{|\q|\leq k_S} \, [dq] \, \frac{(\k\cdot \q)^2}{q^4} g^2\P_s(q) \, \left( e^{-\q\cdot \nabla_{k'}} -1\right)\right\} \nn\\
&& \frac{(2l_h+m_h)! (2r_h+m_h)!}{2^{(l_h+r_h)} m_h! l_h! r_h!} 
g^{2(l_h+r_h+m_h)}(\e)
 \int \,[ d'Q_h]\,  \P(|\k'-\sum_{i=1}^{m_h-1}\k_i|) \nn \\
&& F_{2l_h+m_h} (\k_1,\dots, \k_{m_h}, \q_1, -\q_1, \dots, \q_{l_h}, - \q_{l_h}) \, \nn \\
&& F_{2r_h+m_h} (\k_1,\dots, \k_{m_h}, \p_1, -\p_1, \dots, \p_{r_h}, - \p_{r_h}) \Big|_{\k_{m_h}=\k-\sum\limits_{i=1}^{m_h-1} \k_i} \Bigg|_{\k'=\k} \,. \nn\\ 
\eea
The derivative operator is identical to the one encountered when no hard loops are taken into account. 
Furthermore, the remaining hard loops have
a structure similar to the SPT loop contribution at order $(l_h,r_h,m_h)$ to the power spectrum. It is important to keep in mind that only terms where all the derivatives
$\nabla_{k'}$ act on the wiggly part of the power spectrum contribute to the leading order in $\varepsilon$. This may be used to rewrite the previous result in two different ways, that we discuss below: 

\begin{enumerate}

\item One can use that terms in which the derivatives act on the SPT kernels or on the smooth part of the spectrum give a
contribution that is suppressed compared to the LO in $\varepsilon$, and can therefore be added (or subtracted) without changing the LO. Using this freedom,
one can formally replace $\k_{m_h}=\k-\sum_i \k_i$  by $\k_{m_h}=\k'-\sum_i \k_i$ in (\ref{eq:PlrmIRhard3}). Then the
corresponding expression on the right-hand side of (\ref{eq:PlrmIRhard3}) is precisely identical to the SPT loop contribution
to the power spectrum evaluated with all loop momenta above the IR-cutoff $k_S$,
\be
\label{eq:PlrmIRhard4}
P_{(l_h,r_h,m_h)}^{IR} (\eta;k) =  \exp\left\{ \int_{|\q|\leq k_S} \, [dq] \, \frac{(\k\cdot \q)^2}{q^4} g^2\P_s(q) \, \left( e^{-\q\cdot \nabla_{k'}} -1\right)\right\} 
 P_{(l_h,r_h,m_h)}^{|{\bf Q}_i|>k_S}(\eta;k')  \Bigg|_{\k'=\k} \,.  
\ee
Here the right-hand side is to be computed like the usual SPT loop contributions with the lower cutoff $|{\bf Q}_i|\geq {k_S}$ for all loop momenta. Since it is valid for
all loop contributions, it also holds \emph{e.g.} for the 1-loop,
\be
\label{eq:PlrmIRhard5}
P_{1-loop}^{IR} (\eta;k) =\exp\left\{ \int_{|\q|\leq k_S} \, [dq] \, \frac{(\k\cdot \q)^2}{q^4} g^2\P_s(q) \, \left( e^{-\q\cdot \nabla_{k'}} -1\right)\right\} 
 P_{1-loop}^{|{\bf Q}_i|>{k_S}}(\eta;k')  \Bigg|_{\k'=\k} \,,  
\ee
and analogously for higher loops. We stress again that this expression is valid at LO in $\varepsilon$. Furthermore, the LO in the number of hard loops is recovered from
the above formula by $l_h=r_h=0$, $m_h=1$, which corresponds to evaluating (\ref{eq:PlrmIRhard5}) with the linear instead of the one-loop spectrum. A disadvantage of
this representation is that it is hard to evaluate the derivative operator numerically.

\item An alternative way to rewrite (\ref{eq:PlrmIRhard3}) is to insert the decomposition $\bar P=\P_s+\P_w$ of the initial power spectrum in smooth and wiggly
components. 
Then, at LO in $\varepsilon$, the derivative operator is different from unity only when it acts on the wiggly part of the
spectrum. For example, this is the case when the derivative operators act on the wiggly part of $\P(|\k'-\sum_i \k_i|)$ in  (\ref{eq:PlrmIRhard3}).
From the structure of the measure, one can see that there is also a contribution from the derivative acting on loop power spectra $\P(k_i)$ (with $1\leq i\leq m_h-1$) that arises
due to integration by parts.
Indeed,
\bea
   \nabla_{k'} \int [dk_i] \P(k_i) \P\left(|\k'-\sum \k_j|\right) &=& \int [dk_i] \bigg( \P_s(k_i) \nabla_{k'} \P_w(k'-\sum_i k_i) \nn\\
  && \hspace{-3.5cm} + (\nabla_{k_i} \P_w(k_i)) \P_s\left(|\k'-\sum \k_j|\right) \bigg)\left(1+{O}\left(\frac{k_{osc}}{k_{hard}}\right)\right)\,, 
\eea
where $k_{hard}$ is of order of $k_i$ or $k'$. Due to the symmetry of the loop integrand, this kind of contributions add up a factor $m_h$
to the formula \eqref{eq:PlrmIRhard3} that is,
the eventual contribution to the power spectrum
that is linearly expanded in the wiggly part $P_w$ is given by
\be
\begin{split}
\label{eq:PlrmIRhard6}
P_{(l_h,r_h,m_h)}^{IR, w} &(\eta;k) = \frac{(2l_h+m_h)! (2r_h+m_h)!}{2^{(l_h+r_h)} m_h! l_h! r_h!} g^{2(l_h+r_h+m_h)} \, m_h \int \, [dQ_h^s]\, \\
&  \tilde P_w(\e;|\k-\sum_{i=1}^{m_h-1}\k_i|\Big|k,{k_S})  \, F_{2l_h+m_h} (\k_1,\dots, \k_{m_h}, \q_1, -\q_1, \dots, \q_{l_h}, - \q_{l_h}) \,\\
& F_{2r_h+m_h} (\k_1,\dots, \k_{m_h}, \p_1, -\p_1, \dots, \p_{r_h}, - \p_{r_h}) \Big|_{\k_{m_h}=\k-\sum\limits_{i=1}^{m_h-1} \k_i} \,. 
\end{split}
\ee
where $[dQ_h^s]$ is defined as (\ref{eq:measurehs}), but involving only the smooth parts of the power spectra.
Furthermore, we defined the 
`generalized' IR resummed wiggly PS as follows,
\begin{align}
\tilde P_w(\e;p|k,{k_S})  &\equiv  \exp\left\{ \int_{|\q|<{k_S}} \, [dq] \, \frac{(\k\cdot \q)^2}{q^4} P_s(\eta;q) \, \left( e^{-q\cdot \nabla_{p'}} -1\right)\right\}  P_w(\e;p') \Bigg|_{p'=p} \nn\\
\simeq  \exp&\left\{- 4\pi k^2 \int_0^{k_S} dq \, \P_{s}(\eta;q) \left(f_1(q/k_{osc})+\cos^2({\k,{\bf p}}) f_2(q/k_{osc}) \right) \right\} P_w(\e;p)\,,\nn\\
\end{align}
where in the last line we have evaluated the derivative operator at LO in $\ve$
assuming $\nabla P_w(p)=\hat {\bf p}/k_{osc} P_w(p)$, with $\hat {\bf p}={\bf p}/|\p|$ and
\bea
  f_1(x) &\equiv& \frac13 - \frac12 (1+\partial_x^2) \frac{\sin(x)}{x}\,,\\
  f_2(x) &\equiv& \frac12 (1+3\partial_x^2) \frac{\sin(x)}{x}\,.
\eea
The spectrum $\tilde P_w$ reduces to the LO 
IR resummed wiggly power spectrum for $p=k$, $\tilde P_w(\e;k|k,{k_S})=e^{-g^2\S}P_w(\e;k)$. The
explicit expression for $\tilde P_w$ is suitable to use (\ref{eq:PlrmIRhard6})
for computing the IR resummed power spectrum including hard loops. For example, the IR resummed
one-loop spectrum reads
\bea
\label{eq:PlrmIRhard1loop}
  P_{1-loop}^{IR,w} (\eta;k) &=& 4 \int_{|\p|>{k_S}} [dp] (F_2(\p,\k-\p))^2 P_s(\e;\p) \tilde P_w(\e;|\k-\p|\big|k,{k_S})\nn\\
 && + 6 \tilde P_w(\e;k|k,{k_S}) \int_{|\p|>{k_S}} [d p] F_3(\k,\p,-\p) P_s(\e;p)\,.
\eea
\end{enumerate}

\hiddensubsection{Comparison with TSPT}\label{sec:compare}

We have already seen that the LO IR resummed power spectrum as well
as the NLO corrections in the parameter $\varepsilon$ obtained
in SPT and TSPT are identical. 
This is a valuable
cross check, since the rather involved derivation within the well-established SPT framework
supports the TSPT result for the matter power spectrum.
Let us now discuss the NLO including the first hard corrections (\ref{eq:PlrmIRhard1loop}) and \eqref{eq:nh=1}. 
The SPT contribution at NLO in $\sigma^2_{h}$ is given by
(\ref{eq:PlrmIRhard1loop}). 
This contribution does not match the TSPT result precisely. 
Nevertheless, we will see that the difference between SPT and
TSPT is beyond our power counting, so one can eliminate it 
without changing the order of precision.
The difference can be expressed as
\be
\begin{split}
 & \delta P_{1-loop}^{IR, w} (\eta;k) \\
  &= 4\!\!\!\int\limits_{|\p|>{k_S}}\!\!\! d^3 p 
  (F_2(\p,\k-\p))^2 
  P_s(\e;p) \Big(\tilde P_w(\e;|\k-\p|\Big|k,{k_S})-\tilde P_w(\e;|\k-\p|\Big||\k-\p|,{k_S})\Big)\\
 & \quad- 6 P_s(\e;k) \int d^3 p F_3(\k,\p,-\p) \tilde P_w(\e;p|p,{k_S})\,.
 \end{split}
\ee 
Since the wiggly part appears under integrals in both terms, the difference is 
extremely suppressed due cancellations occurring for
an oscillating integrand, as was argued in Sec.\,\ref{sec:hard}.
In fact, we neglected such contributions also in the TSPT derivation.
Therefore, in this sense, we see that despite substantially different derivations,
the SPT and TSPT NLO IR-resummed power spectra agree.

\section{Cross check of IR resummation within Zel'dovich approximation}\label{app:ZA}

In this appendix we demonstrate how to obtain the IR resummed result for the
power spectrum within the Zel'dovich approximation. Since in ZA a closed-form
expression for the full density power spectrum is known, this derivation serves as
a cross check of the TSPT result, which is applicable also beyond ZA.
Furthermore, it is instructive to see how the leading and
subleading terms can be extracted from the full ZA result
within the power counting scheme pursued in this work.
We discuss in the following the leading IR resummed case, as well as
subleading terms in $\varepsilon$ and in $\sigma^2_{h}$ (see
Sec.~\ref{sec:PCsub} for definitions).

The full density power spectrum in ZA is given by
\be
\label{eq:za}
  P(\eta;k) = \int \frac{d^3z}{(2\pi)^3}\, e^{-i\k\cdot \z} \, e^{-\frac12 k_ik_jA_{ij}(z)} \,,
\ee
with
\be\label{eq:Adef}
  A_{ij}(z) = 4 \int d^3q \, \frac{q_iq_j}{q^4} P^{lin}(\eta;q) \sin^2\left(\frac{(\q\cdot \z)}{2}\right)\,.
\ee
First we expand the linear power spectrum $P^{lin}\equiv g^2(\eta)\P(k)$ in
smooth and a wiggly parts, $\P=\P_s+\P_w$, and expand   
Eq. \eqref{eq:za}
linearly
in $P_w$,
\be
  P(\eta;k)\Big|_{{O}(P_w)} = -\frac12 \int \frac{d^3z}{(2\pi)^3}\, e^{-i\k\cdot \z} \, e^{-\frac12 k_ik_jA^s_{ij}(z)} k_ik_jA_{ij}^w(z)\,,
\ee
where $A_{ij}=A_{ij}^s+A_{ij}^w$ inherits the smooth/wiggly
decomposition of the linear spectrum. 
Next we decompose the $q$-integration in an IR part, with
$|\q|<{k_S}$, and a UV part with 
$|\q|>{k_S}$, 
\be\label{eq:Asplit}
  A_{ij}^s=A_{ij}^{IR}+A_{ij}^{UV}=4 \left[ \int_{|\q|<{k_S}} d^3q \, + \int_{|\q|>{k_S}} d^3q \right] \frac{q_iq_j}{q^4} P_{s}(\eta;q) \sin^2\left(\frac{(\q\cdot \z)}{2}\right)\,.
\ee
We keep the IR contribution in the exponent, but expand the UV contribution perturbatively.
Parametrically, this generates an expansion in $\sigma^2_{h}$,
\be\label{eq:PwexpandZA}
  P(\eta;k)\Big|_{{O}(P_w)} = P^{IR}_w(\eta;k) 
  + P^{IR,NLO_h}_{w}(\eta;k) + \dots \,.
\ee
In addition, in each term we need to expand in powers of ${k_S}/k$, to
match the TSPT and SPT 
computations. 

\hiddensubsection{IR resummed power spectrum at LO in $\sigma^2_{h}$}

Let us first look at the LO term in $\sigma^2_{h}$, which is given by
\be
  P^{IR}_w(\eta;k)  =   -2 \int \frac{d^3z}{(2\pi)^3}\, e^{-i\k\cdot 
 \z} \, e^{-\frac12 k_ik_jA^{IR}_{ij}(z)} \int d^3q \, \frac{(\k\cdot \q)^2}{q^4} P_w(\eta;q)  \sin^2\left(\frac{(\q\cdot \z)}{2}\right) \,.
\ee
In order to bring it into a form that resembles the IR resummed expression derived in TSPT, we rewrite it in the following way
\bea
  P^{IR}_w(\eta;k) &=&   -2 e^{-\frac12 k_ik_jA^{IR}_{ij}(i\nabla_{k'})}  \int \frac{d^3z}{(2\pi)^3}\, e^{-i\k'\cdot \z} \, \int d^3q \, 
\frac{(\k\cdot \q)^2}{q^4}
   P_w(\eta;q)  \sin^2\left(\frac{(\q\cdot \z)}{2}\right)\Big|_{k'=k}\nn\\
&=& -2 e^{-\frac12 k_ik_jA^{IR}_{ij}(i\nabla_{k'})} \, \int d^3q \, \frac{(\k\cdot \q)^2}{q^4} P_w(\eta;q) \nn\\
&& {} \times \left(-\frac14\delta^{(3)}(\q-\k') - \frac14\delta^{(3)}(\q+\k') + \frac12 \delta^{(3)}(\k')\right)\Big|_{\k'=\k}\nn\\
&=& e^{-\frac12 k_ik_jA^{IR}_{ij}(i\nabla_{k'})} \, 
\frac{(\k\cdot \k')^2}{(k')^4}
P_w(\eta;k')\Big|_{\k'=\k}\,.
\eea
In the last step we used the symmetry of the integrand under $\q\to -\q$, and dropped the last delta function since we are
only interested in modes with $k>0$.
The exponential differential operator in front precisely agrees with the one introduced in Eq.\,(\ref{Slead}), so that we obtain
\be
  P^{IR}_w(\eta;k)  =  e^{-g^2\S} \, \frac{(\k\cdot \k')^2}{(k')^4}
   P_w(\eta;k')\Big|_{\k'=\k}\,.
\ee
Note that we did not yet perform the expansion in ${k_S}/k$, so that
this is the full expression for the contribution to 
the power spectrum from IR modes below ${k_S}$ within ZA. The expansion can be recovered by noticing that the $k'$-derivatives
contained in the operator $\S$ can act either on the wiggly power
spectrum $P_w(k',\eta)$ or on the factor $(\k\cdot \k')^2/(k')^4$. 
In the former case $\nabla_{k'}\to {O}(1/k_{osc})$ while in the
latter $\nabla_{k'}\to {O}(1/k)$. Therefore the leading 
result in $\ve$ is obtained when \emph{all} the derivatives act
exclusively on $P_w$, 
\be
  P^{IR,LO}_{w}(\eta;k)  =  e^{-g^2 \S}  P_w(\eta;k')\Big|_{\k'=\k}\,.
\ee
As expected, this agrees precisely with the result obtained in TSPT in
Eq.\,(\ref{Irressim}), and also with the perturbative SPT derivation 
leading to Eq.\,(\ref{eq:PIRwleading}).

The first sub-leading terms in  ${k_S}/k$ can be taken into account by
considering the possibility that \emph{at most one} 
derivative acts on  the factor $(\k\cdot \k')^2/(k')^4$. It is useful
to recall Euler's rule in the form, 
\be
  F(i\nabla_{k'})f(k')g(k') = [F(i\nabla_{k'})f(k')]g(k') + [F'(i\nabla_{k'})f(k')][i\nabla_{k'}]g(k') + {O}(\nabla^2 g)\,,
\ee
where $F'(z)\equiv dF/dz$. Applying this identity to 
$f=P_w$, $g(k')=(\k\cdot \k')^2/(k')^4$, and $F=e^{-g^2\S}$,
one finds that the sub-leading terms agree precisely with the NLO$_s$
terms contained\footnote{The simplest way to extract the NLO$_s$ terms
from (\ref{irres3}) is to use the standard SPT expression for the
1-loop power spectrum and expand the relevant kernels at soft loop momenta
(see Eqs.~(\ref{eq:Fnexpansion}), (\ref{eq:FnSub})).} 
in (\ref{irres3}) for $\varkappa=0$, as expected
for the density power spectrum in ZA.

\hiddensubsection{IR resummed power spectrum at NLO in $\sigma^2_{h}$}

To obtain the correction to the previous results at first order in $\sigma^2_{h}$
we insert (\ref{eq:Asplit}) into (\ref{eq:PwexpandZA}) and expand to linear
order in $A^{UV}$. This gives
\begin{equation}
 P_{w}^{IR,NLO_h}(\eta;k) = \frac14 \int \frac{d^3z}{(2\pi)^3}\, e^{-i\k\cdot \z} \, e^{-\frac12 k_ik_jA^{IR}_{ij}(z)} \left(k_mk_nA_{mn}^{UV}(z)\right)\left(k_pk_qA_{pq}^w(z)\right) \,.
\end{equation}
Inserting the explicit form for $A^{UV}$ and $A^w$ from (\ref{eq:Adef}), along with pulling the part containing $A^{IR}$ in front as before, gives
\bea
  P_{w}^{IR,NLO_h}(\eta;k) &=& 4  e^{-\frac12
    k_ik_jA^{IR}_{ij}(i\nabla_{k'})} \int \frac{d^3z}{(2\pi)^3}\,
  e^{-i\k'\cdot \z} \!\!\int\limits_{|\p|>{k_S}}\!\! 
d^3p \, \frac{(\k\cdot \p)^2}{p^4} P_{s}(\eta;p) \sin^2\left(\frac{(\p\cdot \z)}{2}\right) \nn\\
  && {} \times \int d^3q \, \frac{(\k\cdot \q)^2}{q^4} P_{w}(\eta;q) \sin^2\left(\frac{(\q\cdot \z)}{2}\right)\bigg|_{\k'=\k} \,.
\eea
The integration over $z$ can now be performed trivially, and yields a sum of delta functions. Using the symmetry of the integrand w.r.t sign flips in
$p$ and $q$, one obtains
\bea
  P_{w}^{IR,NLO_h}(\eta;k) &=& e^{-\frac12
    k_ik_jA^{IR}_{ij}(i\nabla_{k'})} 
\int_{|\p|>{k_S}} d^3p \, 
\frac{(\k\cdot \p)^2}{p^4}
  P_{s}(\eta;p) 
  \int d^3q \, \frac{(\k\cdot \q)^2}{q^4}
  P_{w}(\eta;q) \nn\\
  && {} \times \left(\delta^{(3)}(\p+\q-\k')-\delta^{(3)}(\p-\k')-\delta^{(3)}(\q-\k')+\delta^{(3)}(\k')\right)\bigg|_{\k'=\k} \,.
\eea
The last term does not contribute for any $k\not=0$, and one obtains the following three contributions
\bea
  P_{w}^{IR,NLO_h}(\eta;k) &=& e^{-\frac12 k_ik_jA^{IR}_{ij}(i\nabla_{k'})} \Bigg\{ \int\limits_{|\p|>{k_S}} d^3p \, 
  \frac{(\k\cdot \p)^2}{p^4}
   \frac{(\k\cdot (\k'-\p))^2}{(\k'-\p)^4} P_{s}(\eta;p)  P_{w}(\eta;|\k'-\p|) \nn\\
  && {} - \frac{(\k\cdot \k')^2}{(k')^4}
  P_{w}(\eta;k') \int_{p>{k_S}} d^3p \, \frac{(\k\cdot \p)^2}{p^4}
   P_{s}(\eta;p) \nn\\
  && {} - \frac{(\k\cdot \k')^2}{(k')^4} 
  P_{s}(\eta;k')\theta(k-{k_S}) \int d^3q \, 
  \frac{(\k\cdot \q)^2}{q^4}
  P_{w}(\eta;q) \Bigg\}\Bigg|_{\k'=\k} \,.
\eea
So far we have not performed the expansion in $\ve$. The leading order in this expansion corresponds to keeping only terms where the derivative operator in
front of the expressions acts directly on the wiggly power spectrum. Therefore, at LO in ${k_S}/k$, one can replace $k'\to k$ except when $k'$ appears
in the argument of $P_w$. In addition, one can express the integrands in terms of SPT kernels, that are in ZA given by
\be
 F_n^{ZA}(\k_1,\dots,\k_n) = \frac{1}{n!}\frac{(\k\cdot \k_1)}{k_1^2}\cdots\frac{(\k\cdot \k_n)}{k_n^2}\,,
\ee
where $\k\equiv \sum_i \k_i$. Altogether, the contribution at LO in $\ve$ and NLO in $\sigma^2_{h}$ is given by
\bea
  P_{w}^{IR,NLO_h}(\eta;k) &=& e^{-\frac12 k_ik_jA^{IR}_{ij}(i\nabla_{k'})} \Bigg\{ 4 \int_{p>{k_S}} d^3p \, \left(F_2^{ZA}(\k-\p,\p)\right)^2 
  P_{s}(\eta;p)  P_{w}(\eta;|\k'-\p|) \nn\\
  && {} - 6 P_{w}( \eta;k')  \int_{p>{k_S}} d^3p \, F_3^{ZA}(\k,\p,-\p) P_{s}(\eta;p) \nn\\
  && {} - 6 P_{s}(\eta;k) \int d^3q \, F_3^{ZA}(\k,\q,-\q) P_{w}(\eta;q) \Bigg\}\Bigg|_{\k'=\k} \,.
\eea
The first term resembles closely the standard expression for 
$P_{22}$
expanded to
first order in the wiggly part $P_w$ when inserting the decomposition $P^{lin}=P_s+P_w$, and evaluated with the IR cutoff ${k_S}$.
The second and third terms correspond to a similar expansion of $2P_{13}$
although one should
note that the last term contains a $q$-integration over all wavenumbers. Furthermore, the derivative operator does not act
on $P_w(q)$, but only on $P_w(|\k'-\p|)$ and on $P_w(k')$. 

The first two lines agree precisely with the corresponding SPT result (\ref{eq:PlrmIRhard1loop}), evaluated in ZA.
The last line contains the wiggly part inside the integrand and is 
therefore strongly suppressed, see Sec.\,\ref{sec:hard}.
Therefore, the exact ZA result agrees 
with both the SPT and TSPT IR resummed power spectra when evaluated in the Zel'dovich approximation.



\begin{thebibliography}{99}

\bibitem{Eisenstein:1998tu}
  D.~J.~Eisenstein, W.~Hu and M.~Tegmark,
  Astrophys.\ J.\  {\bf 504} (1998) L57
  [astro-ph/9805239].

\bibitem{Eisenstein:2005su}
  D.~J.~Eisenstein {\it et al.} [SDSS Collaboration],
  Astrophys.\ J.\  {\bf 633} (2005) 560
  [astro-ph/0501171].

\bibitem{Percival:2009xn}
  W.~J.~Percival {\it et al.} [SDSS Collaboration],
  Mon.\ Not.\ Roy.\ Astron.\ Soc.\  {\bf 401} (2010) 2148
  [arXiv:0907.1660 [astro-ph.CO]].

\bibitem{Anderson:2013zyy}
  L.~Anderson {\it et al.} [BOSS Collaboration],
  Mon.\ Not.\ Roy.\ Astron.\ Soc.\  {\bf 441} (2014) no.1,  24
  [arXiv:1312.4877 [astro-ph.CO]].

\bibitem{Delubac:2014aqe}
  T.~Delubac {\it et al.} [BOSS Collaboration],
  Astron.\ Astrophys.\  {\bf 574} (2015) A59
  [arXiv:1404.1801 [astro-ph.CO]].

\bibitem{Slepian:2015hca}
  Z.~Slepian {\it et al.},
  ``The large-scale 3-point correlation function of the SDSS BOSS DR12 CMASS galaxies,''
  arXiv:1512.02231 [astro-ph.CO].

\bibitem{Bernardeau:2001qr}
  F.~Bernardeau, S.~Colombi, E.~Gaztanaga and R.~Scoccimarro,
  Phys.\ Rept.\  {\bf 367} (2002) 1
  [astro-ph/0112551].

\bibitem{Baldauf:2015xfa}
  T.~Baldauf, M.~Mirbabayi, M.~Simonovic and M.~Zaldarriaga,
  Phys.\ Rev.\ D {\bf 92} (2015) 4,  043514
  [arXiv:1504.04366 [astro-ph.CO]].

\bibitem{Crocce:2007dt}
  M.~Crocce and R.~Scoccimarro,
  Phys.\ Rev.\ D {\bf 77} (2008) 023533
  [arXiv:0704.2783 [astro-ph]].


\bibitem{Eisenstein:2006nk}
  D.~J.~Eisenstein, H.~j.~Seo, E.~Sirko and D.~Spergel,
  Astrophys.\ J.\  {\bf 664} (2007) 675
  [astro-ph/0604362].

\bibitem{Eisenstein:2006nj}
  D.~J.~Eisenstein, H.~j.~Seo and M.~J.~White,
  Astrophys.\ J.\  {\bf 664} (2007) 660
  [astro-ph/0604361].

\bibitem{Blas:2013bpa} 
  D.~Blas, M.~Garny and T.~Konstandin,
  JCAP {\bf 1309}, 024 (2013)
  [arXiv:1304.1546 [astro-ph.CO]].
 
\bibitem{Scoccimarro:1995if} 
  R.~Scoccimarro and J.~Frieman,
  Astrophys.\ J.\ Suppl.\  {\bf 105}, 37 (1996)
  [astro-ph/9509047].

 

\bibitem{Jain:1995kx}
  B.~Jain and E.~Bertschinger,
  Astrophys.\ J.\  {\bf 456} (1996) 43
  [astro-ph/9503025].

\bibitem{Kehagias:2013yd}
  A.~Kehagias and A.~Riotto,
  Nucl.\ Phys.\ B {\bf 873} (2013) 514
  [arXiv:1302.0130 [astro-ph.CO]].
  
\bibitem{Peloso:2013zw}
  M.~Peloso and M.~Pietroni,
  JCAP {\bf 1305} (2013) 031
  [arXiv:1302.0223 [astro-ph.CO]].

\bibitem{Sugiyama:2013gza}
  N.~S.~Sugiyama and D.~N.~Spergel,
  JCAP {\bf 1402} (2014) 042
  [arXiv:1306.6660 [astro-ph.CO]].
  
\bibitem{Senatore:2014via} 
  L.~Senatore and M.~Zaldarriaga,
  JCAP {\bf 1502}, no. 02, 013 (2015)
  [arXiv:1404.5954 [astro-ph.CO]].
 
\bibitem{Burden:2014cwa}
  A.~Burden, W.~J.~Percival, M.~Manera, A.~J.~Cuesta, M.~V.~Magana and S.~Ho,
  Mon.\ Not.\ Roy.\ Astron.\ Soc.\  {\bf 445} (2014) 3,  3152
  [arXiv:1408.1348 [astro-ph.CO]].
  
\bibitem{Noh:2009bb}
  Y.~Noh, M.~White and N.~Padmanabhan,
  Phys.\ Rev.\ D {\bf 80} (2009) 123501
  [arXiv:0909.1802 [astro-ph.CO]].

\bibitem{Tassev:2012hu}
  S.~Tassev and M.~Zaldarriaga,
  JCAP {\bf 1210} (2012) 006
  [arXiv:1203.6066 [astro-ph.CO]].
  
\bibitem{Matsubara:2007wj}
  T.~Matsubara,
  Phys.\ Rev.\ D {\bf 77} (2008) 063530
  [arXiv:0711.2521 [astro-ph]].


\bibitem{Padmanabhan:2008dd}
  N.~Padmanabhan, M.~White and J.~D.~Cohn,
  Phys.\ Rev.\ D {\bf 79} (2009) 063523
  [arXiv:0812.2905 [astro-ph]].
 
 
\bibitem{Blas:2015qsi} 
  D.~Blas, M.~Garny, M.~M.~Ivanov and S.~Sibiryakov,
  arXiv:1512.05807 [astro-ph.CO].

\bibitem{Kim:2011ab}
  J.~Kim, C.~Park, G.~Rossi, S.~M.~Lee and J.~R.~Gott, III,
  J.\ Korean Astron.\ Soc.\  {\bf 44} (2011) issue,  217
  [arXiv:1112.1754 [astro-ph.CO]].

\bibitem{Gorbunov:2011zzc} 
  D.~S.~Gorbunov and V.~A.~Rubakov,
  Hackensack, USA: World Scientific (2011) 489 p

\bibitem{Seo:2007ns} 
  H.~J.~Seo and D.~J.~Eisenstein,
  Astrophys.\ J.\  {\bf 665}, 14 (2007)
  [astro-ph/0701079].
  
\bibitem{Peebles:1980}
  P.J.E.~Peebles,
  ``The Large-scale Structure of the Universe,''
  Princeton University Press (1980).
 
  
\bibitem{Goroff:1986ep} 
  M.~H.~Goroff, B.~Grinstein, S.~J.~Rey and M.~B.~Wise,
  Astrophys.\ J.\  {\bf 311}, 6 (1986).
  
\bibitem{Blas:2014hya} 
  D.~Blas, M.~Garny, T.~Konstandin and J.~Lesgourgues,
  JCAP {\bf 1411}, no. 11, 039 (2014)
  [arXiv:1408.2995 [astro-ph.CO]].
  
  
 \bibitem{CrSc1}
  M.~Crocce and R.~Scoccimarro,
  Phys.\ Rev.\  D {\bf 73} (2006) 063519
  [arXiv:astro-ph/0509418].
 
\bibitem{Blas:2011rf}
  D.~Blas, J.~Lesgourgues and T.~Tram,
  JCAP {\bf 1107} (2011) 034
  [arXiv:1104.2933 [astro-ph.CO]].

\bibitem{Eisenstein:1997jh}
   D.~J.~Eisenstein and W.~Hu,
   Astrophys.\ J.\  {\bf 511} (1997) 5
   [astro-ph/9710252].
   
\bibitem{Anselmi:2012cn} 
  S.~Anselmi and M.~Pietroni,
  JCAP {\bf 1212}, 013 (2012)
  [arXiv:1205.2235 [astro-ph.CO]].
  
\bibitem{Tassev:2013zua} 
  S.~Tassev,
  JCAP {\bf 1406}, 012 (2014)
  [arXiv:1311.6316 [astro-ph.CO]].

\bibitem{Bellini:2015oua}
  E.~Bellini and M.~Zumalacarregui,
  Phys.\ Rev.\ D {\bf 92} (2015) 6,  063522
  [arXiv:1505.03839 [astro-ph.CO]].

\bibitem{Weinberg:2012es} 
  D.~H.~Weinberg, M.~J.~Mortonson, D.~J.~Eisenstein, C.~Hirata, A.~G.~Riess and E.~Rozo,
  Phys.\ Rept.\  {\bf 530}, 87 (2013)
  [arXiv:1201.2434 [astro-ph.CO]].

\bibitem{Smith:2007gi} 
  R.~E.~Smith, R.~Scoccimarro and R.~K.~Sheth,
  Phys.\ Rev.\ D {\bf 77}, 043525 (2008)
  [astro-ph/0703620 [ASTRO-PH]].

\bibitem{Seo:2009fp} 
  H.~J.~Seo {\it et al.},
  Astrophys.\ J.\  {\bf 720}, 1650 (2010)
  [arXiv:0910.5005 [astro-ph.CO]].
  
\bibitem{Mehta:2011xf}
  K.~T.~Mehta, H.~J.~Seo, J.~Eckel, D.~J.~Eisenstein, M.~Metchnik, P.~Pinto and X.~Xu,
  Astrophys.\ J.\  {\bf 734} (2011) 94
  [arXiv:1104.1178 [astro-ph.CO]].
  


\bibitem{Prada:2014bra} 
  F.~Prada, C.~G.~Scoccola, C.~H.~Chuang, G.~Yepes, A.~A.~Klypin, F.~S.~Kitaura and S.~Gottlober,
  arXiv:1410.4684 [astro-ph.CO].

 
\bibitem{Sherwin:2012nh}
  B.~D.~Sherwin and M.~Zaldarriaga,
  Phys.\ Rev.\ D {\bf 85} (2012) 103523
  [arXiv:1202.3998 [astro-ph.CO]].
 
\bibitem{Padmanabhan:2009yr} 
  N.~Padmanabhan and M.~White,
  Phys.\ Rev.\ D {\bf 80}, 063508 (2009)
  [arXiv:0906.1198 [astro-ph.CO]].

\bibitem{Xu:2010tf} 
  X.~Xu {\it et al.},
  Astrophys.\ J.\  {\bf 718}, 1224 (2010)
  [arXiv:1001.2324 [astro-ph.CO]].

\bibitem{Matsubara:2016wth}
  T.~Matsubara and V.~Desjacques,
  arXiv:1604.06579 [astro-ph.CO].
  
\bibitem{ZaldaICTP}   
M.~Zaldarriaga,  in the {\it First ICTP Advanced School on Cosmology}, 2015. 
\url{http://indico.ictp.it/event/a14277/}.
  
\bibitem{Blas:2013aba} 
  D.~Blas, M.~Garny and T.~Konstandin,
  JCAP {\bf 1401}, no. 01, 010 (2014)
  [arXiv:1309.3308 [astro-ph.CO]].
  
\bibitem{Sugiyama:2013pwa}
  N.~S.~Sugiyama and T.~Futamase,
  Astrophys.\ J.\  {\bf 769} (2013) 106
  [arXiv:1303.2748 [astro-ph.CO]].



  
\end{thebibliography}
\end{document}